\documentclass{aa}  
\usepackage{xcolor}
\usepackage{graphicx}
\usepackage{longtable}
\usepackage{amsmath}
\usepackage{mwe}
\usepackage{fontawesome}
\usepackage{lscape}

\definecolor{dgreen}{rgb}{0.0, 0.5, 0.0}
\usepackage{txfonts}
\usepackage{hyperref}
\hypersetup{
    colorlinks=true,
    linkcolor=blue,
    filecolor=blue,
    citecolor=blue,
    urlcolor=blue,
}

\begin{document}

   \title{LIDA - The Leiden Ice Database for Astrochemistry}

   \author{W. R. M. Rocha\inst{1,3},
          M. G. Rachid\inst{1},
          B. Olsthoorn\inst{2},
          E. F. van Dishoeck\inst{3},
          M. K. McClure\inst{3},
          \and
          H. Linnartz
          \inst{1}
          }

   \institute{Laboratory for Astrophysics, Leiden Observatory, Leiden University, P.O. Box 9513, NL 2300 RA Leiden, The Netherlands.\\
    \email{rocha@strw.leidenuniv.nl}
          \and
             Nordita, KTH Royal Institute of Technology and Stockholm University, Hannes Alfv{\'e}ns v{\"a}g 12, SE-114 21 Stockholm, Sweden
         \and
             Leiden Observatory, Leiden University, PO Box 9513, NL 2300 RA Leiden, The Netherlands
             }

   \date{Received ZZZZ; accepted YYYY}

 
  \abstract
   {High quality vibrational spectra of solid-phase molecules in ice mixtures and for temperatures of astrophysical relevance are needed to interpret infrared observations toward protostars and background stars. Such data are collected worldwide, by several laboratory groups, in support of existing and upcoming astronomical observations. Over the last 25 years, the Laboratory for Astrophysics at Leiden Observatory has provided more than 1100 (high resolution) spectra of diverse ice samples.}
   {Timely with the recent launch of the {\it James Webb} Space Telescope, we have fully upgraded the Leiden Ice Database for Astrochemistry (LIDA) adding recently measured spectra. The goal of this manuscript is to describe what options exist to get access to and work with a large collection of IR spectra, and the UV/vis to mid-infrared refractive index of H$_2$O ice. This also includes astronomy-oriented online tools to support the interpretation of IR ice observations.}
   {LIDA is based on open-source Python software, such as {\texttt{Flask}} and {\texttt{Bokeh}} for generating the web pages and graph visualization, respectively, Structured Query Language (SQL) for searching ice analogues within the database and {\texttt{Jmol}} for three-dimensional molecule visualization. The database provides the vibrational modes of molecules known and expected to exist as ice in space. These modes are characterized by using density functional theory with the \texttt{ORCA} software. The infrared data in the database are recorded via transmission spectroscopy of ice films condensed on cryogenic substrates. The real UV/vis refractive indices of H$_2$O ice are derived from interference fringes created from the simultaneous use of a monochromatic HeNe laser beam and a broadband Xe-arc lamp, whereas the real and imaginary mid-IR values are theoretically calculated. LIDA not only provides information on fundamental ice properties but also offers online tools. The first tool, SPECFY, is directly linked to the data in the database to create a synthetic spectrum of ices towards protostars. The second tool allows the upload of external files and the calculation of mid-infrared refractive index values.}
   {LIDA provides an open-access and user-friendly platform to search, download and visualize experimental data of astrophysically relevant molecules in the solid-phase, as well as to provide the means to support astronomical observations, in particular, those that will be obtained with the {\it James Webb} Space Telescope. As an example, we analyse the spectrum of the protostar AFGL~989 using the resources available in LIDA and derive the column densities of H$_2$O, CO and CO$_2$ ices.}
   {}

\authorrunning{Rocha et al.}
   \maketitle
%

\section{Introduction}

Infrared (IR) spectroscopy is a diagnostic tool used to characterize chemical structures of molecules, and distinguish their functional groups \citep[e.g.,][]{Coblentz1905, BALKANSKI1989729}. For this reason a number of laboratories around the world have been focusing on providing laboratory based IR data of interstellar ice analogues for a range of different ice compositions and temperatures \citep[e.g.,][]{Hagen1979, Strazzulla1984, Schmitt1989, Grim1989, Hudgins1993, Boudin1998, Palumbo1998, Schutte1999, Caro2002, Oberg2009, Pilling2010, Vinogradoff2015, Scheltinga2018, Urso2020, Scheltinga2021, Rachid2020, Rachid2021, Potapov2021}. IR spectra directly represent the molecular geometry of a molecule and as such can act as a molecular fingerprint. In the gas phase and at very high resolution, such rovibrationally resolved spectra are unique, although overlap may still occur. In the solid state, however, interactions with the ice matrix prohibit molecules to (freely) rotate, and cause spectra to broaden and shift with respect to the unperturbed gas phase value. Additionally, spectral overlaps are more common. The amount of broadening and shifting depends on ice composition (both ice constituents and concentration) and ice temperature, as well as other parameters such as the level of ice porosity. In dedicated laboratory studies all these parameters can be derived under fully controlled conditions. Examples can be found in \citet{Oberg2007}.

IR spectroscopy is also the technique widely used to detect solid-phase molecules in the interstellar medium  \citep[ISM, e.g.,][]{Gillett1973, Schutte1996, Pontoppidan2003, Gibb2004, Boogert2008, Zasowski2009, Bottinelli2010, Boogert2013,  Penteado2015, Perotti2020, Rocha2021, Onaka2021}. The light of a protostar, edge-on disks or background star passes through the circumstellar material, and absorption features in the IR are seen in the protostellar spectral energy distribution (SED). The correct interpretation of those absorption bands is only possible upon comparison with the spectra of ice analogues measured in the laboratory. With this methodology, important discoveries have been done through observations of space- and ground-based telescopes, such as the {\it Infrared Space Observatory} (ISO), the {\it Spitzer} Space Telescope/Infrared Spectrograph (IRS) and the Infrared Spectrometer And Array Camera mounted on the Very Large Telescope (VLT/ISAAC). Up to date, the molecules securely identified in ices are H$_2$O, CO, CO$_2$, NH$_3$, CH$_4$ and CH$_3$OH \citep{Oberg2011, Boogert2015}, and the isotopologues $^{13}$CO and $^{13}$CO$_2$ \citep{Boogert2002_isotoplogue}. Except in the cases of CO and $^{13}$CO, which have only one vibrational mode, these molecules were identified in astrophysical ices by the detection of multiple absorption bands across the IR spectrum. These identifications in ices allowed to study the solid-phase chemistry in different astrophysical environments. For example, amorphous water ice is predominantly found towards background stars, and low-mass protostars \citep[][]{Smith1989, Boogert2008}, whereas some fraction of crystalline water ice was found in the circumstellar material of high-mass protostars \citep{Dartois2002}. CO is also an important discriminator of the ice environment, and astronomical observations indicate that it does not only exist in the pure form, but can also be mixed with CO$_2$, H$_2$O or CH$_3$OH \citep[e.g.,][]{Pontoppidan2003, Cuppen2011}. In the case of CO$_2$ ice, the bending mode around 15~$\mu$m provides a diagnostic of heating and segregation of polar and apolar molecules in ices \citep[e.g.,][]{Ehrenfreund1996, Pontoppidan2008, Isokoski2013}. Among the list of molecules identified in ices, CH$_3$OH (methanol) belongs to the group of the so-called complex organic molecules (COMs), which in astrochemistry is defined as organic molecules containing six or more atoms \citep[e.g., C$_x$H$_y$Y$_z$, with Y = O, N, P, S;][]{Herbst2009}. A number of small molecules have been tentatively identified in ices, for which only one vibrational mode could be assigned from astronomical observations. This list also includes sulfur-bearing molecules \citep[notably, SO$_2$;][]{Boogert1997} and ions \citep[notably, OCN$^-$;][]{Schutte1997}.

Many different COMs have been identified in the gas phase through radio and submillimeter surveys \citep[e.g.,][]{Blake1987, Jorgensen2012, McGuire2016, Belloche2020, vanGelder2020, McGuire2021, Jorgensen2020, Nazari2021, Rivilla2021, Brunken2022}, but astronomical observations have not been able to unambiguously identify frozen COMs larger than CH$_3$OH due to low spectral resolution or sensitivity. Nevertheless, tentative detections of CH$_3$CHO (acetaldehyde) and CH$_3$CH$_2$OH (ethanol) ice have been reported in the literature \citep{Schutte1999_weak, Oberg2011, Scheltinga2018, Rocha2015, Rocha2021}. Consistent with these tentative detections, several laboratory experiments have shown that such molecules can be formed in ices. Some examples are interstellar ice analogues processed by UV radiation \citep[e.g.,][]{Bernstein1995, MunozCaro2003, Oberg2009, Meinert2016, Oberg2016, Nuevo2018, Ishibashi2021, Bulak2021}, electron bombardment \citep[e.g.,][]{Brown1982, Materese2015, Mifsud2021}, X-rays \citep[e.g.,][]{Pilling2015, Ciaravella2019}, cosmic-rays \citep[e.g.,][]{Hudson2001, Domaracka2010, Pilling2010}, and via thermal processing \citep[e.g.,][]{Danger2011, theule2013thermal}. Other mechanisms excluding the presence of energetic triggers, such as atom addition reactions that are more representative for dark clouds conditions, have also been shown to result in the formation of COMs \citep{Watanabe2002, Fuchs2009, theule2013thermal, Linnartz2015, Fedoseev2017, Ioppolo2021}.

Apart from IR spectroscopy, the complex refractive index (CRI) of ice samples is important for the interpretation of astronomical observations. CRI is given by a complex number, $\tilde{m} = n + ik$, where $n$ and $k$, are the real and imaginary parts and are associated with scattering and absorption effects, respectively. In protostellar environments, CRI has been used to evaluate the effect of icy grain sizes and shapes in the spectral features of ices \citep[e.g.,][]{Ehrenfreund1997, Boogert2002, Pontoppidan2005, Boogert2008, Rocha2015, Perotti2020, Dartois2022}. For example, \citet{Boogert2008} observed a dependence of the libration mode of H$_2$O ice peak position with the size of spherical grains. Better fits of this band are obtained when small grains are adopted in the models. Similarly, CRI values have been used to interpret the absorption band at 3~$\mu$m, associated to the O$-$H stretching mode of H$_2$O \citep[e.g.,][]{Smith1989, Dartois2001}. In the solar system, the CRI also play a crucial role in the simulation of reflected light due to icy surfaces to interpret spectral observations. \citep[e.g.,][]{Clark2012, dalle2015}. And, finally, the CRI may be used to construct opacities for a dust grain size distribution model \citep{Weingartner2001}, which can be used with a radiative transfer code to calculate self-consistently the temperature and density distributions of dusty astronomical objects, e.g. protoplanetary disks \citep{Dalessio2006}.

The advances in the identification of molecules in both gas and solid-phase have been strongly supported by atomic and molecular data in open-access databases. In fact, electronic databases have become an essential tool in the context of astrochemistry, given the large amount of data that is produced by laboratory experiments. In particular, the astrochemical community targeting gas-phase chemical species is well served with multiple databases. For example, the Cologne Database for Molecular Spectroscopy\footnote{\url{https://cdms.astro.uni-koeln.de/}} \citep[CDMS;][]{Muller2001, Muller2005, Endres2016} and the Jet Propulsion Laboratory\footnote{\url{https://spec.jpl.nasa.gov/}} \citep[JPL;][]{Pickett1998, Pearson2010} databases provide catalogues with transition frequencies, energy levels and line strengths for atoms and molecules in the gas-phase of astrophysical and atmospheric interest. Collisional rate coefficients are available through the Leiden Atomic and Molecular Database (LAMDA)\footnote{\url{https://home.strw.leidenuniv.nl/~moldata/}} for non-LTE excitation \citep{Schoier2005, Tak2020}. Similarly, BASECOL contains a repository of collisional ro-vibrational excitation data of molecules by colliding with different agents such as atoms, ions, molecules or electrons \citep{Dubernet2006, Dubernet2013}. More oriented to chemical reactions, the UMIST Database for Astrochemistry\footnote{\url{http://udfa.ajmarkwick.net/}} \citep[UDfA;][]{McElroy2013} contains the reaction rates of more than 6000 gas-phase reactions. In a similar vein, the Kinetic Database for Astrochemistry\footnote{\url{https://kida.astrochem-tools.org/}} \citep[KIDA;][]{Wakelam2012} has provided reaction rate coefficients for a massive number of chemical species for astrochemical studies. The photodissociation and photoionization values of gas-phase molecules relevant for astrophysics are available online\footnote{\url{https://home.strw.leidenuniv.nl/~ewine/photo/index.html}}, and described by \citet{Heays2017, vanDishoeck2006, vanDishoeck1988}. The properties of gas-phase polycyclic aromatic hydrocarbons (PAHs) are widely available through the NASA Ames PAH IR Spectroscopy Database\footnote{\url{https://www.astrochemistry.org/pahdb/}} \citep{Bauschlicher2010, Boersma2014, Mattioda2020}. 

The astrochemistry community working with solid-phase materials has also been served with databases. The refractive index of refractory materials is available via the Database of Optical Constants for Cosmic Dust\footnote{\url{https://www.astro.uni-jena.de/Laboratory/OCDB/index.html}} \citep{Henning1999, Jager2003}. Likewise, IR spectra of binary ice mixtures and refractive indexes of pure ices can be found at the webpage of the Cosmic Ice Laboratory\footnote{\url{https://science.gsfc.nasa.gov/691/cosmicice/}} from NASA \citep[e.g.,][]{Moore2010, Knez2012, Gerakines2020} and at Databases of the Astrophysics \& Astrochemistry Laboratory\footnote{\url{http://www.astrochem.org/databases.php}} that contain measurements by \citet{Hudgins1993}. A database of refractive indices of ice samples irradiated by heavy ions is also available at LASA (Laboratório de Astroquímica e Astrobiologia da Univap) webpage\footnote{\url{https://www1.univap.br/gaa/nkabs-database/data.htm}} with calculations performed by \citet{Rocha2014}, \citet{Rocha2018}, and \citet{Rocha2020}. Infrared refractive indices of CO and CO$_2$ ices are available from the Experimental Astrophysics Laboratory at Catania Astrophysical Observatory website\footnote{\url{http://www.ct.astro.it/lasp/optico.html}} \citep{Baratta1998}. Finally, we also mention the Solid Spectroscopy Hosting Architecture of Databases and Expertise\footnote{\url{https://www.sshade.eu/}} \citep[SSHADE;][]{Schmitt2018}, that contains a compilation of spectral and photometric data obtained by various spectroscopic techniques over the whole electromagnetic spectrum from gamma to radio wavelengths, through X-rays, UV, Vis, IR, and millimeter ranges. The data are not limited to ices, but also contain measurements of liquids, minerals, rocks, organic and carbonaceous materials. 

Similarly to many of the databases mentioned above, the Leiden Database for Ices has served the astronomical community since the '90s, but until recently no COM spectra were included, and the spectral resolution of the data was around 1$-$2~cm$^{-1}$ \citep[e.g.,][]{Gerakines1996, Ehrenfreund1996, Ehrenfreund1997}. Additionally, the data were fragmented into several databases targeting specific ice samples. To continue supporting the interpretation of ice observations with current and future telescopes, in particular the {\it James Webb} Space Telescope (JWST), we have fully upgraded the Leiden Ice Database for Astrochemistry (LIDA; \url{https://icedb.strw.leidenuniv.nl/}). In particular, LIDA is a deliverable of the Early Release Science program ICE AGE\footnote{\url{http://jwst-iceage.org/}} (PI: Melissa McClure; Co-PIs: Adwin Boogert, Harold Linnartz). In LIDA all data are now available at one central location, and appealing features are included, such as a search capability and dynamic data visualization. Additionally, online tools are included in LIDA to support JWST data analysis or to prepare observing blocks, by deriving integration times based on expected column densities. LIDA covers the most abundant solid-phase species observed in the ISM, which are listed in Table~\ref{icedb_list}, along with information about their secure, tentative or non-identification in the solid-phase in astrophysical environments from previous observations. JWST has the technical potential to enlarge the inventory of ice identifications in space, and several programs (ERS, Guaranteed Time Observations - GTO and General Observer - GO) will search for new ice features toward protostars and background stars, using high spatial and spectral resolution observing modes. Moreover, JWST will shed light on the conundrum of the formation of COMs in ices. For this purpose, comparison with spectra of COMs in astrophysically relevant ice matrices at high spectral resolution is needed. Such data are required for a range of different physical conditions, e.g., mixing ratios, temperatures and porosity levels, as these differences affect the spectral appearance of the ice absorption bands. 

This manuscript systematically guides (new) users through LIDA. Section~\ref{data_db} and Appendix give an overview of the data available in the 2022 version of LIDA and describe the type of data and how these were achieved. Section~\ref{struct_db} provides information about the database structure, namely, the relational design, web interface, and visualization tools. In Section~\ref{on_tools} we introduce the computational online tools dedicated to supporting JWST data analysis. An application ilustrating the potential of LIDA is also shown. Section~\ref{future} points out the upgrades on LIDA that are intended for the next years. A summary of this work is provided in Section~\ref{summary}. 



\begin{table*}
\caption{\label{icedb_list} List of molecules with relevant data on LIDA and their solid-phase (tentative or non) detection in the ISM.}
\renewcommand{\arraystretch}{1.0}
\centering 
\begin{tabular}{lllccccc}
\hline\hline
Chemical & Chemical & Name & & Notes on LIDA & & & Solid-phase$^{b,c}$\\
\cline{4-7}
structure$^a$ & formula & & IR spectrum & UV/vis-mid IR & Heating & UV irr. & detection/Ref.\\
\hline
& & & &\\
\vspace{0.1cm}
\raisebox{-.5\height}{\includegraphics[height=0.25in]{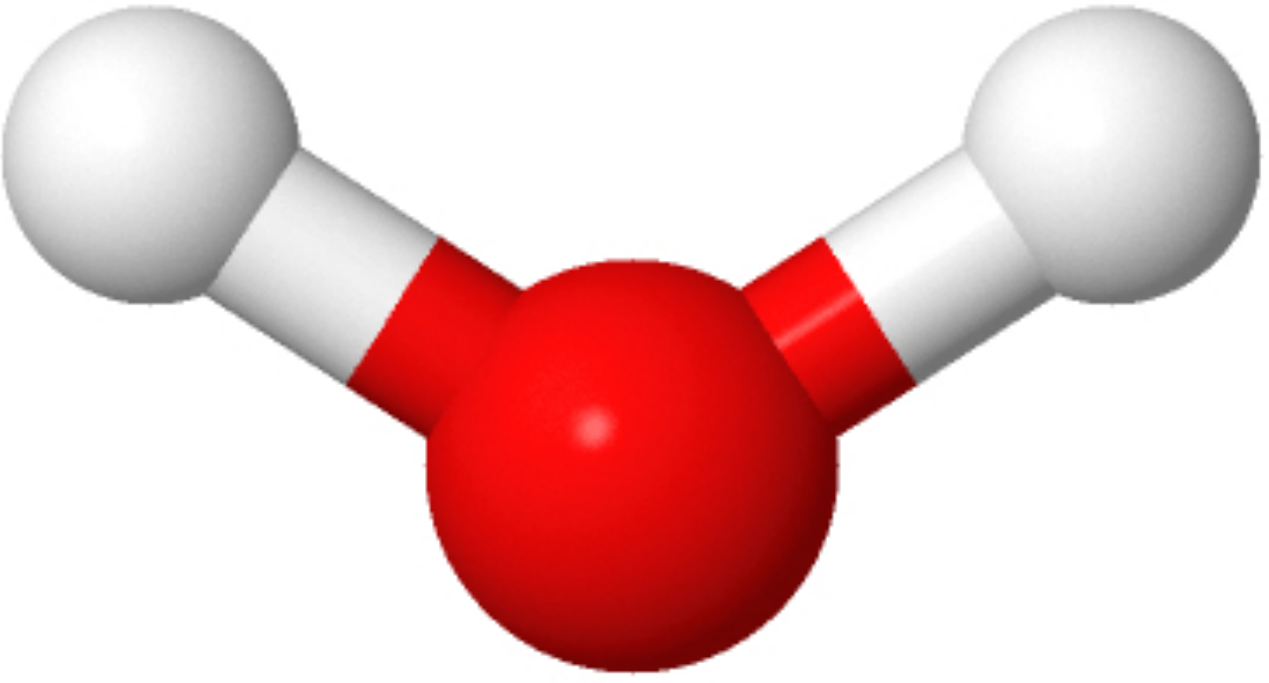}} & H$_2$O & Water & yes & yes & yes & yes & \textcolor{dgreen}\faCheckCircle  $\hspace{0.1cm}$ / [1]\\
\raisebox{-.5\height}{\includegraphics[height=0.25in]{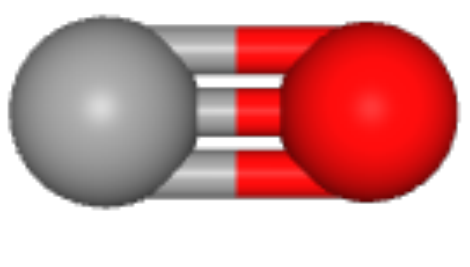}} & CO & Carbon monoxide & yes & no & yes & yes & \textcolor{dgreen}\faCheckCircle  $\hspace{0.1cm}$ / [2]\\
\raisebox{-.5\height}{\includegraphics[height=0.25in]{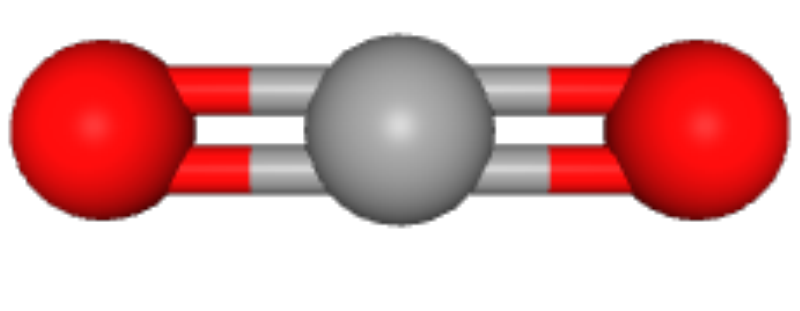}} & CO$_2$ & Carbon dioxide & yes & no & yes & yes & \textcolor{dgreen}\faCheckCircle  $\hspace{0.1cm}$ / [3]\\
\raisebox{-.5\height}{\includegraphics[height=0.45in]{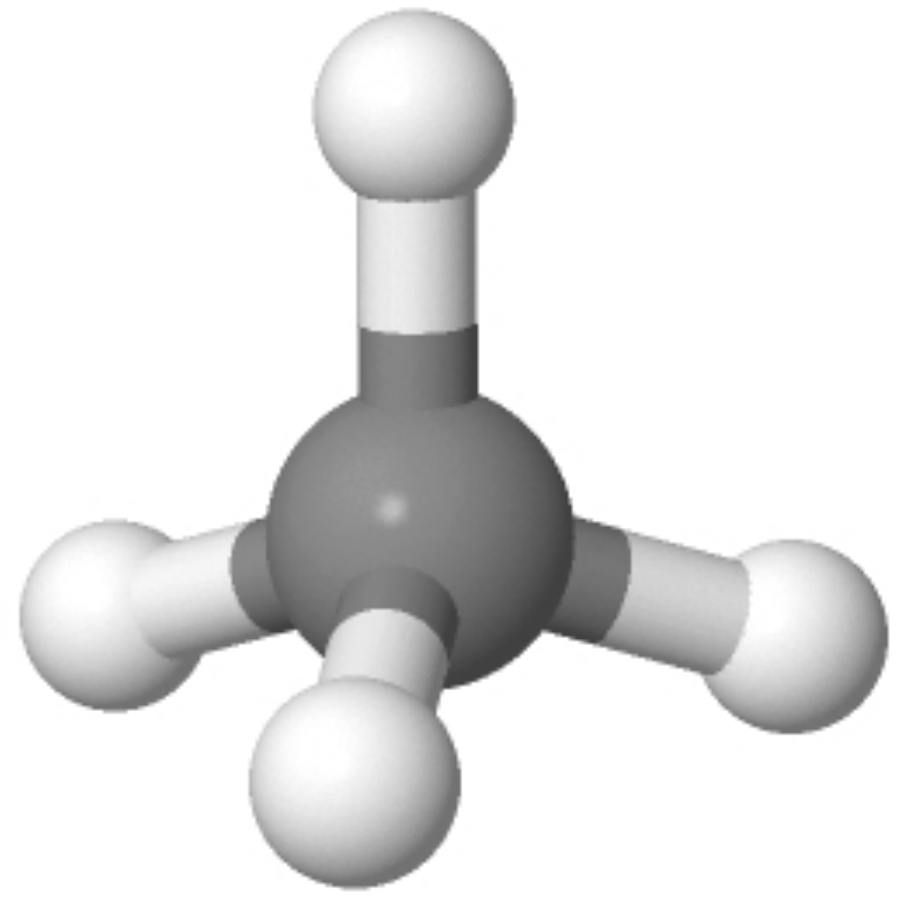}} & CH$_4$ & Methane & yes & no & yes & yes & \textcolor{dgreen}\faCheckCircle  $\hspace{0.1cm}$ / [4]\\
\raisebox{-.5\height}{\includegraphics[height=0.35in]{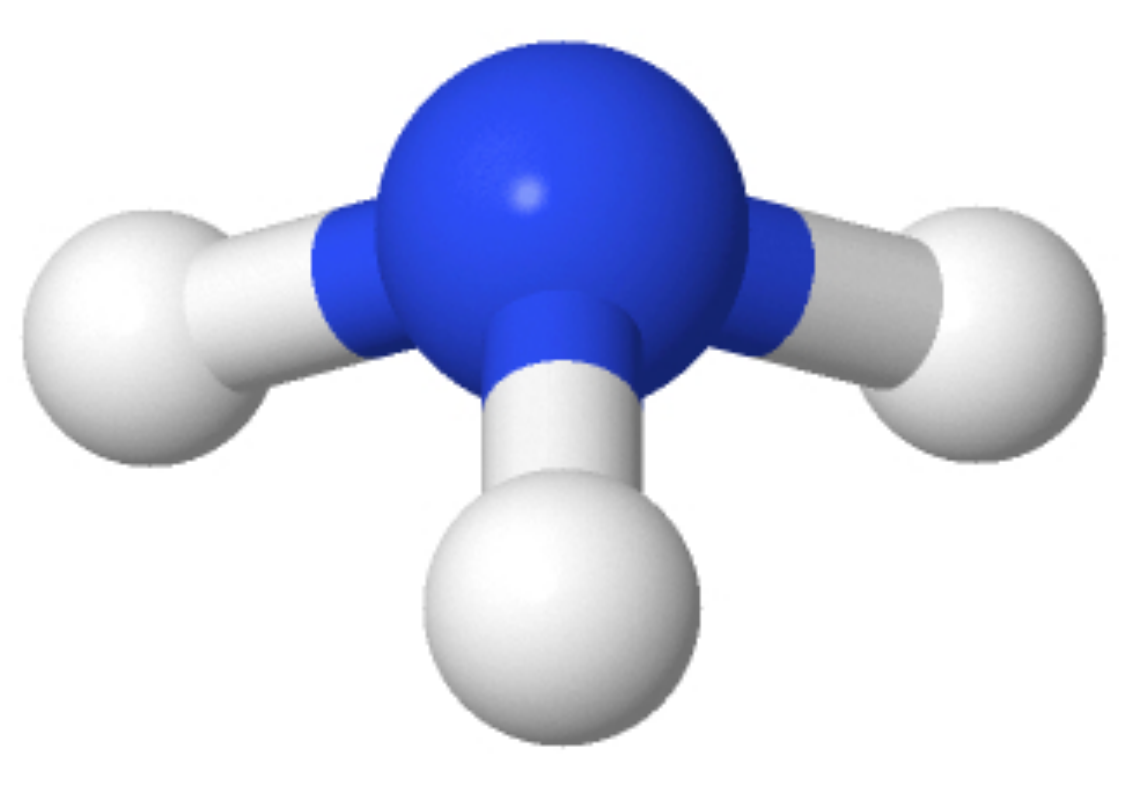}} & NH$_3$ & Ammonia & yes & no & yes & yes & \textcolor{dgreen}\faCheckCircle  $\hspace{0.1cm}$ / [5]\\
\raisebox{-.5\height}{\includegraphics[height=0.5in]{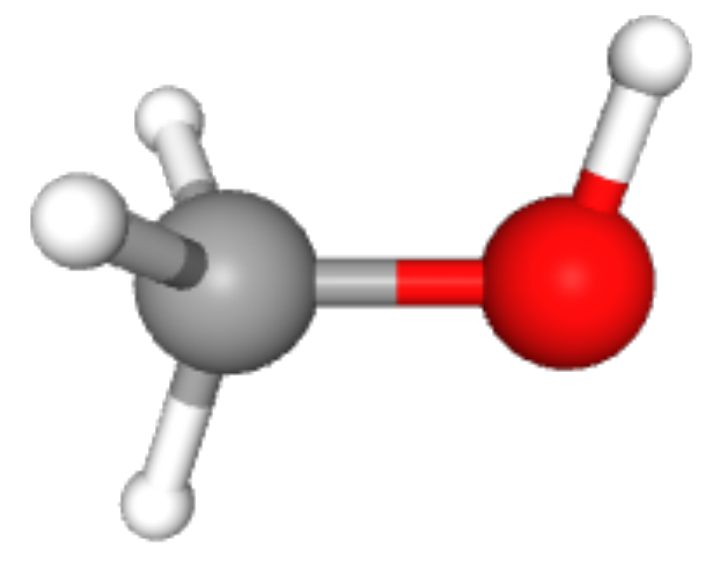}} & CH$_3$OH  & Methanol & yes & no & yes & yes & \textcolor{dgreen}\faCheckCircle  $\hspace{0.1cm}$ / [6]\\
\raisebox{-.5\height}{\includegraphics[height=0.52in]{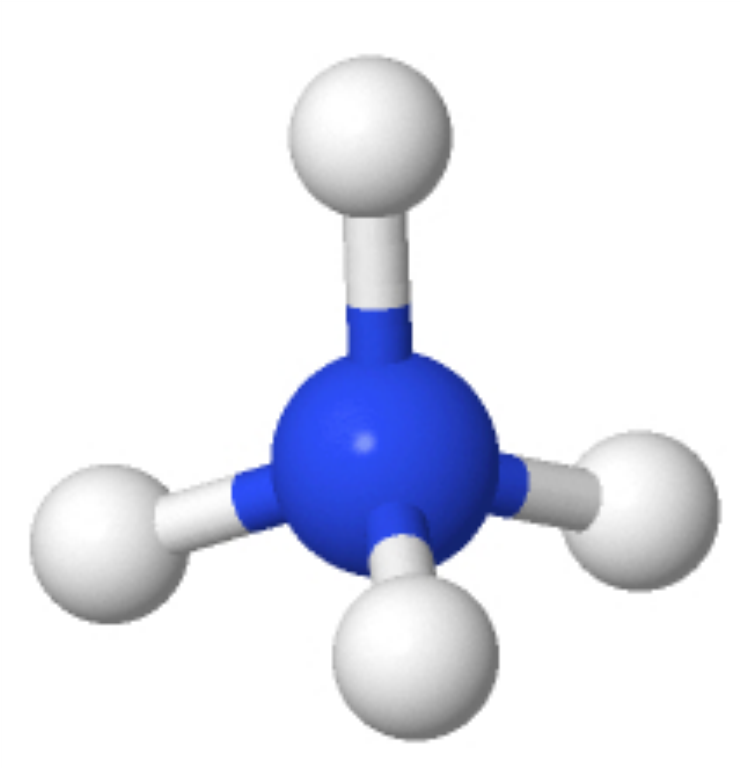}} & NH$_4^+$  & Ammonium ion & yes & no & no & no & \textcolor{orange}\faExclamationTriangle $\hspace{0.1cm}$ / [7]\\
\raisebox{-.5\height}{\includegraphics[height=0.55in]{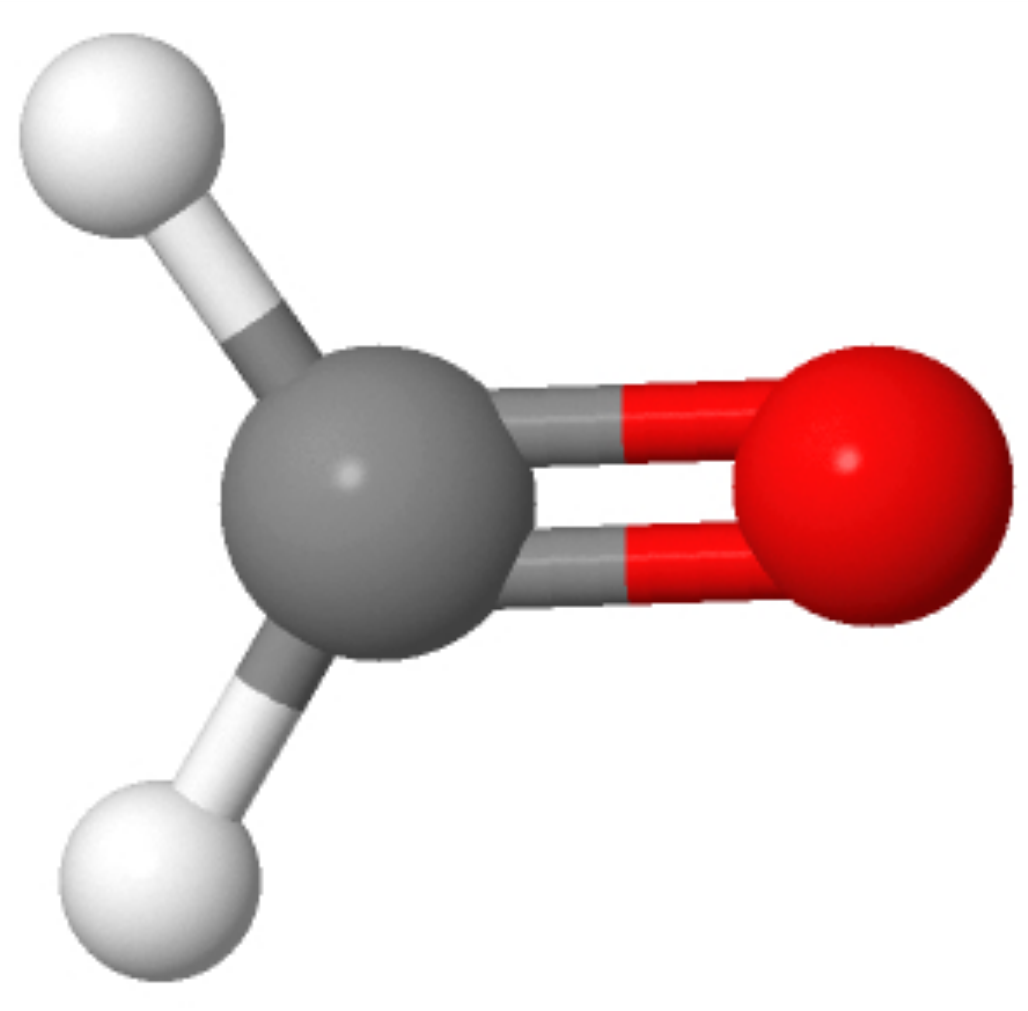}} & H$_2$CO  & Formaldehyde & yes & no & no & no & \textcolor{orange}\faExclamationTriangle $\hspace{0.1cm}$ / [8]\\
\raisebox{-.5\height}{\includegraphics[height=0.25in]{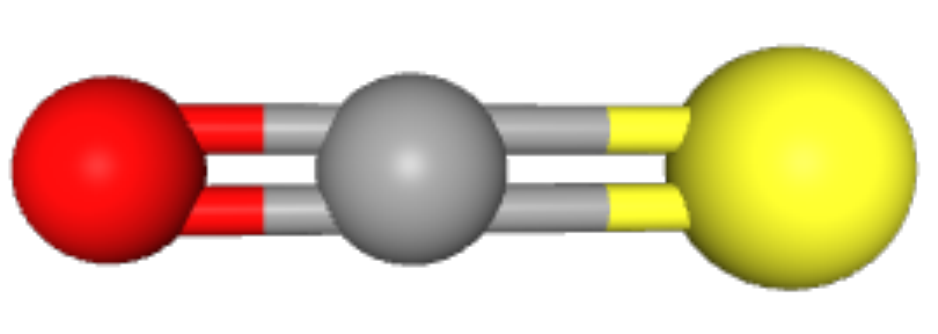}} & OCS  & Carbonyl sulfide & yes & no & yes & no & \textcolor{orange}\faExclamationTriangle $\hspace{0.1cm}$ / [9]\\
\raisebox{-.5\height}{\includegraphics[height=0.3in]{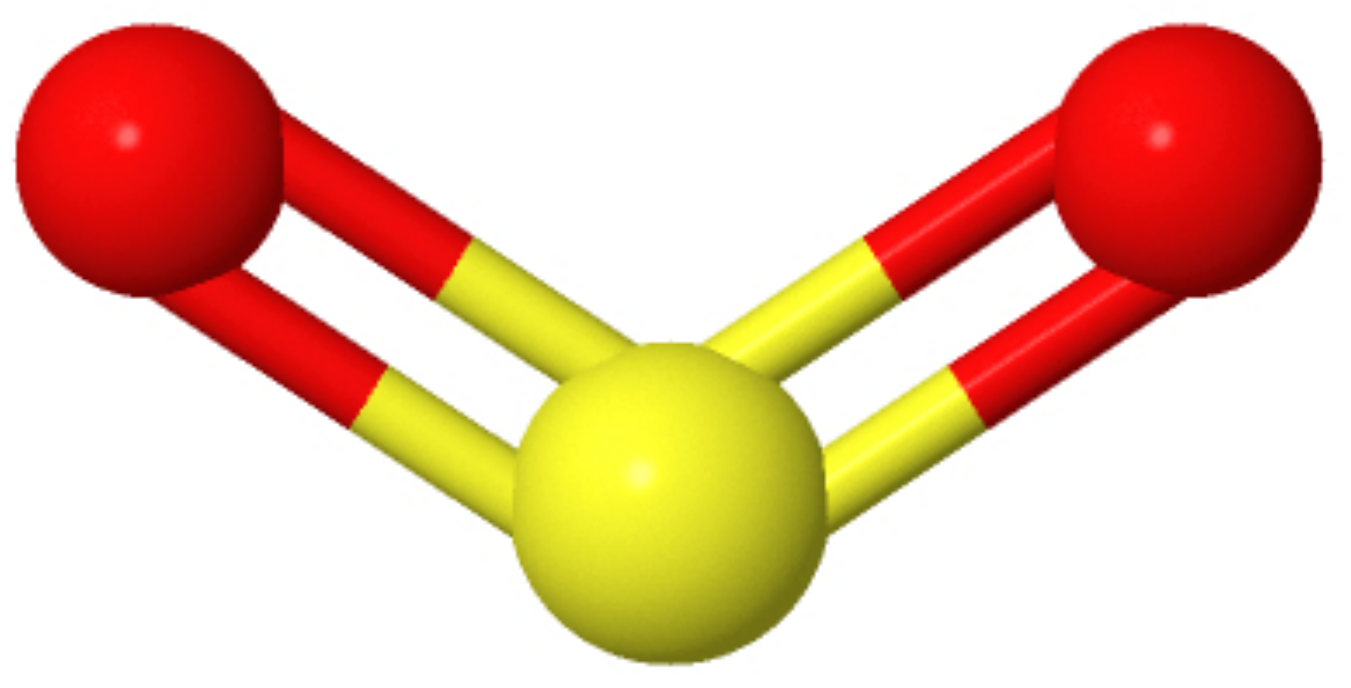}} & SO$_2$  & Sulfur dioxide & yes & no & no & no & \textcolor{orange}\faExclamationTriangle $\hspace{0.1cm}$ / [10]\\
\raisebox{-.5\height}{\includegraphics[height=0.23in]{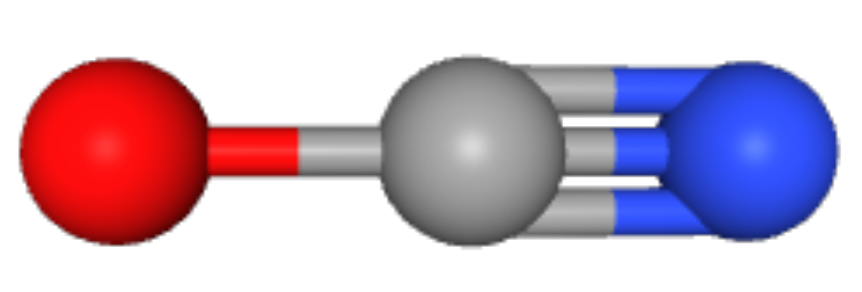}} & OCN$^-$  & Cyanate ion & yes & no & no & no & \textcolor{orange}\faExclamationTriangle $\hspace{0.1cm}$ / [11]\\
\raisebox{-.5\height}{\includegraphics[height=0.6in]{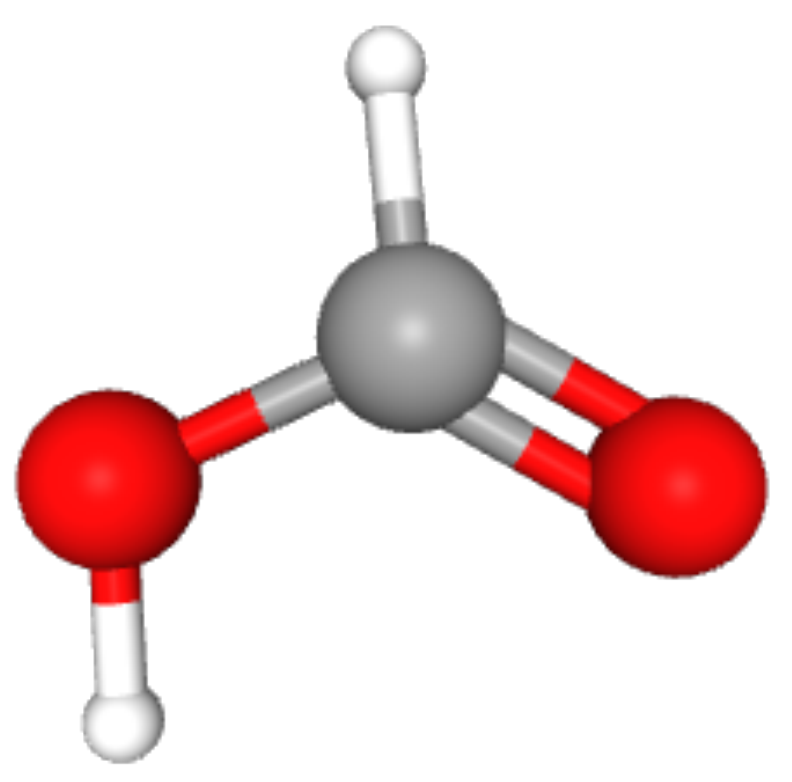}} & HCOOH  & Formic acid & yes & no & yes & no & \textcolor{orange}\faExclamationTriangle $\hspace{0.1cm}$ / [7]\\
\raisebox{-.5\height}{\includegraphics[height=0.5in]{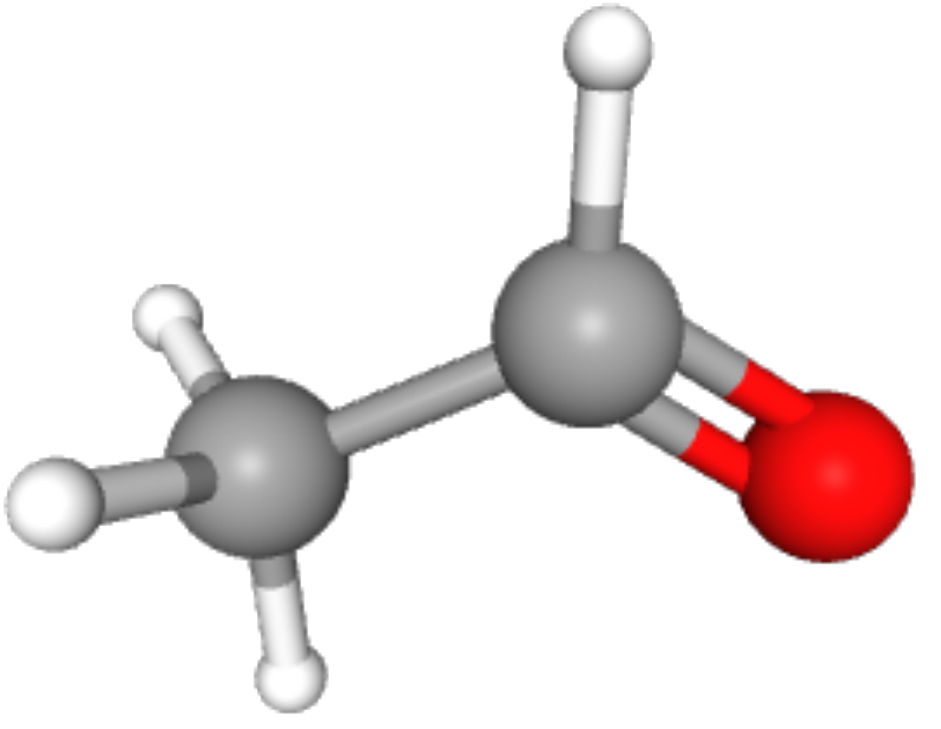}} & CH$_3$CHO  & Acetaldehyde & yes & no & yes & no & \textcolor{orange}\faExclamationTriangle $\hspace{0.1cm}$ / [12]\\
\raisebox{-.5\height}{\includegraphics[height=0.5in]{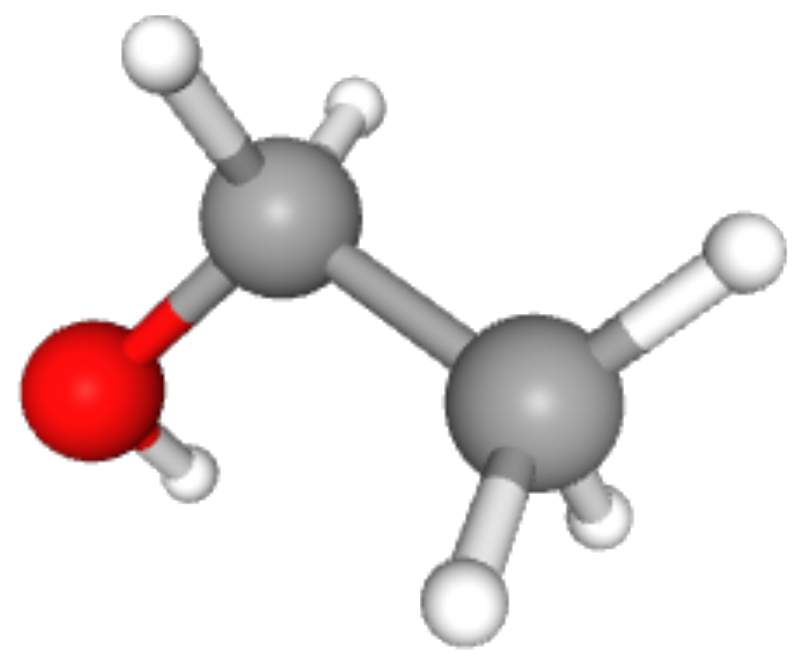}} & CH$_3$CH$_2$OH  & Ethanol & yes & no & yes & no & \textcolor{orange}\faExclamationTriangle $\hspace{0.1cm}$ / [13]\\
\raisebox{-.5\height}{\includegraphics[height=0.5in]{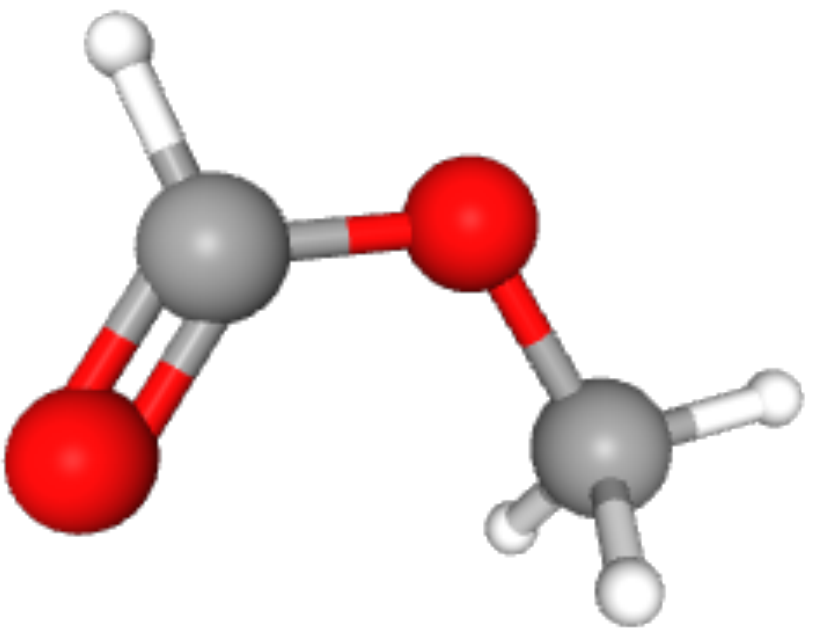}} & CH$_3$OCHO & Methyl formate & yes & no & yes & no & \textcolor{orange}\faExclamationTriangle $\hspace{0.1cm}$ / [14]\\
\raisebox{-.5\height}{\includegraphics[height=0.4in]{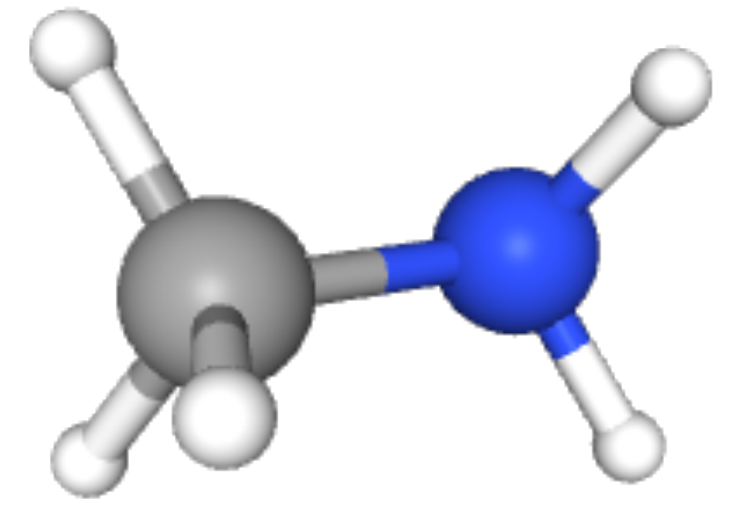}} & CH$_3$NH$_2$ & Methylamine & yes & no & yes & no & \textcolor{orange}\faExclamationTriangle $\hspace{0.1cm}$ / [15]\\
\raisebox{-.5\height}{\includegraphics[height=0.4in]{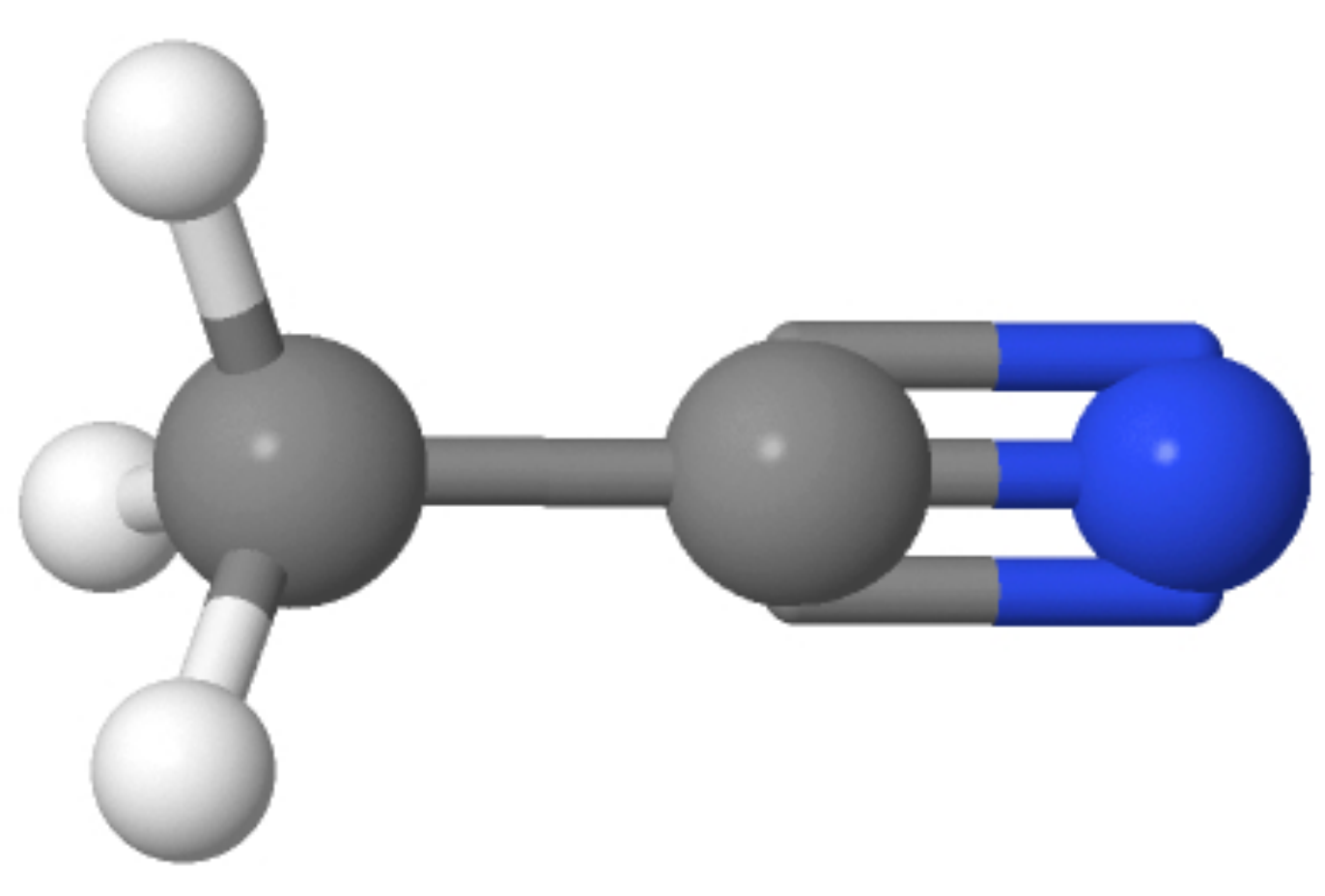}} & CH$_3$CN & Acetonitrile & yes & no & yes & no & \textcolor{orange}\faExclamationTriangle $\hspace{0.1cm}$ / [16]\\
\raisebox{-.5\height}{\includegraphics[height=0.35in]{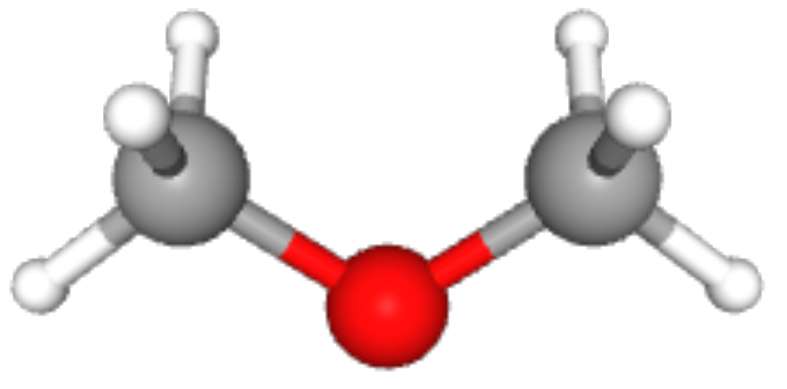}} & CH$_3$OCH$_3$  & Dimethyl Ether & yes & no & yes & no & \textcolor{red}\faTimesCircleO $\hspace{0.1cm}$ / ...\\
\raisebox{-.5\height}{\includegraphics[height=0.5in]{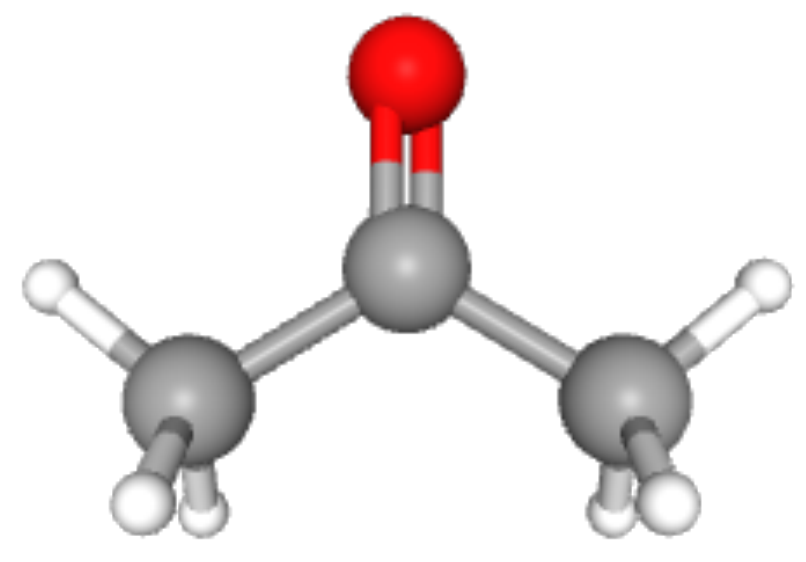}} & CH$_3$COCH$_3$ & Acetone & yes & no & yes & no & \textcolor{red}\faTimesCircleO $\hspace{0.1cm}$ / ...\\
\raisebox{-.5\height}{\includegraphics[height=0.23in]{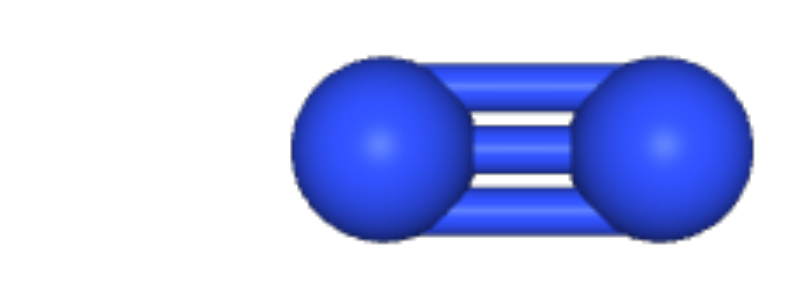}} & N$_2$ & Nitrogen & yes & no & yes & no & \textcolor{red}\faTimesCircleO $\hspace{0.1cm}$ / ...\\
\raisebox{-.5\height}{\includegraphics[height=0.20in]{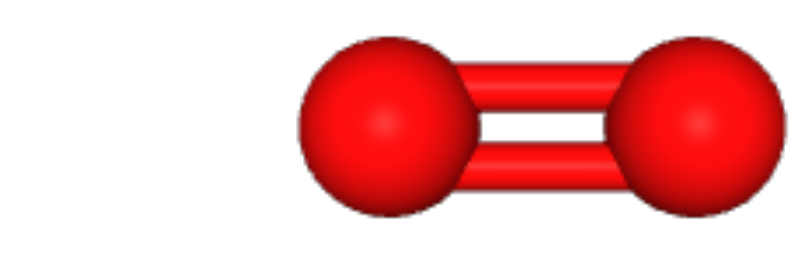}} & O$_2$ & Oxigen & yes & no & yes & no & \textcolor{red}\faTimesCircleO $\hspace{0.1cm}$ / ...\\
\hline
\end{tabular}
\tablefoot{\footnotesize
$^a$Taken from \url{https://pubchem.ncbi.nlm.nih.gov/}.
$^b$ Symbols for detection $-$ \textcolor{dgreen}\faCheckCircle: secure; \textcolor{orange}\faExclamationTriangle: tentative; \textcolor{red}\faTimesCircleO: no. $^c$ All these molecules have also been detected in the gas-phase, except the ions NH$_4^+$ and OCN$^-$ \citep[see 2021 census in][]{McGuire2021}. References of first observations: [1] \citet{Gillett1973}, [2] \citet{Lacy1984}, [3] \citet{deGraauw1996CO2}, [4] \citet{Lacy1991}, [5] \citet{Lacy1998}, [6] \citet{Grim1989}, [7] \citet{Schutte1996}, [8] \citet{Keane2001}, [9] \citet{Palumbo1995}, [10] \citet{Boogert1997}, [11] \citet{Schutte1997}, [12] \citet{Schutte1999_weak}, [13] \citet{Oberg2011}, [14] \citet{Scheltinga2018}, [15] \citet{Rachid2021}, [16] \citet{Rachid2022}
}
\end{table*}


\section{Data in the database}
\label{data_db}
In this section, we provide details on the experimental techniques used to measure the data available through LIDA. In summary, LIDA contains mid-infrared (mid-IR) spectra of ice samples ($\sim$4000$-$500~cm$^{-1}$; 2.5$-$20~$\mu$m) and UV/visible to mid-IR refractive indices of water ice in the 0.25$-$20~$\mu$m range. The IR ice spectra are available for pure and mixed ices for different settings. The H$_2$O ice refractive indices ($n$ and $k$) are available for ices deposited at different temperatures.

\subsection{Experimental setups and ice grow techniques}
\begin{figure*}
   \centering
   \includegraphics[width=\hsize]{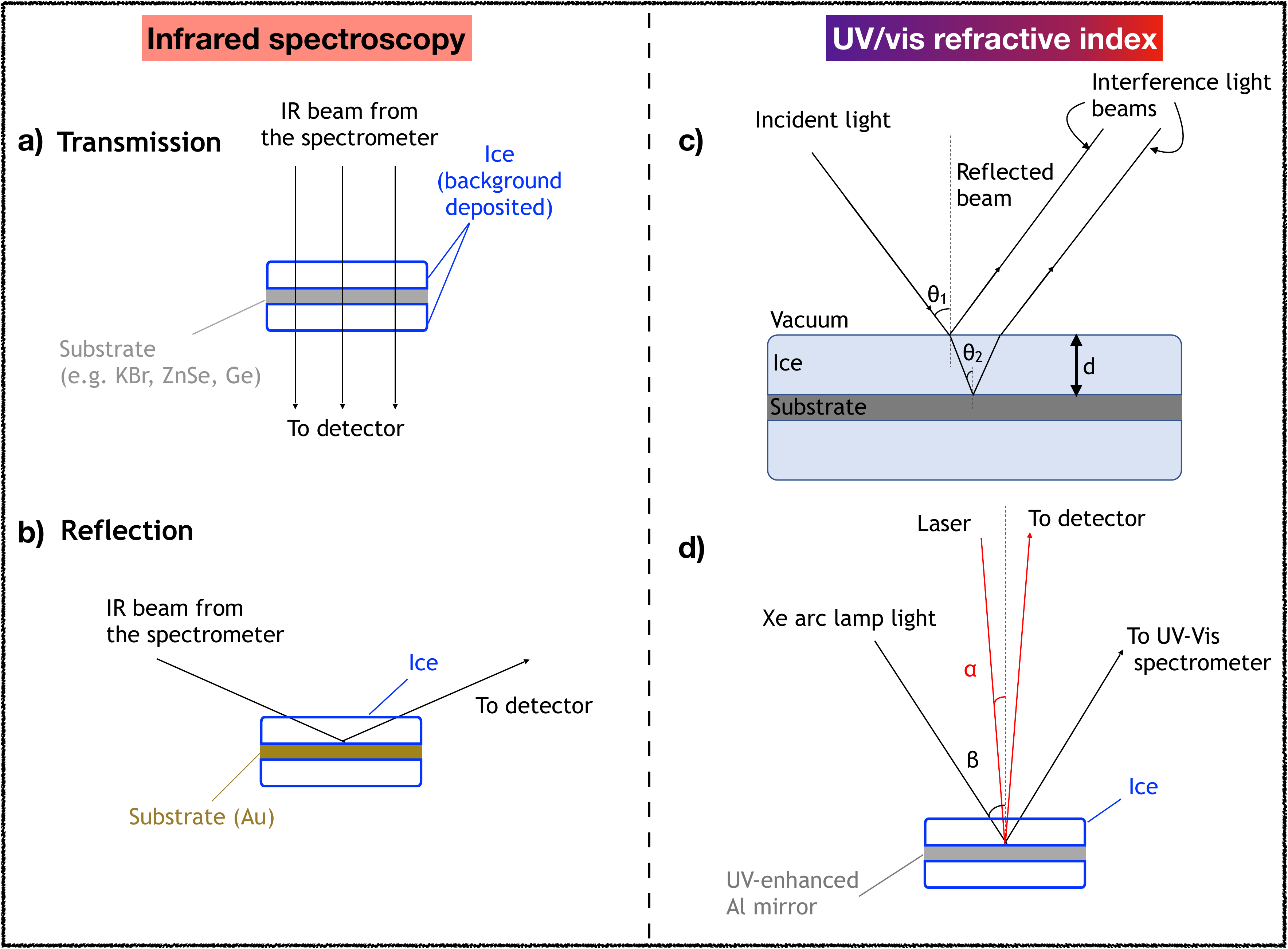}
      \caption{Overview of experimental techniques used to obtain the data hosted in LIDA. The left panel shows two infrared spectroscopic techniques to record the spectrum of ices, namely, transmission mode ($a$) and reflection-absorption mode ($b$). The right panel illustrates the technique used to measure UV-vis refractive index values. Cartoon~$c$ provides details of the diffraction and reflection phenomena that generate a interference pattern during the ice growth. Cartoon~$d$ shows the incidence and reflection of the monochromatic and broad band light beams when interacting with the ice.}
         \label{exp_tecniques}
   \end{figure*}

Between 1990 and 2020, the majority of the IR data in Leiden were recorded with our HV-setup, a regular high vacuum setup (10$^{-7}$~mbar) in which the broadband light of a Fourier Transform IR spectrometer (best resolution 0.1~cm$^{-1}$) is transmitted through a cryogenically cooled substrate covered with ice that is grown under fully controlled laboratory conditions (Figure~\ref{exp_tecniques}a). The transmitted beam is focused into a detector and processed with a Fourier transform to provide the transmitted light per wavelength. In the '90s, this setup was also equipped with a microwave discharge hydrogen flow lamp that was used to irradiate ice samples with a flux of $\sim$10$^{15}$ photons cm$^{-2}$ s$^{-1}$ dominated by Lyman-$\alpha$ emission, in order to study radical or ionic ice constituents or species formed upon irradiation \citep{Gerakines1996}. The HV setup has been regularly upgraded and details are available from \citet{Oberg2009}. From 2020, a new setup has been used, IRASIS or InfraRed Absorption Setup for Ice Spectroscopy, that uses the same measurement principle but operates at substantially lower pressures (10$^{-9}$~mbar) to minimize contaminants. Moreover, laser interferometry has been incorporated to perform thickness measurements in order to derive experimental absorption cross-sections. All these data are recorded in transmission. Details about IRASIS are available from \citet{Rachid2021}. In the nearby future, also a quartz crystal micro-balance will be incorporated. A few spectra available from LIDA have been recorded using RAIRS (Reflection Absorption IR) spectroscopy (Figure~\ref{exp_tecniques}b), but they are not included in the collection of ice spectra presented in this paper. The reader can find more details about those experiments in \citet{vanBroekhuizen2005phd}, \citet{Fuchs2006}, \citet{Oberg2009} and \citet{Fayolle2011, Ligterink2018} and about CRYOPAD (CRYOgenic Photoproduct Analysis Device), a setup that uses RAIRS and is dedicated to study the impact of Vacuum UV irradiation on ice samples. 

Apart from ice spectra we also present real UV/vis refractive index measurements at cryogenic temperatures using our Optical Absorption Setup for Ice Spectroscopy \citep[OASIS;][]{kofman2019, He2022}. The base pressure on OASIS is around 6$\times$10$^{-8}$~mbar. In this setup, a light source impinges on the growing ice and the reflected beam creates an interference pattern (Figure~\ref{exp_tecniques}c). Specifically, a Xe arc lamp and a HeNe laser (632~nm) are the light sources for the interference technique (Figure~\ref{exp_tecniques}d). The light from the Xe arc lamp strikes the growing ice at 45 degrees and is reflected toward the aperture of an Andor 303i Shamrock spectrometer. In the spectrometer, the light is dispersed and collected onto a CCD (Andor iDus DV420 OE), which allows recording the interference pattern at different wavelengths in the 250 and 750~nm region. The HeNe beam strikes the ice at a small angle ($\sim$ 3$^{\circ}$) and is recorded by a photodetector. The interference pattern is later used to derive the refractive index of the ice (see Section~\ref{sec_uvvis}). So far, the experiments on OASIS have targeted measurements of pure ices. Next, this setup will be used to measure the refractive index of binary ice mixtures as well.

In IRASIS, OASIS and CRYOPAD, a single gas/vapour component or a gaseous mixture is introduced into the chamber through a controllable leak valve and deposited onto a cold substrate. Usually, the substrate used in transmission spectroscopy is one of the following materials: potassium bromide - KBr, zinc selenide - ZnSe, or germanium - Ge, whereas gold (Au) is used for RAIRS. An UV-enhanced aluminium mirror is used as a substrate in refractive index experiments. In most of the data available from LIDA, the ices are background deposited, which means that the gas inlet does not point toward the sample, allowing the molecules to impinge onto the substrate coming from random directions and stick on both sides of the substrate. This is more representative of the way molecules interact with an icy dust grain in space and generally causes the ice to be somewhat more porous. In the case of mixed ices, the samples can be prepared in a separate mixing system or by admitting the individual gas/vapor components in the chamber through different dosing lines. In either case, the molecules are considered to be homogeneously mixed before freezing out onto the cooled substrate. The ice thickness is often given in number of monolayers or Langmuir, where one monolayer (1L) corresponds to 10$^{15}$~cm$^{-2}$ \citep{Langmuir1938}. In IR spectroscopy experiments, the ice thickness can be as thin as a few monolayers. In some cases, when ice mixtures are used, the ice has to be thicker ($\sim$ 3000 monolayers) in order to allow the detection of the less abundant molecular component in the sample or guarantee that the deposition of background gases during the measurement is negligible \citep[see][]{Scheltinga2018}. In experiments to measure the ice refractive index, the ice thickness is generally much thicker ($\sim$45000~ML) because the technique requires to record several fringes in the interference pattern. It is worth mentioning that the shape and position of the IR bands are not affected by the ice thickness nor the underlying substrate used in the experiments performing transmission spectroscopy.

\subsection{Absorbance spectrum}
\label{abs_sec}
The majority of the IR absorbance spectra in LIDA has been measured using transmission spectroscopy. The ground principle to measure the absorbance spectrum is that the incident radiation is attenuated when crossing the ice sample due to the intrinsic properties of the material. The intensity of the transmitted light at each wavelength ($I_{\lambda}$) is calculated with Lambert-Beer's law, which is given by:
\begin{equation}
    I_{\lambda} =  I_{\lambda}^0 \mathrm{exp}\left(-\alpha_{\lambda} r \ell \right)
    \label{transmittance}
\end{equation}
where $I_{\lambda}^0$ is the incident light intensity, $\alpha_{\lambda}$ is the wavelength-dependent absorption coefficient, $r$ is the concentration in the sample, and $\ell$ is the effective radiation path within the ice. The absorbance is derived from Equation~\ref{transmittance} as shown below:
\begin{equation}
    Abs_{\lambda} = \mathrm{-log_{10}}\left( \frac{I_{\lambda}}{I_{\lambda}^0} \right)
            = 0.434 \alpha r \ell
            \label{abs_eq}
\end{equation}
where the absorbance is directly proportional to the molecular concentration and the radiation path in the ice. In transmission spectroscopy, the substrate is transparent to the IR light, and the absorption bands observed in the IR spectra are due to the molecules in the ice sample.

When RAIRS is used, the absorbance is no longer obtained from Equation~\ref{abs_eq}. The absorbance spectrum is calculated as a function of the reflected light, and depending on the ice thickness, the geometry of the light path, and the setup itself, the spectrum can change substantially. Briefly, in RAIRS, the IR light shines onto a reflective gold (Au) surface at grazing angles ($\sim$ 90$^{\circ}$ w.r.t. surface normal) and is reflected towards the detector (Figure \ref{exp_tecniques}b). Upon specular reflection, the s-polarized light becomes negligible, and only the p-polarized component interacts with the molecules. In this way, RAIRS has an additional selection rule for absorption, which imposes that the vibrational motions have a component orthogonal to the reflection surface \citep{palumbo2006}. RAIRS comes with the disadvantage that spectra cannot be directly compared with astronomical data, as in the case of spectra recorded in transmission. On the other hand, RAIRS has the advantage of increasing the signal-to-noise ratio of the data.

Either transmission or RAIRS can record data of pure or mixed ices before and after warm-up, or processed by UV radiation. The ice spectrum is taken by averaging a certain number of scans, allowing a higher signal-to-noise ratio, and typical spectral resolutions are within 0.5~cm$^{-1}$ and 2.0~cm$^{-1}$, whereas 0.1~cm$^{-1}$ spectra can be recorded if needed. Likewise, the absorbance accuracy is a characteristic of the IR spectrometer, and is around 1\%. Warm-up experiments are performed by depositing the ice at low temperature ($\sim$10~K), followed by a slow increase of the substrate temperature (e.g., 25~K per hour) while IR spectra are continuously taken. In the cases where experiments with UV ice processing are performed, the absorbance spectrum is taken after the irradiation process. The recorded IR absorbance spectrum often shows a curved baseline which needs to be corrected. Typically, a low-order polynomial function is used to flatten the ice spectrum and perform corrections to remove artefacts from the IR spectra. The baseline correction is made by interpolating a function for wavelengths where there is no absorption and subtracting it from the original signal. The spectra contained in the database have been previously baseline corrected using a polynomial or linear function to set the data to zero absorbance where there are no absorption features. When available, the non-baseline corrected spectrum is offered for download. The original spectra (raw data) are not offered for download to avoid publication of data in the literature that have not been treated correctly. This requires appropriate knowledge on how to deal with these datasets.

For astrochemical applications, the absorbance spectrum is often converted to an optical depth scale. The optical depth of experimental data is given by \citet{dHendecourt1986}: 
\begin{equation}
    \tau_{\rm{\lambda}}^{\rm{lab}} = 2.3 \cdot Abs_{\lambda}.
    \label{od_eq}
\end{equation}

Ultimately, $\tau_{\lambda}^{\rm{lab}}$ is used to calculate the column density of the ice sample from the equation below:
\begin{equation}
    N_{\rm{ice}} = \frac{1}{\mathcal{A}} \int_{\nu_1}^{\nu_2} \tau_{\lambda}^{\rm{lab}} d\nu,
    \label{CD_eq}
\end{equation}
in which $\mathcal{A}$ is the band strength of the vibrational modes associated with the absorption features. Most of the $\mathcal{A}$ values in the literature have been derived for pure ice samples \citep[e.g.,][]{Gerakines1995, Gerakines1996, Kerkhof1999, Bouilloud2015, Hudson2017bs}. However, \citet{Oberg2007} and \citet{Bouwman2007} show that variation in the chemical composition of the ice leads to changes in the band strength of solid-phase molecules. These changes are often reported as relative values with respect to the pure ice, because information such as the ice density is unknown when the molecular concentrations change within the ice. In Table~\ref{ice_bs}, we compile $\mathcal{A}$ values from the literature for pure ices, which were then used to derive the column densities of the ices in LIDA. These values are also used to derive the column densities of most of the ice mixtures. Otherwise, we use tabulated values from \citet{Oberg2007} to derive the column densities of H$_2$O:CO$_2$, and the values from \citet{Scheltinga2018}, \citet{Rachid2020}, \citet{Scheltinga2021}, \citet{Rachid2021} to derive the column densities of ice-containing COMs.

\begin{table*}
\caption{\label{ice_bs} List of vibrational transitions and band strengths of the molecules in pure ices presented in the literature.}
\centering 
\begin{tabular}{lccccc}
\hline\hline
Molecule & $\lambda \; [\mu \mathrm{m}]$ & $\nu \; \mathrm{[cm^{-1}]}$ & Identification & $\mathcal{A} \; \mathrm{[cm \; molec^{-1}]}$ & References\\
\hline
H$_2$O          & 3.01    & 3,322 & O$-$H stretch & $\mathrm{2.2 \times 10^{-16}}$ & \citet{Bouilloud2015}\\
H$_2$O          & 6.00    & 1,666 & bend & $\mathrm{1.1 \times 10^{-17}}$ & \citet{Bouilloud2015} \\
H$_2$O          & 13.20    & 760 & libration & $\mathrm{3.2 \times 10^{-17}}$ & \citet{Bouilloud2015} \\
CO$_2$        & 4.27    & 2,341      & CO stretch & $\mathrm{1.3 \times 10^{-16}}$ & \citet{Bouilloud2015}\\
CO$_2$        & 15.27    & 660, 665      & bend & $\mathrm{1.2 \times 10^{-17}}$ & \citet{Bouilloud2015}\\
CO        & 4.67    & 2,141      & CO stretch & $\mathrm{1.4 \times 10^{-17}}$ & \citet{Bouilloud2015}\\
NH$_3$          & 2.96   & 3,376 & NH$_3$ asym-stretch & $\mathrm{2.3 \times 10^{-17}}$ & \citet{Bouilloud2015}\\
NH$_3$          & 6.15   & 1,624 & NH$_3$ def. & $\mathrm{5.6 \times 10^{-18}}$ & \citet{Bouilloud2015}\\
NH$_3$          & 9.01   & 1,109 & NH$_3$ umbrella & $\mathrm{2.1 \times 10^{-17}}$ & \citet{Bouilloud2015}\\
NH$_4^+$          & 6.85   & 1,460 & bend & $\mathrm{4.4 \times 10^{-17}}$ & \citet{Schutte2003}\\
OCN$^-$          & 9.01   & 1,109 & CN stretch & $\mathrm{1.3 \times 10^{-16}}$ & \citet{vanBroekhuizen2005}\\
SO$_2$          & 7.60    & 1,320 & SO$_2$ stretch & $\mathrm{3.4 \times 10^{-17}}$ & \citet{Boogert1997}\\
OCS             & 4.93    & 2,025 & CO stretch & $\mathrm{3.4 \times 10^{-17}}$ & Rachid et al. (in prep.)\\
CH$_4$          & 3.32    & 3,010 & CH$_4$ deformation & $\mathrm{1.1 \times 10^{-17}}$ & \citet{Bouilloud2015}\\
CH$_4$          & 7.67    & 1,303 & CH$_4$ deformation & $\mathrm{8.4 \times 10^{-18}}$ & \citet{Bouilloud2015}\\
H$_2$CO           & 3.45    & 2,891 & CH$_2$ a-stretch     & $\mathrm{4.7 \times 10^{-18}}$ & \citet{Bouilloud2015}\\
H$_2$CO           & 3.53    & 2,829 & CH$_2$ s-stretch     & $\mathrm{1.3 \times 10^{-17}}$ & \citet{Bouilloud2015}\\
H$_2$CO           & 5.79    & 1,725 & C$=$O stretch & $\mathrm{1.6 \times 10^{-17}}$ & \citet{Bouilloud2015}\\
CH$_3$OH        & 3.55    & 2,816 & O$-$H stretch & $\mathrm{1.0 \times 10^{-16}}$ & \citet{Bouilloud2015}\\
CH$_3$OH        & 6.85    & 1,460 & O$-$H bend & $\mathrm{6.5 \times 10^{-18}}$ & \citet{Bouilloud2015}\\
CH$_3$OH        & 8.85    & 1,130 & CH$_3$ rock & $\mathrm{1.8 \times 10^{-18}}$ & \citet{Bouilloud2015}\\
CH$_3$OH        & 9.74    & 1,128 & C$-$O stretch & $\mathrm{1.8 \times 10^{-17}}$ & \citet{Bouilloud2015}\\
HCOOH           & 5.85    & 1,708 & C$=$O stretch     & $\mathrm{5.4 \times 10^{-17}}$ & \citet{Bouilloud2015}\\
HCOOH           & 7.25    & 1,384 & OH bend     & $\mathrm{2.6 \times 10^{-18}}$ & \citet{Schutte1999_weak}\\
HCOOH           & 8.22    & 1,216 & C$-$O stretch     & $\mathrm{2.9 \times 10^{-17}}$ & \citet{Bouilloud2015}\\
HCOOH           & 9.31    & 1,074 & CH bend     & $\mathrm{3.1 \times 10^{-19}}$ & \citet{Bouilloud2015}\\
HCOOH           & 10.76    & 929 & OH bend     & $\mathrm{6.4 \times 10^{-17}}$ & \citet{Bouilloud2015}\\
CH$_3$CHO       & 5.80    & 1,723 &  C$-$O stretch  & $\mathrm{1.3 \times 10^{-17}}$ & \citet{Schutte1999_weak}\\
CH$_3$CH$_2$OH  & 9.17    & 1,090 &  CH$_3$ rock   & $\mathrm{7.3 \times 10^{-18}}$ & \citet{Hudson2017}\\
CH$_3$CH$_2$OH  & 9.51    & 1,051 &  CO stretch   & $\mathrm{1.4 \times 10^{-17}}$ & \citet{Hudson2017}\\
CH$_3$CH$_2$OH  & 11.36    & 880 &  CC stretch   & $\mathrm{3.2 \times 10^{-18}}$ & \citet{Hudson2017}\\
CH$_3$OCH$_3$   & 8.59    & 1,163 & COC stretch + CH$_3$ rock     & $\mathrm{9.8 \times 10^{-17}}$ & \citet{Scheltinga2018}\\
CH$_3$OCH$_3$   & 10.85    & 921 & COC stretch     & $\mathrm{5.0 \times 10^{-18}}$ & \citet{Scheltinga2018}\\
CH$_3$COCH$_3$   & 5.84    & 1,710 & C$=$O stretch     & $\mathrm{2.7 \times 10^{-17}}$ & \citet{Hudson2018}\\
CH$_3$COCH$_3$   & 7.05    & 1,417 & CH$_3$ a-stretch     & $\mathrm{9.2 \times 10^{-18}}$ & \citet{Hudson2018}\\
CH$_3$COCH$_3$   & 7.33    & 1,363 & CH$_3$ s-stretch     & $\mathrm{1.4 \times 10^{-17}}$ & \citet{Hudson2018}\\
CH$_3$COCH$_3$   & 8.14    & 1,228 & CCC a-stretch     & $\mathrm{7.3 \times 10^{-18}}$ & \citet{Hudson2018}\\
CH$_3$COCH$_3$   & 18.79    & 532 & CO deformation     & $\mathrm{2.1 \times 10^{-18}}$ & \citet{Hudson2018}\\
CH$_3$OCHO   & 5.80    & 1723 & C$=$O stretch     & $\mathrm{5.0 \times 10^{-17}}$ & \citet{Modica2010}\\
CH$_3$OCHO   & 8.25    & 1211 & C$-$O stretch     & $\mathrm{2.9 \times 10^{-17}}$ & \citet{Modica2010}\\
CH$_3$OCHO   & 8.58    & 1165 & CH$_3$ rock     & $\mathrm{2.0 \times 10^{-17}}$ & \citet{Modica2010}\\
CH$_3$OCHO   & 10.98    & 910 & O$-$CH$_3$ stretch     & $\mathrm{4.8 \times 10^{-18}}$ & \citet{Modica2010}\\
CH$_3$OCHO   & 13.02    & 768 & OCO deformation     & $\mathrm{1.2 \times 10^{-18}}$ & \citet{Modica2010}\\
CH$_3$NH$_2$   & 3.47    & 2881 & CH$_3$ a-stretch     & $\mathrm{2.6 \times 10^{-18}}$ & \citet{Rachid2021}\\
CH$_3$NH$_2$   & 3.58    & 2791 & CH$_3$ s-stretch     & $\mathrm{3.8 \times 10^{-18}}$ & \citet{Rachid2021}\\
CH$_3$NH$_2$   & 6.76    & 1478 & CH$_3$ a-deformation     & $\mathrm{1.1 \times 10^{-18}}$ & \citet{Rachid2021}\\
CH$_3$NH$_2$   & 6.87    & 1455 & CH$_3$ a-deformation     & $\mathrm{7.0 \times 10^{-19}}$ & \citet{Rachid2021}\\
CH$_3$NH$_2$   & 7.024    & 1420 & CH$_3$ s-deformation     & $\mathrm{2.0 \times 10^{-19}}$ & \citet{Rachid2021}\\
CH$_3$NH$_2$   & 8.63    & 1159 & CH$_3$ rock     & $\mathrm{1.5 \times 10^{-18}}$ & \citet{Rachid2021}\\
CH$_3$CN   & 3.33    & 3001 & CH$_3$ a-stretch     & $\mathrm{1.5 \times 10^{-18}}$ & \citet{Rachid2022}\\
CH$_3$CN   & 3.40    & 2940 & CH$_3$ s-stretch     & $\mathrm{5.3 \times 10^{-19}}$ & \citet{Rachid2022}\\
CH$_3$CN   & 4.44    & 2252 & CN stretch     & $\mathrm{1.9 \times 10^{-18}}$ & \citet{Rachid2022}\\
CH$_3$CN   & 7.09    & 1410 & CH$_3$ a-deformation     & $\mathrm{1.90 \times 10^{-18}}$ & \citet{Rachid2022}\\
CH$_3$CN   & 7.27    & 1374 & CH$_3$ s-deformation     & $\mathrm{1.2 \times 10^{-18}}$ & \citet{Rachid2022}\\
CH$_3$CN   & 9.60    & 1041 & CH$_3$ rock     & $\mathrm{1.6 \times 10^{-18}}$ & \citet{Rachid2022}\\
CH$_3$CN   & 10.88    & 919 & CC stretch     & $\mathrm{3.5 \times 10^{-19}}$ & \citet{Rachid2022}\\
\hline
\end{tabular}
\end{table*}

When an ice column density is derived from RAIRS data, a correction must be performed on the band strength values from the literature. For spectra measured with RAIRS, the ice column density is given by:
\begin{equation}
    N_{\rm{ice}} = \frac{1}{R\mathcal{A}} \int_{\nu_1}^{\nu_2} \tau_{\lambda}^{\rm{lab}} d\nu
    \label{CD_eq_rairs}
\end{equation}
where $R$ is the correction factor. Specifically, in RAIRS experiments the path-length of the light beam is longer across the ice than in transmission IR spectroscopy. Consequently, band strengths measured with RAIRS are no longer the same of those measured in transmission. This correction depends on the molecule and individual calibration experiments are performed. Since this paper only presents results from transmission spectroscopy, the $R$ values are not provided. Future upgrades of LIDA will include RAIRS data and their respective correction factors for the ice column density calculation.


Spectra of mixed ices are shown in the database, with the ratio between the molecule given in the label. For example, in a H$_2$O:CO$_2$ (10:1) ice there are 10 molecules of H$_2$O for each molecule of CO$_2$ in the ice. Layered ices can also be made by depositing a certain number of monolayers (1~ML $\sim$ 10$^{15}$~molecules cm$^{-2}$) of a pure molecule on the substrate followed by a number of ML of another pure or mixed ice. When this is the case, the mixture is named as ``CO over CO$_2$'', which means that pure CO was deposited on top of a pure CO$_2$ ice. Similarly, ``CO under CO$_2$'' means that CO was deposited in the bottom layer, followed by CO$_2$ deposition at the top layer.

\begin{figure}
   \centering
   \includegraphics[width=\hsize]{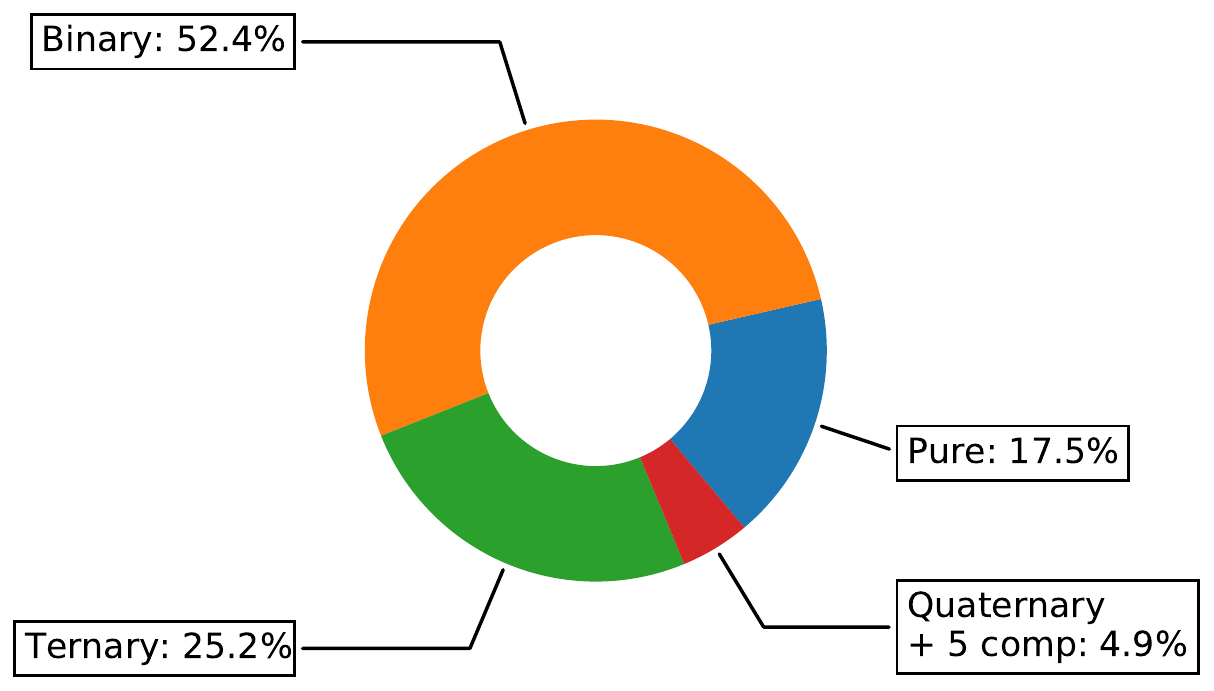}
      \caption{Piechart displaying the percentage division of ice analogues in LIDA from pure samples to 5 components mixtures.}
         \label{piechart}
   \end{figure}

In Figure~\ref{piechart}, we show the fraction of pure and mixed samples hosted in LIDA. The majority of the IR spectra are of binary ice samples that account for 52.4\%. Most of these samples are mixtures of simple molecules (e.g., H$_2$O, CO, CO$_2$). Recently, binary ices containing COMs (e.g., CH$_3$CHO, CH$_3$CH$_2$OH, CH$_3$OCH$_3$, CH$_3$NH$_2$, CH$_3$OCHO, CH$_3$COCH$_3$, CH$_3$CN) have been included in the database. The next larger group (25.2\%) is of ice samples containing three compounds, which may include a COM. Pure ice samples make up the third group (17.5\%) and contain simple and complex molecules, as well as ions (NH$_4^+$ and OCN$^-$). We note, however, that these ions are formed via warm-up of HNCO:NH$_3$ ice \citep{Novozamsky2001}. Moreover, some of the pure ice samples were exposed to UV radiation with experimental details given in \citet{Gerakines1996}. Finally, the groups of quaternary and five-component ice samples are combined and account for 4.9\% of all spectra in the database. A full list with all ice analogues in the database is presented in Table~\ref{analogue_list} of Appendix~\ref{Laboratory_data_list}.

\subsection{UV/visible and mid-IR refractive index}
\label{sec_uvvis}
When light shines upon the ice surface, part of it refracts into the ice, and part is specularly reflected by the surface (Figure \ref{exp_tecniques}c,d). The refracted beam is reflected in the ice-substrate interface and eventually emerges back into the vacuum. The phase difference ($\Delta$) between the light rays that pass through the ice and the ones reflected by the surface is related to their optical path difference ($\delta$), which is given by:
\begin{subequations}
\begin{align}
\Delta &= \frac{2 \pi}{\lambda} \delta,\\
\delta &= 2nd \rm{cos}(\theta_2),
\label{phase-difference}
\end{align}
\end{subequations}
where $\lambda$ is the wavelength of the incoming light, $n$ is the real part of the ice refractive index at wavelength $\lambda$, $d$ is the ice thickness, and $\theta_2$ is refraction angle (see Figure~\ref{exp_tecniques}c), i.e., the angle between the refracted light and the normal plane perpendicular to the ice. The incident angle $\theta_1$ is related to refraction angle $\theta_2$ by Snell's law. When $\delta/\lambda$ is an even number, $\Delta$ is a multiple of 2$\pi$, resulting in constructive interference of the light beams. Conversely, when $\delta/\lambda$ is an odd number, the interference is destructive. Consequently, the intensity of the resulting beam reflected by the ice surface is an oscillation pattern with the form:



\begin{equation}
\begin{split}
     I(t) & = A + B \; \mathrm{cos}[\Delta (t)] \\ 
          &= A + B \; \mathrm{cos}\bigg( \frac{4 \pi n d(t) \; \mathrm{cos}(\theta_2)}{\lambda} \bigg ),
\end{split}
\label{intensity}
\end{equation}
where A and B are constants. Thus, the intensity of the interference pattern carries information about both the refractive index and the rate at which the ice thickness increases during deposition. Since each of these parameters is unknown, they cannot be derived from a single interference measurement. However, by recording the interference pattern of growing ice employing two different incident angles or wavelengths and employing Equation \ref{intensity}, both the ice refractive index and the growth rate can be derived. By recording the interference pattern of growing ice using two light beams of the same wavelength ($\lambda$) but different angles ($\alpha$, $\beta$), the refractive index expression can be derived from the frequency of the oscillations (Equation \ref{intensity}) and Snell's law:

\begin{equation}
    n_{uv-vis} = \sqrt{\frac{\sin^2 \alpha-(P_{\beta}/P_{\alpha})^2 \sin^2 \beta}{1 - (P_{\beta}/P_{\alpha})^2}},
\label{refractive}
\end{equation}
where $P_{\rm{\alpha}}$ and $P_{\rm{\beta}}$  are the periods of the interference patterns generated by the light beams striking the ice at angles $\alpha$ and $\beta$, respectively. For more details about the derivation of Equation~\ref{refractive}, see for example \cite{tempelmeyer1968refractive}, \citet{beltran2015double} and \citet{He2022}. 

While Equation~\ref{refractive} provides the ice refractive index in the UV-vis range, the refractive index in the mid-IR can be calculated using the Kramers-Kronig relations \citep{Kronig1926, Kramers1927}, which is given by:
\begin{equation}
    n(\nu) = n_{\rm{670nm}} + \frac{2}{\pi} \mathcal{P} \int_{\nu_1}^{\nu_2} \frac{\nu' k(\nu')}{\nu'^2 - \nu^2}d\nu'
    \label{n-value}
\end{equation}
where $n_{\mathrm{670nm}}$ is the refractive index of the sample at 670~nm and is within the UV-vis range for which the refractive index was derived (through Equation~\ref{refractive}), $\nu$ is the wavenumber corresponding to the peak of the band and $\nu'$ is the wavenumber before and after the $\nu$ value. The Cauchy principal value $\mathcal{P}$ is used to overcome the singularity when $\nu = \nu'$. The term ``$k$'' corresponds to the imaginary part of CRI, and is given by:
\begin{equation}
    k = \frac{1}{4\pi \nu d} \cdot \left( 2.3 \times Abs_{\nu} + \mathrm{ln} \left| \frac{\tilde{t}_{01}\tilde{t}_{02}}{1 + \tilde{r}_{01} \tilde{r}_{12} e^{2i\tilde{x}}} \right|^2 \right)
    \label{k-value}
\end{equation}
where $Abs_{\nu}$ is the absorbance spectrum value (Equation~\ref{abs_eq}), $d$ is the thickness of the ice sample, and $\tilde{t}_{01}$, $\tilde{t}_{02}$, $\tilde{r}_{01}$, $\tilde{r}_{12}$ are the Fresnel coefficients. The sub-label 0, 1 and 2 refers to regions of vacuum, ice sample, and substrate respectively. The refractive index of the substrate is implicit in the terms $\tilde{t}_{02}$ and $\tilde{r}_{12}$. Finally, the term $\tilde{x}$ is given by $\tilde{x} = 2\pi \nu d \tilde{m}$; $\tilde{m}$ is the CRI.

To determine the real and imaginary refractive index in the mid-IR, LIDA provide tools to solve Equations~\ref{n-value} and \ref{k-value} numerically in a iterative procedure. Specifically, LIDA uses the Maclaurin's formula described in \citet{Ohta1988} to obtain the real refractive index, and subsequently, the imaginary refractive index is derived. This methodology was also employed in other computational codes dedicated to calculate the CRI values of ice samples \citep{Rocha2014, Gerakines2020}.

In the current version of LIDA, we present the H$_2$O ice refractive index in the UV/vis, measured on the OASIS setup, and mid-IR optical constants calculated with the online tool available in LIDA (see Section~\ref{refrac_index}. In a follow-up paper, the refractive indexes of pure H$_2$O shown in this database will be systematically compared with the literature values (Rocha et al. {\it in prep.}). The refractive index values of other molecules (e.g., CO, CO$_2$, N$_2$, CH$_4$, CH$_3$OH) have been in part already measured and will be included in future LIDA upgrades. These will also include astronomically relevant ice mixtures.

\section{Features of the database}
\label{struct_db}
The upgraded LIDA is an extendable platform, designed to host IR spectra and UV/vis refractive indices of ice samples, as well as to support the upload of new datasets that will be obtained in future experiments. Access to these data is performed with dynamical and interactive visualization software, that is also linked to online tools to perform astronomy-oriented calculations. Additionally, all data are available for download in a standard ascii format. In the next subsections, we provide details about different aspects of the database. More information describing the software and approaches used to construct the database are given in Appendix~\ref{DB_design} and interactive documentation is available at \url{https://leiden-ice-database.readthedocs.io}. 




\subsection{User interface}
\label{us_interface}
The user interface of LIDA shows four sub-modules, namely, (i) spectral data, (ii) optical constants, (iii) online tools, and (iv) further information and a contact form. Access to these sub-modules is performed via the navigation bar at the top of the web interface. All IR ice spectra are available in the sub-module named {\it Spectral data}, which currently counts for more than 1100 ice spectra related to over 150 different ice samples. In the future, new data will be added, and the option exists to add previously recorded data that are currently scattered over the literature. The {\it Optical constants} section contains for now only the real refractive index of H$_2$O ice at different temperatures. However, more data from ongoing experiments will be added, which includes measurements of N$_2$, CO, CO$_2$, CH$_4$ and CH$_3$OH. LIDA is also equipped with {\it Online tools} focused on astronomy-oriented calculations. Finally, the user can visualize the {\it credits}, and {\it contact} the developers and scientific managers of the database. To render the database user interface, we used common web-technologies such as HTML (HyperText  Markup Language), CSS (Cascading Style Sheets) and JavaScript (JS). A list of all software used to develop LIDA is available at \url{https://icedb.strw.leidenuniv.nl/Credit}.

\subsection{Searching capability and metadata}
\label{searching_cap}
The IR spectra and the UV/vis optical constants of the ice analogues in LIDA can be searched via a {\it search box} by accessing the tabs {\it Spectral data} and {\it Optical constants}, respectively, in the navigation bar. The searching capability uses SQLAlchemy\footnote{\url{https://www.sqlalchemy.org/}}, a python SQL (Structured Query Language) toolkit, and Object Relational Mapper to enable the searching in \texttt{Flask} applications.    

To find a specific analogue in the database, either the chemical formula or the molecule name can be used. For example, the user can type \texttt{water} or \texttt{H$_2$O} to search for water ice spectra in the database. Searching for ice mixtures is possible by providing a list of the chemical formulas separated by space bar (e.g., \texttt{H2O CO2 CH3OH}). LIDA can also be used to search for molecules sharing common chemical structures. For example, when the query is \texttt{CO}, a list of all the molecules containing a carbon-oxygen bond (both simple and double) will be displayed on the web interface (e.g., CH$_3$OH, CH$_3$CHO, HCOOH). Search for more specific structures, such as functional groups, is also possible. As an example, if the query is \texttt{COOH}, a list with samples containing molecules that carry a carboxylic acid functional group will be returned (e.g., HCOOH). LIDA also supports searching by the type of ice processing. For example, water ice thermally processed can be searched with \texttt{H$_2$O category=warm-up}. Similarly, an energetically processed ice can be searched with \texttt{H$_2$O category=irradiation}. Finally, the user can also search a spectrum by the author who published the data with the command \texttt{H$_2$O author={{\"O}berg}}.

By searching for a specific ice sample, the user can also visualize the metadata. For example, information such as spectral resolution, deposition temperature, ice thickness and publication is visible. All spectra hosted in LIDA are available for download in ascii format (e.g., \texttt{.txt} extension), which is a standard format that can be imported to several software and computational codes. This feature offers the download of a single spectrum or all spectra related to an ice analogue.

\subsection{Data visualization}
\label{visual}
The data in LIDA are plotted interactively with \texttt{Bokeh}\footnote{\url{https://docs.bokeh.org/}}, a Python library for interactive visualization \citep{bokeh2018}. This software provides several control buttons by default to support the interactive inspection of the plots. More details can be accessed via the \texttt{Bokeh} documentation.

\begin{figure*}
   \centering
   \includegraphics[width=\hsize]{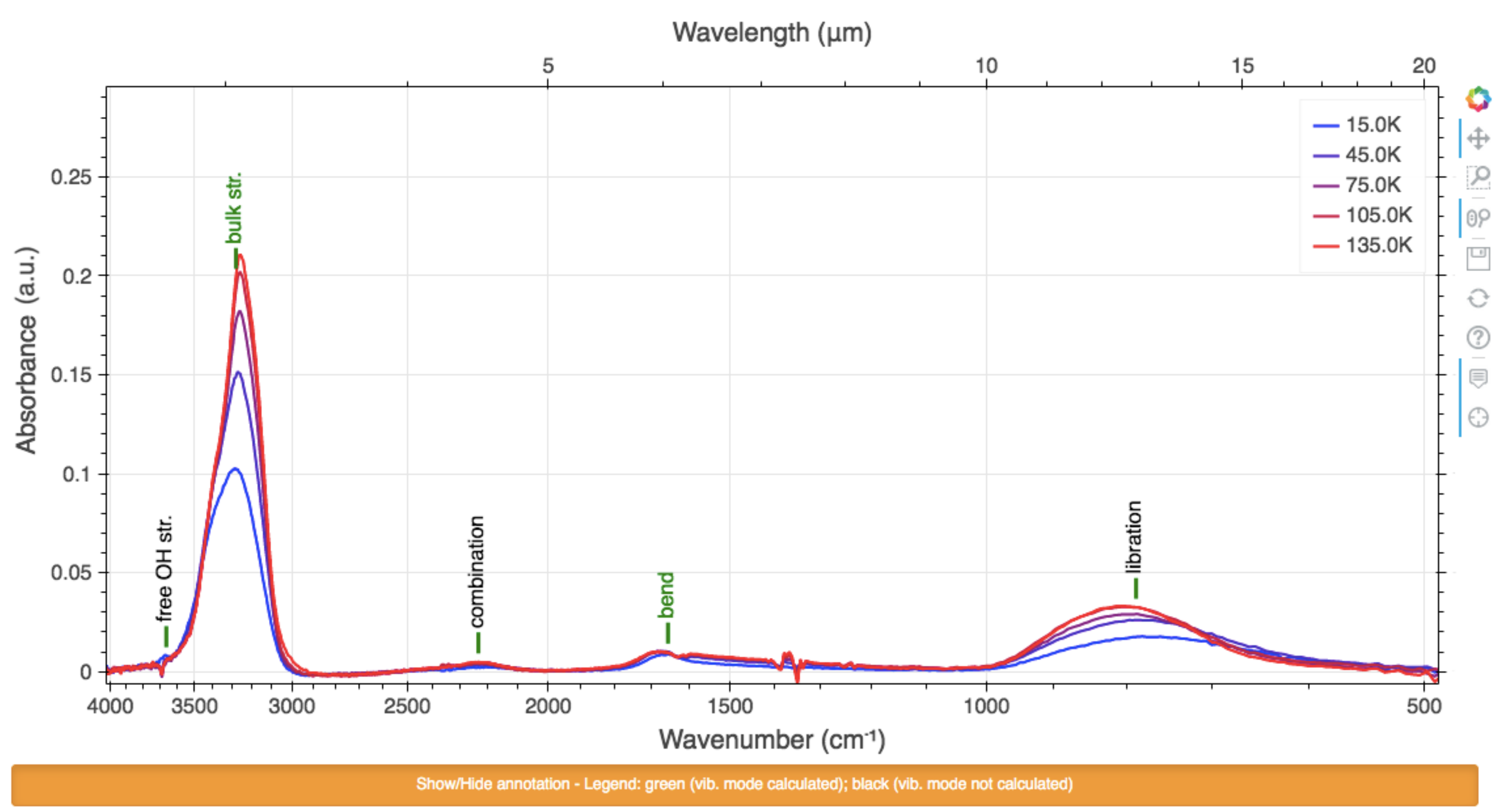}
      \caption{Screenshot of the spectrum visualization window showing the IR spectrum of H$_2$O ice at different temperatures given by the colour code. The annotations of the water vibrational modes are shown in green. They can be hidden by clicking on the yellow toggle below the plot. It also describes the annotation colour code - green means vibrational mode is calculated, and black indicates the vibrational mode is not calculated. The hover set at the position around 3000~cm$^{-1}$ displays the information of the spectral data point, i.e., wavenumber in cm$^{-1}$ (bottom X-axis), Wavelength in $\mu$m (top X-axis), and Absorbance (Y-axis). The toolbar is placed on the right side of the plot.}
         \label{spectrum}
   \end{figure*}

As an example of the data visualization in LIDA, Figure~\ref{spectrum} shows the IR absorbance spectrum of pure H$_2$O ice at temperatures of 15, 45, 75, 105, and 135~K \citep{Oberg2007}. The colour of each spectrum is linked to the temperature for which the data was recorded, i.e., blue and red are the lowest and highest temperature, respectively. The spectral visualization in LIDA also contains the annotations for the vibrational modes of the molecule. Four spectral features are indicated for H$_2$O ice. The feature around 3800~cm$^{-1}$ (2.63~$\mu$m) corresponds to the free OH stretching or dangling bond. This band is often observed in amorphous water ice, and decreases upon compaction after ion irradiation of the ice as shown by \citet{Palumbo2006_oh}. However, \citet{Bossa2014,Bossa2015} suggest that the OH dangling bond is only partially a suited tracer of ice porosity, as a non detection does not fully exclude that an ice is still somewhat porous. After the ice is warmed-up, the dangling bond is no longer observed in this water ice spectrum. The most prominent feature is the absorption band around 3300~cm$^{-1}$ (3~$\mu$m), which refers to the OH bulk stretching in the ice. This band is broad and relatively symmetric at low temperatures, whereas it becomes narrow and sharp at higher temperatures. This variation in the shape of the band is due to the phase transition of water ice from the amorphous to the crystalline structure. The water bending mode is observed at 1666~cm$^{-1}$ (6~$\mu$m). The effect of the temperature on this feature is flattening of the band during warm-up. Finally, the libration water band is observed around 800~cm$^{-1}$. The peak position of this band is also sensitive to the physical conditions of the ice, and is blue-shifted at higher temperatures.

\begin{figure*}
   \centering
   \includegraphics[width=\hsize]{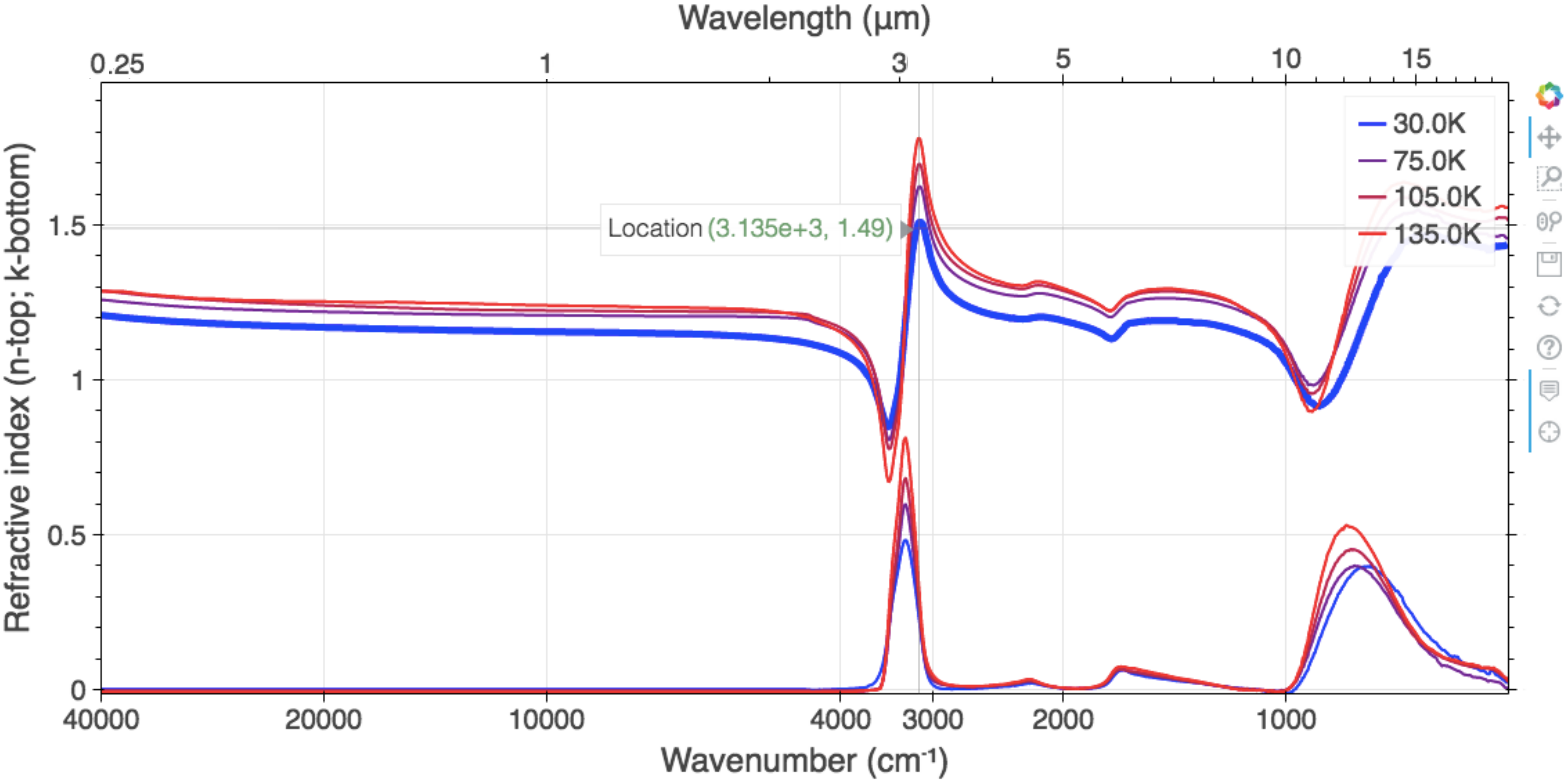}
      \caption{Screenshot showing the UV/vis and mid-IR refractive index of pure H$_2$O at different temperatures. The colour is associated with the temperature of the ice. The hover shows the wavenumber in cm$^{-1}$ and refractive index values as indicated in this figure.}
         \label{spectrum_refacindex}
   \end{figure*}

In Figure~\ref{spectrum_refacindex}, we display the UV/vis and mid-IR refractive index (0.25$-$20~$\mu$m) of pure H$_2$O ice at 30, 50, 100 and 150~K. The UV/vis was measured on the the OASIS setup \citep{He2022}, whereas the mid-IR values were calculated using the refractive index calculator available via LIDA (see Section~\ref{refrac_index}). The water ice refractive index shows a clear dependence with the temperature. In particular, the real refractive index at 670~nm is adopted as 1.29 and 1.32 for the amorphous and crystalline phases, respectively \citep[e.g.,][]{Warren1984, Hudgins1993, Mastrapa2008, Mastrapa2009} which are higher than the values presented in this paper.



\subsection{3D molecule viewer}
\label{3dview}

The 3D molecule viewer aims to provide complementary information about the molecules in the ice analogues available through LIDA. The viewer is built with \texttt{Jmol}\footnote{\url{http://jmol.sourceforge.net/}}, an open-source Java package for visualization of chemical structures in 3D \citep{jmol}. The web rendering of the viewer is done via \texttt{JSmol}, an interactive browser object that is written in JavaScript and utilizes \texttt{HTML5}.

\begin{figure*}
   \centering
   \includegraphics[width=\hsize]{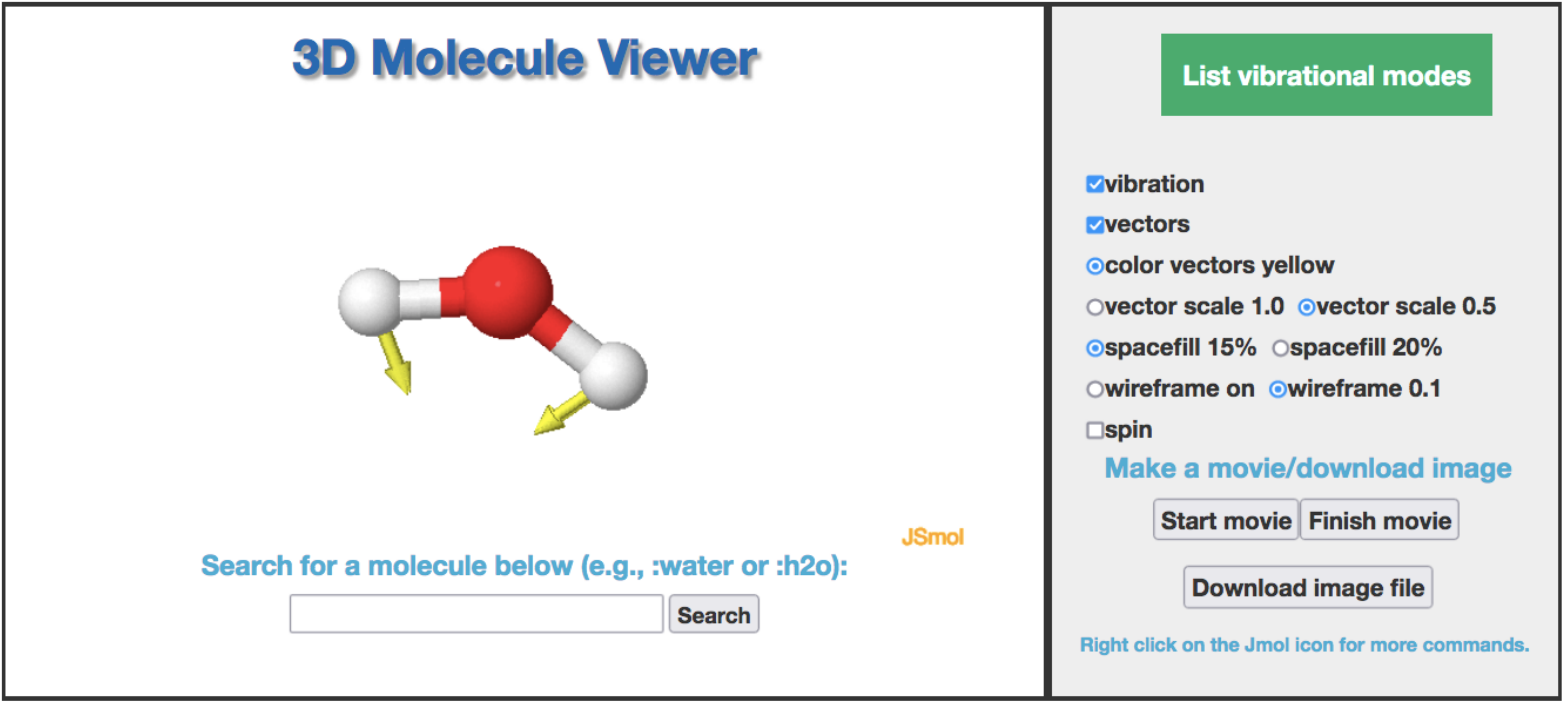}
      \caption{Screenshot of the three-dimensional (3D) molecule viewer embedded in LIDA using the \texttt{JSmol} package. The viewer is connected to external public databases of molecule structures and can be viewed using the searching bar below the molecules. A few dedicated controls can be used to move and animate the molecule. The calculated vibrational modes can be listed and  animated using the green button. As proof of concept, the yellow arrows indicate the vibrational motion of the water bending mode seen at 1643~cm$^{-1}$ in Figure~\ref{spectrum}. Buttons for making a movie and download the image file as PNG are also provided.}
         \label{mol3d}
   \end{figure*}

\texttt{JSmol} has several built-in functions that are also available in this tool, such as measurements of distances and angles, visualization of vibrational modes, animations, orbitals, and surfaces. As a 3D viewer, the molecule can be rotated to different angles, and change the type of the bonds to wireframes. A few dedicated controls are available in the viewer of LIDA, for example, {\it spin} to rotate the molecule; {\it vibration} to show an animation of the vibrational modes; {\it vectors} to show the direction of the vibration modes of the functional groups. All these capabilities are important for a better understanding of the spectroscopic properties of the molecules available in LIDA. It should be noted, that in the ice environment, molecular rotations are quenched and vibrations are hindered depending on the ice matrix. Furthermore, the ice geometry changes with the variation of the temperature and upon irradiation, which also affects the molecular vibrations.  

With \texttt{JSmol} linked to LIDA, one can animate the normal vibrational modes of the molecules when visualizing their IR ice spectra. This is performed by reading ``.xyz'' via \texttt{JSmol}, which contains information about the molecular geometry in Cartesian coordinates, as well as the normal frequencies of the vibrational modes. The default \texttt{JSmol} buttons to control the vibrational modes animations are disabled when the ``.xyz'' is not available yet in LIDA. In Section~\ref{vibmodes} we provide further details about the calculation of the vibrational modes used in the database. This viewer only shows one molecule per ice analogue. This means that for an ice mixture such as H$_2$O:CH$_3$CH$_2$OH, only the H$_2$O molecule is immediately displayed in the viewer. To allow the user to visualize other molecules (e.g., CH$_3$CH$_2$OH), the 3D Molecule Viewer is linked to PubChem\footnote{\url{https://pubchem.ncbi.nlm.nih.gov/}} \citep{Pubchem2011, Pubchem2021} that is a comprehensive database of freely accessible chemical information maintained by the National Center for Biotechnology Information (NCBI). Searching for a molecule is as simple as typing \texttt{:ethanol} to visualize the 3D shape of CH$_3$CH$_2$OH. The colon symbol ``:'' provides the key to connect with the PubChem database. These databases contain detailed information on several molecules, that can help the user to understand different aspects of the molecular properties. 

Figure~\ref{mol3d} shows an example of the 3D molecule viewer, that displays a screenshot of the bending mode animation of the H$_2$O molecule.

\subsection{Vibrational modes calculation}
\label{vibmodes}
The vibrational modes of the molecules in LIDA are calculated with the ORCA\footnote{\url{https://orcaforum.kofo.mpg.de}} software \citep{Orca2012, Orca2018, Orca2020} that contains a wide variety of quantum chemistry methods for different purposes. In the 2022 release of LIDA, the aim of the calculation of the vibrational modes is to show the animation of the vibrational modes, and, therefore, the focus is not on the accuracy of the vibrational frequencies. That have to be taken from experimental values. For the calculations, it is assumed that a molecule is isolated, not in a matrix surrounded by other molecules, and in the electronic ground-state. In addition, ORCA considers that all vibrations are strictly harmonic. The consequence of such approaches is that the wavenumbers of some vibrational modes calculated with ORCA deviate from the wavenumbers of the absorption bands observed in experimental IR spectra or may even not be present. The numerical error in the calculation of vibrational frequencies with ORCA may be as large as 50~cm$^{-1}$, although it is considerably lower in most of the cases. Nonetheless, vibrational mode assignments are correct and can be used as a tool to visualize the animation of the molecular motions.

For the molecule geometry optimization and calculation of the vibrational modes we adopt the Density Functional Theory (DFT) with the functional B3LYP that stands for ``Becke, 3$-$parameter, Lee-Yang-Parr'' \citep{Becke1993, Stephens1994}. The input geometry of the molecules is taken from the PubChem database. The vibrational frequencies calculated for the molecules in the database can be visualized in the 3D molecule viewer described in Section~\ref{3dview}. Additionally, the modes with calculated frequencies are indicated in green in the annotations of the spectrum visualization. Rotational transitions are not available in these files because they are quenched in the ice environment.  

\section{Online tools and applications}
\label{on_tools}
In this section, we introduce two new online tools focused on the creation of synthetic spectra using the laboratory data from the database and the derivation of the CRI at IR wavelengths of ice samples. These tools also have an intuitive graphical user interface that makes it easier to use and download the output results. The details are given in the subsections below. 

\subsection{SPECFY}
\label{synt_spec}
\texttt{SPECFY} is an online tool available through LIDA to construct synthetic spectra of protostars containing ice absorption bands. This tool uses Python \texttt{Flask} for rendering the web-page and \texttt{JavaScript} for showing the absorbance spectra in a drop-down menu to be used by \texttt{SPECFY}. The web interface of \texttt{SPECFY} is shown in Appendix~\ref{Specfy}. Next subsections describe the tool and show practical applications of how to use \texttt{SPECFY} to interpret astronomical observations.

\subsubsection{Synthetic spectra}
To construct a synthetic spectrum with multiple ice features, \texttt{SPECFY} performs a linear combination of experimental data in LIDA which is available via a drop-down menu in the web interface. The linear combination is given by:
\begin{equation}
    \tau_{\rm{\lambda}}^{\rm{tot}} =  \sum_{i=0}^{n} w_i \tau_{\rm{\lambda,i}}^{\rm{lab}},
    \label{tot_tau}
\end{equation}
where $w_{i}$ is the weighting factor used to increase or decrease the intensity of the ice bands, and $\tau_{\rm{\lambda,i}}^{\rm{lab}}$ is calculated with Equation~\ref{od_eq}. The weighting factor $w_{i}$ is calculated by the following equation:
\begin{equation}
    w_i = \frac{N_{\rm{ice}}^{\rm{inp}}}{N_{\rm{ice}}^{\rm{lab}}},
\end{equation}
where $N_{\rm{ice}}^{inp}$ is the input ice column density provided by the user in LIDA, and $N_{\rm{ice}}^{lab}$ is the ice column density of the sample itself, which is calculated with Equation~\ref{CD_eq}. For example, if the user requires a column density of $10^{18}$~cm$^{-2}$ and the experimental spectrum has a column density of $10^{17}$~cm$^{-2}$, the selected spectrum will be multiplied by a factor of 10 in Equation~\ref{tot_tau}. It is worth noting that all experimental data is interpolated during the linear combination to ensure consistency of the method and avoid spectral range variations of the input data.

Besides the ice spectra hosted in LIDA, the template amorphous silicate spectrum of the galactic center source GCS~3 taken from \citet{Kemper2004} is also available to be combined with the ices. This spectrum was observed with ISO towards the Galactic Center, and has been used as a template to remove the silicate features observed toward protostars in previous works \citep[e.g.,][]{Boogert2008, Bottinelli2010}. In LIDA, this silicate spectrum is important for synthetic spectrum calculations because it makes it possible to check the effects of the Si$-$O bands when blended with ice absorption features. However, we stress that no mixing rule such as Maxwell Garnett \citep{Garnett1904, Garnett1906} and Bruggeman \citep{Bruggeman1935, Bruggeman1936} theories is assumed in this procedure. In practice, \texttt{SPECFY} assumes isolated materials. Additionally, this tool does not include secondary effects of grain size and geometry nor scattering processes that might affect the shape of the ice bands. Those features will be included in future work dedicated to improving \texttt{SPECFY}.

\begin{figure*}
   \centering
   \includegraphics[width=15cm]{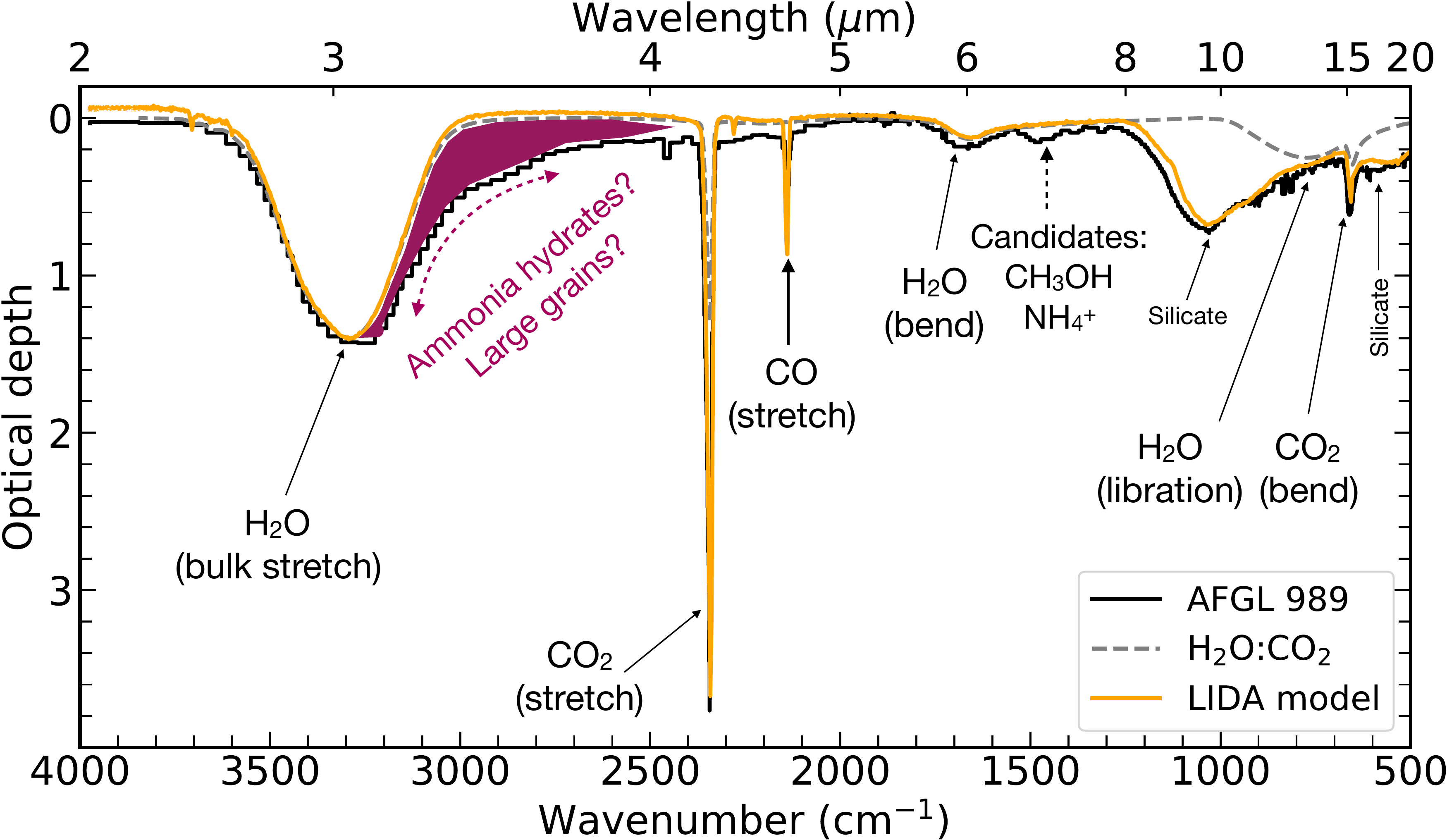}
      \caption{AFGL~989 $vs.$ LIDA model. LIDA model using the \texttt{SPECFY} online tool (orange) superposed to the ISO spectrum of the protostar AFGL~989 (black) in optical depth scale taken from \citet{Gibb2004}. The synthetic spectrum in optical depth scale is composed by the linear combination of three ice spectra (Pure H$_2$O, H$_2$O:CO$_2$ (10:1), and CO:CO$_2$ (2:1)) and silicate template from GCS~3 source. The dominant ice spectrum is H$_2$O:CO$_2$ (10:1) shown by the grey dashed line. The assignments of a few bands are indicated. The red hatched highlights an infrared absorption excess attributed to ammonia hydrates and large grains (see text).}
         \label{synhtetic_afgl}
   \end{figure*}

The combined ice spectrum can be used to match observational data. As an example, we create a synthetic spectrum using the parameters described in Table~\ref{ss_par}. The results are shown in Figure~\ref{synhtetic_afgl} and the outputs in the web interface of \texttt{SPECFY} are displayed in Figure~\ref{synhtetic}. The LIDA model in optical depth scale is constructed with \texttt{SPECFY} by combining ice and silicate spectra with different input column densities. The ice components in this combination are composed of pure H$_2$O at 15~K, and the mixtures H$_2$O:CO$_2$ (10:1) and CO:CO$_2$ (2:1). These three ice samples comprise the most abundant ice molecules observed toward protostars \citep{Oberg2011, Boogert2015}. Superposed to the LIDA model, we display the spectrum of the protostar AFGL~989, observed with {\it ISO} \citep{Gibb2004}. The good agreement between the model and the strong bands in observations show that \texttt{SPECFY} is a useful tool to model astronomical data. This solution is not necessarily unique to the AFGL~989 spectrum, but this methodology provides the means to help in the quantification of the ice column densities as well as with the interpretation of astronomical observations.

The H$_2$O:CO$_2$ ice mixture dominates the absorption profile of the band at 3~$\mu$m, but it cannot explain entirely the absorption excess of the spectral red wing region of AFGL~989. The nature of this strong absorption profile is under debate, but is often attributed to scattering due to large grains \citep[e.g.,][]{Boogert2000} and ammonia hydrates \citep[H$_2$O:NH$_3$; e.g.,][]{Merrill1976, Dartois2002}. The water ice bending and libration modes are also observed around 6~$\mu$m and 13.6~$\mu$m. Likewise, the CO$_2$ bands at 4.27~$\mu$m and around 15~$\mu$m are not entirely modelled by the carbon dioxide fraction in the H$_2$O:CO$_2$ mixture. Additional CO$_2$ is added by the CO:CO$_2$ ice mixture. A fraction of carbon monoxide is expected to coexist in the same ice matrix of carbon dioxide as indicated in astronomical observations \citep{Pontoppidan2008, Poteet2013}. Although this combination matches relatively well the two CO$_2$ bands, it results in a higher CO ice peak at 4.67~$\mu$m. Finally, the absorption profile of the silicate is relatively well reproduced with the amorphous silicate of GCS~3. Similar to the unclear origin of the absorption excess around 3.3~$\mu$m, other strong absorptions are observed at 6~$\mu$m usually associated with organic refractory material \citep{Gibb2002, Boogert2008} and at 6.85~$\mu$m, that has been attributed to CH$_3$OH \citep[e.g.,]{Bottinelli2010} and NH$_4^+$ \citep[e.g.,][]{Keane2001, Schutte2003, Mate2009, Mate2012}. This exercise shows that the resources available in LIDA can be used to analyse the spectra of protostars and obtain ice column densities. 


Next, the optical depth spectrum can be converted to a flux scale in Jy units by adopting the continuum SEDs of different protostars. We compiled and added to LIDA the continuum SED of seven protostars as calculated by \citet{Gibb2004} and \citet{Boogert2008}, which are listed in Table~\ref{SEDcont}. The sources are representative of objects Class I and Class II and have spectral data obtained with ground- and space-based telescopes. Except in the cases of Elias 29 and AFGL~989, which were observed with the ISO/short-wavelength spectrometer (SWS) in the entire range between 2 and 30~$\mu$m, all sources have coverage of 2.5$-$5~$\mu$m (except 4.0$-$4.4~$\mu$m) and 5$-$30~$\mu$m. The former interval is based on the VLT/ISAAC observations summarized in \citet{Pontoppidan2003} and \citet{vanBroekhuizen2005} or Keck NIRSPEC \citep{McLean1998} observations. The latter range is constrained by space-based observations with the Infrared Spectrograph (IRS) of the {\it Spitzer} Space Telescope. Despite the careful SED determination by \citet{Gibb2004} and \citet{Boogert2008}, inaccuracies may still occur, and this must be taken into account when using these data. Once the continuum SED is known, it can be used to convert the ice experimental spectra from optical depth to a flux scale. The conversion to the synthetic spectrum in flux scale is performed by:
\begin{equation}
    F_{\lambda}^{\rm{synth}} = F_{\lambda}^{\rm{cont}} \rm{exp}(-\tau_{\lambda}^{lab})
    \label{flux_scale}
\end{equation}
where $F_{\lambda}^{\rm{cont}}$ is the continuum SED of the protostar.

\begin{table*}
\caption{\label{SEDcont} Continuum SEDs available in the SPECFY tool compiled from \citet{Gibb2004} and \citet{Boogert2008}.}
\renewcommand{\arraystretch}{1.2}
\centering 
\begin{tabular}{lcccc}
\hline\hline
Protostar & Continuum Model (Jy) & Continuum method$^a$ & Observations\\
\hline
\multicolumn{4}{c}{\bf{Class I (disk, envelope)}}\\
\hline
Elias~29 & B2000 & Blackbody & ISO/SWS\\
AFGL~989 & G2004 & Polynomial + Blackbody & ISO/SWS\\
CrA~IRS7~A & B2008 & Polynomial & ISAAC/VLT \& {\it Spitzer}/IRS\\
CrA~IRS7~B & B2008 & Polynomial & ISAAC/VLT \& {\it Spitzer}/IRS\\
IRAS~23238+7401 & B2008 & Polynomial & NIRSPEC/Keck \& {\it Spitzer}/IRS\\
L1014~IRS & B2008 & Polynomial & NIRSPEC/Keck \& {\it Spitzer}/IRS\\
\hline
\multicolumn{4}{c}{\bf{Transition from Class I to Class II (disk, tenuous envelope)}}\\
\hline
DG~Tau~B & B2008 & Polynomial & NIRSPEC/Keck \& {\it Spitzer}/IRS\\
CRBR~2422.8-3423 & B2008 & Polynomial & NIRSPEC/Keck \& {\it Spitzer}/IRS\\
\hline
\end{tabular}
\tablefoot{$^a$Polynomial: low-order ($\le$ 3) polynomial function. Blackbody: a single or multiple blackbody curves to fit the local continuum adjacent to the ice absorption features.}
\end{table*}

\begin{figure}
   \centering
   \includegraphics[width=\hsize]{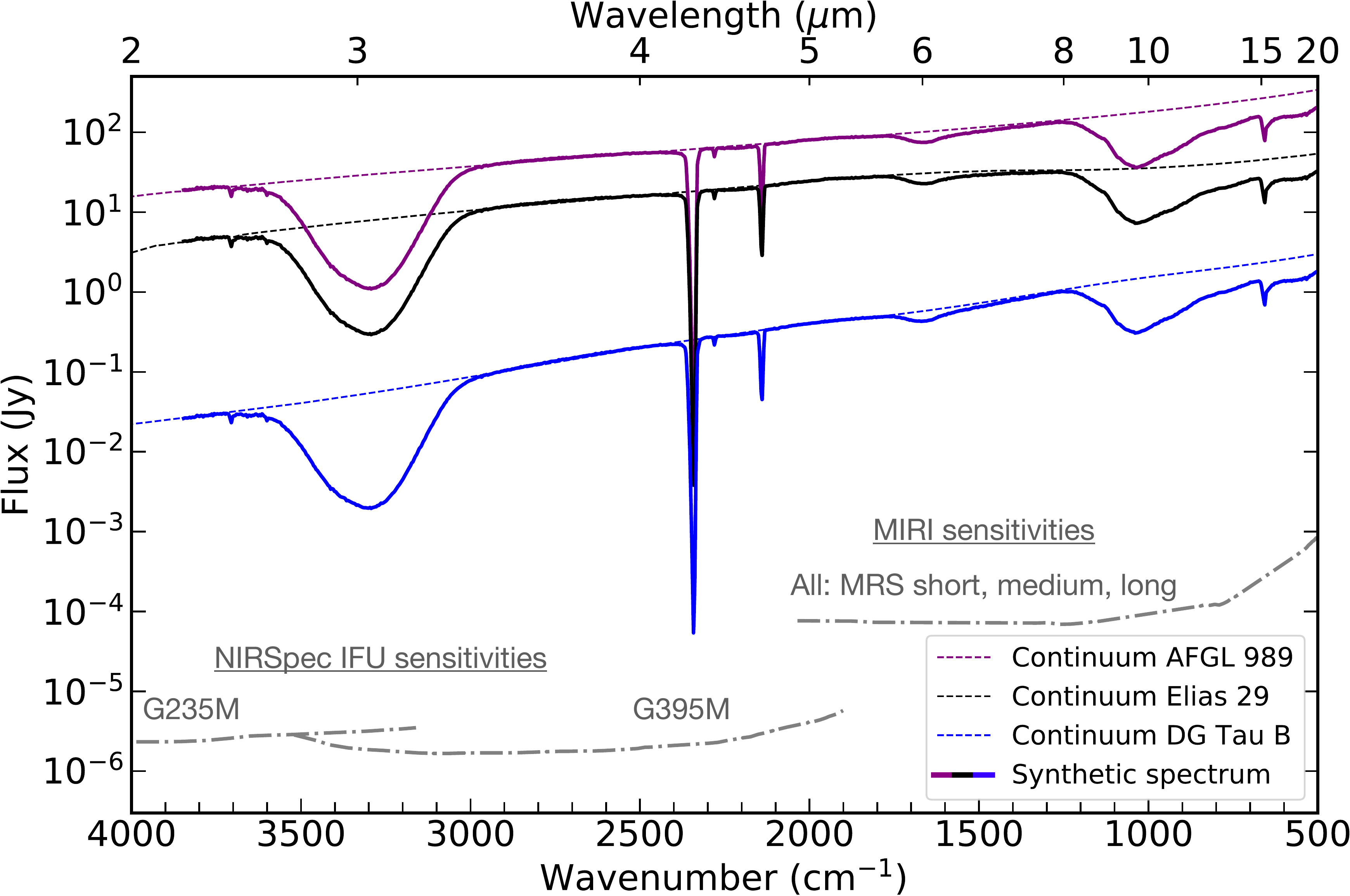}
      \caption{Output spectra (solid lines) from SPECFY in Jy showing the effect of selecting different continuum SED templates (dashed lines). The sensitivities of JWST/NIRSpec medium resolution and MIRI for 10000~s integration time are shown for comparison by the grey dot-dashed lines.}
         \label{synhtetic_cont}
   \end{figure}

Figure~\ref{synhtetic_cont} shows three synthetic spectra using the continuum templates from AFGL~989, Elias~29 and DG Tau B, which represent three protostar categories, i.e., a high mass protostar, a low mass protostar, and a protoplanetary disk, respectively. The continuum applied to the optical depth model is displayed in Figure~\ref{synhtetic_afgl} and the output in the web interface displayed in Figure~\ref{synhtetic}. The effect of the continuum in this example is characterized by different flux intensities and by changing the slope of the protostar SED. Additionally, Figure~\ref{synhtetic_cont} shows the sensitivity limits for the filters G235M and G395M of JWST/Near-Infrared Spectrometer integral field unit (NIRSpec/IFU) and all filters of the Mid-Infrared Instrument at Medium Resolution Spectroscopy (MIRI/MRS). These values represent the minimum detectable signal corresponding to signal-to-noise ratio of 10 obtained with an on-source integration time of 10000 seconds \citep[][]{Glasse2015, Pontoppidan2016}. This comparison shows that ices can be easily detected with JWST toward sources with continuum SED similar to AFGL~989, Elias~29 and DG Tau B. With this feature in LIDA, one can generate input spectra for the JWST Time Exposure Calculator\footnote{\url{https://jwst.etc.stsci.edu/}} (ETC) that can be used in future proposals cycles.



\begin{table*}
\caption{Selected ice spectra and continuum model to construct a synthetic protostar spectrum. As an example, see Figure~\ref{synhtetic_afgl}.}
\label{ss_par}      
\centering
\setlength{\tabcolsep}{3pt} 
\renewcommand{\arraystretch}{1.3} 
\begin{tabular}{c c c c}        
\hline\hline                 
\multicolumn{4}{c}{\underline{Spectrum selection}}\\
Analogue & $T$ (K) & $N_{\rm{ice}}^{\rm{inp}}$ (cm$^{-2}$) & Reference\\
\hline
Pure H$_2$O & 15 & 1.4 $\times$ $10^{17}$ & \citet{Oberg2007} \\
H$_2$O:CO$_2$ (10:1) & 10 & 5.3 $\times$ $10^{18}$ & \citet{Ehrenfreund1997}\\
CO:CO$_2$ (2:1) & 15 & 9.5 $\times$ $10^{17}$ & \citet{vanBroekhuizen2006}\\
Silicate GCS~3 & ... & 1.0 $\times$ $10^{20}$ & \citet{Kemper2004}\\
\hline
\multicolumn{4}{c}{\underline{Continuum selection}}\\
Object & Continuum model & Flux unit & Reference\\
\hline
Elias~29 & B2008 & Jansky & \citet{Boogert2008}\\
AFGL~989 & G2004 & Jansky & \citet{Gibb2004}\\
DG~Tau~B & B2008 & Jansky & \citet{Boogert2008}\\
\hline
\end{tabular}
\tablefoot{All data is interpolate in the range between 2.6 and 20~$\mu$m.}
\end{table*}

\subsubsection{Functional groups in protostellar spectra}
\begin{figure*}
   \centering
   \includegraphics[width=10cm]{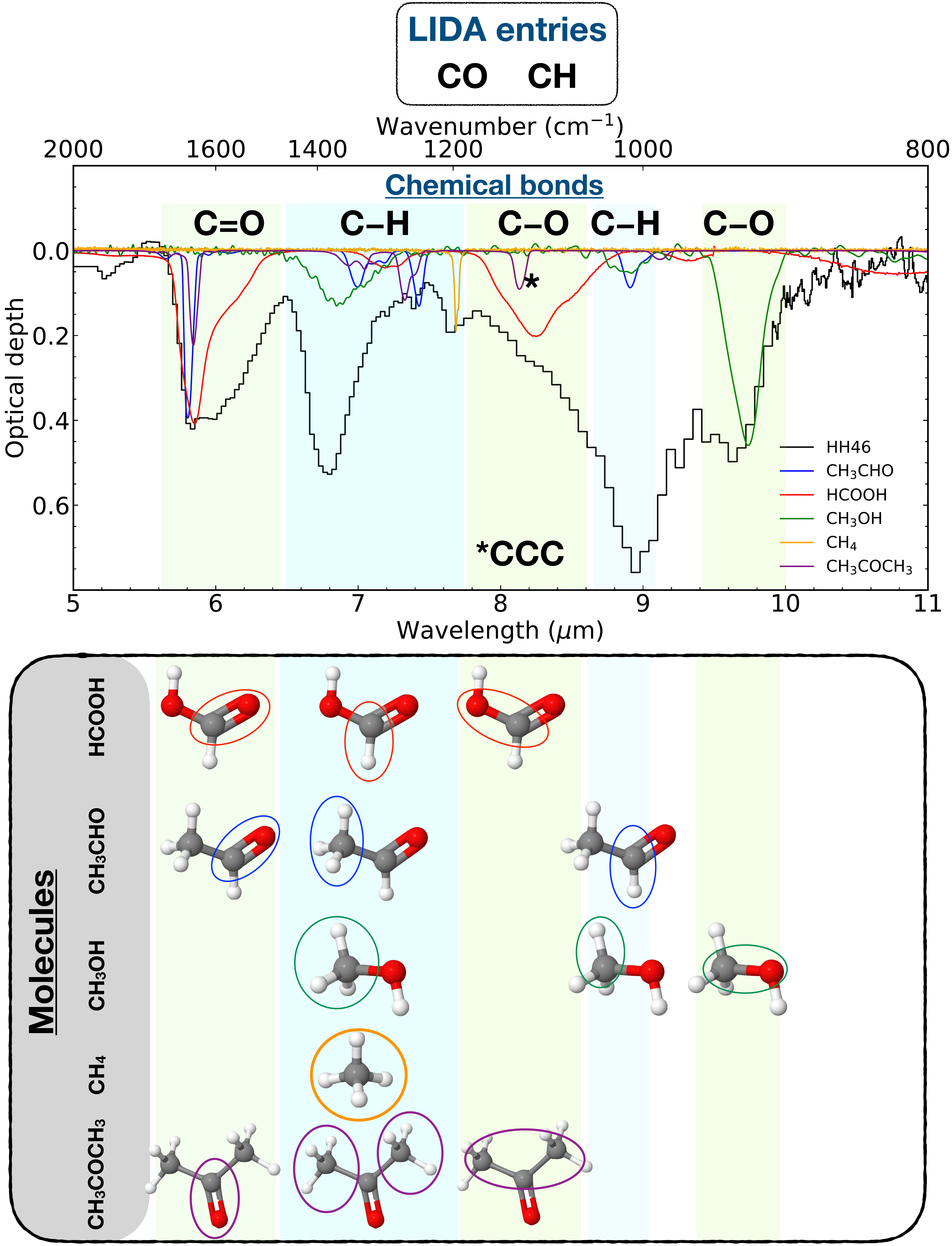}
      \caption{Illustration on how to use LIDA to interpret astronomical observations {\it Top:} LIDA entries to search for molecules sharing CO and CH chemical bonds. {\it Middle:} Selected experimental data scaled to the water-silicate subtracted spectra of the protostar HH46. {\it Bottom:} Molecules representing the pure ices used to match the HH46 spectrum. The ellipses indicate which part of a molecule is responsible for a specific absorption band, and the colours follow the same colour code used in the middle panel.}
         \label{func_group}
   \end{figure*}

LIDA also has the capability of searching for molecules containing similar associations of atoms and some functional groups, as described in Section~\ref{searching_cap}. Once these are chosen, one can select them from the dropdown menu in \texttt{SPECFY} to construct a model spectrum for comparison with the observations. A practical example is given in Figure~\ref{func_group}. We use two separate entries (\texttt{CO} and \texttt{CH}) in the ``Spectral data'' field of LIDA, to search for molecules sharing carbon-oxygen bonds (e.g., carbonyl-bearing molecules, alcohols) and carbon-hydrogen bonds as shown in the top panel of Figure~\ref{func_group}. From the \texttt{CO} entry, several ice analogues are found, including HCOOH, CH$_3$OH, CH$_3$CHO and CH$_3$COCH$_3$. Similarly, the \texttt{CH} entry returns the same molecules because they contain CO and CH chemical bonds. In addition, LIDA also finds CH$_4$ based on the query request.

The vibrational modes of functional groups containing a carbonyl group, C$-$O and C$-$H bonds have been assigned in the spectra of protostars \citep[e.g.,][]{Gibb2004, Boogert2015}. To illustrate how LIDA can further support astronomical data interpretation, we show in the middle panel of Figure~\ref{func_group}, the experimental spectra of HCOOH, CH$_3$CHO, CH$_3$OH, CH$_3$COCH$_3$ and CH$_4$ scaled to the spectra of the low-mass protostar HH46 \citep{Boogert2008}. The HH46 spectrum is subtracted of the water ice and silicate. The chemical bonds associated with the absorption bands are indicated in the green and blue shaded areas. The bottom panel of Figure~\ref{func_group} highlights the chemical bonds of the molecules contributing to the absorption bands towards HH46. The parameters used to scale laboratory data to the observations are given in Table~\ref{hh46_par}. This exercise shows that LIDA can be used to identify the chemical bonds related to different absorption bands, and provide upper limit column densities for ices. Figure~\ref{func_group} also highlights the blending of bands at different spectral regions. For example, the C$=$O stretching modes of HCOOH, CH$_3$CHO and CH$_3$COCH$_3$ lie almost at the same wavelength, which hints for the need of high sensitivity and spectral resolution observational data that will be provided by JWST. Clearly, this is an important tool to explore the contribution of different functional groups and chemical bonds to the overall absorption profile of features observed in interstellar ice spectra. It should be noted that such synthetic spectra allow reproducing observed data, but do not provide a necessarily unique solution. Other public codes, such as the \texttt{ENIIGMA} fitting tool \citep{Rocha2021} have the goal of quantifying the degeneracy of those fits when a large data-set of inputs is taken into account.

\begin{table*}
\caption{Ice spectra selected from LIDA entries \texttt{CO} and \texttt{CH} and their column densities after manually scaling to HH46 spectrum shown in Figure~\ref{func_group}.}
\label{hh46_par}      
\centering
\setlength{\tabcolsep}{3pt} 
\renewcommand{\arraystretch}{1.5} 
\begin{tabular}{l c c c}        
\hline\hline                 
Analogue & $T$ (K) & $N_{\rm{ice}}^{\rm{inp}}$ (cm$^{-2}$) & Reference\\
\hline
Pure HCOOH & 15 & 1.9 $\times$ $10^{17}$ & \citet{Bisschop2007}\\
Pure CH$_3$CHO & 15 & 6.1 $\times$ $10^{17}$ & \citet{Scheltinga2018}\\
Pure CH$_3$COCH$_3$ & 15 & 1.7 $\times$ $10^{17}$ & \citet{Rachid2020}\\
Pure CH$_3$OH & 15 & 7.7 $\times$ $10^{17}$ & \citet{Fraser2004}\\
Pure CH$_4$ & 15 & 1.4 $\times$ $10^{17}$ & \citet{Fraser2004}\\
\hline
\end{tabular}
\end{table*}

\subsection{Infrared refractive index calculator}
\label{refrac_index}

In this section, we introduce the refractive index online calculator which is publicly available through LIDA. The web interface of this tool is shown in Figure~\ref{icenk_page} of Appendix~\ref{app_specfy}. This tool uses the approach adopted in \citet{Rocha2014} for the \texttt{NKABS} code and is briefly described below.
   
\begin{figure}
   \centering
   \includegraphics[width=\hsize]{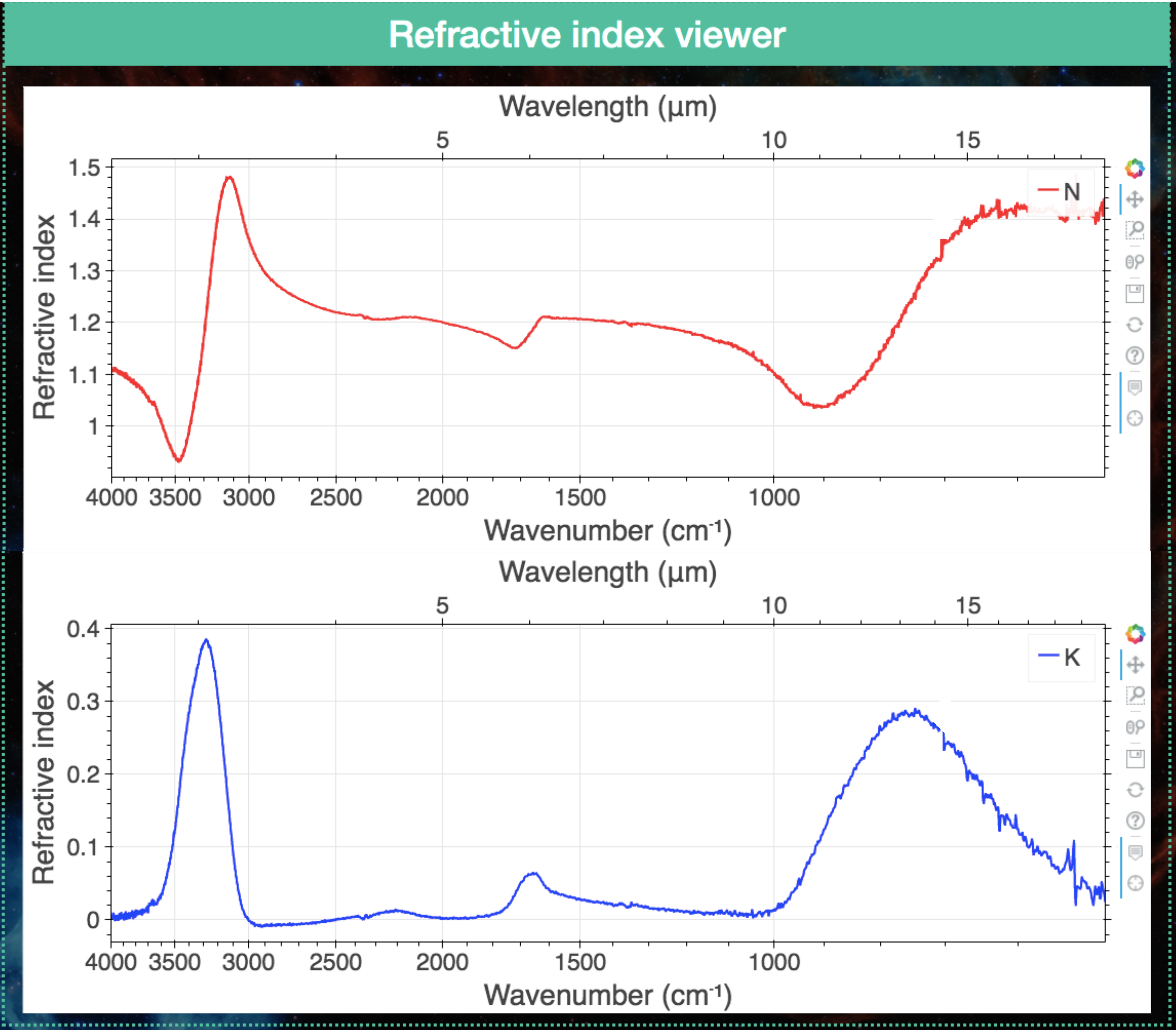}
      \caption{Screenshot of the refractive index viewer showing the wavelength-dependent CRI of H$_2$O ice at 30~K. The top and bottom panels show the real and imaginary  refractive index values, respectively.}
         \label{nk_viewer}
   \end{figure}

The goal of the tool is to calculate the real ($n$) and imaginary ($k$) parts of CRI ($\tilde{m}$) from the absorbance spectrum (Equation~\ref{abs_eq}) of the ice sample as a function of the wavenumber ($\nu$, in units of cm$^{-1}$). The input experimental data is the absorbance spectrum in an \texttt{ascii} format. Other input parameters are required before starting the calculation. They are the thickness of the ice sample ($d$) in $\mu$m, the refractive index of the sample around 670~nm ($n_0$), or at the wavelength of the HeNe laser used in the experiments, the real refractive index of the substrate, and the MAPE (Mean Average Percentage Error), that is used as stop criterion.



Equations~\ref{n-value} and \ref{k-value} are solved interactively, and new values of $k$ are used to calculate new values of $n$. Subsequently, new $n$ improves $k$, until the convergence criteria is reached. The numerical implementation of these equations is described in \citet{Rocha2014}, and follows the procedure presented in \citet{Ohta1988} to solve the Kramers-Kronig equation. As an example, we calculate the CRI values of pure H$_2$O at 30, 75, 105 and 135~K. The H$_2$O ice IR spectra used as input are taken from \citet{Oberg2007}, namely, \texttt{Pure H$_2$O (3000~ML)}\footnote{\url{https://icedb.strw.leidenuniv.nl/data/14}}. Table~\ref{nkabs_par} list the $n_0$ values, the number of iterations used by the tool and the final MAPE.

The data from OASIS and from the theoretical calculation cover two spectral ranges, i.e., 0.25 to 0.7~$\mu$m and 2 to 20~$\mu$m, respectively. The interval between 0.7 and 2~$\mu$m is not available in the experimental spectrum. We deal with this missing data using different approaches to determine $n$ and $k$. For $k$ values we extrapolate the imaginary refractive index at 2~$\mu$m (10$^{-4}$) until 0.25~$\mu$m. In the case of $n$, we use a low order polynomial to link the water ice $n$ values from \citet{He2022} to the data starting at 2~$\mu$m. The caveat in this approach is that we do not take into account the water ice absorption bands in the interval between 0.7 and 2~$\mu$m. However, the absorption features between 1.4 and 1.8~$\mu$m for amorphous and crystalline water ice are very weak \citep{Mastrapa2008}. For example, the $k$ values calculated by \citet{Mastrapa2008} range from 10$^{-5}$ to 10$^{-3}$ which is close to the value used in our extrapolation (10$^{-4}$). Similarly, the variation in $n$ is 0.2\% between the lowest and highest values.

The results are visualized in the ``Refractive index viewer'' shown in Figure~\ref{nk_viewer} for the data at 30~K. The download of the data files and plots is also available. A terminal-based version of this tool is available for download in the GitHub\footnote{\url{https://github.com/leiden-laboratory-for-astrophysics/refractive-index}} repository of LIDA. Both Linux and executable files for Windows platforms can be used for highly resolved spectral data that demands high computational efficiency. 

\begin{table}
\caption{Input parameters used to calculate the mid-IR water ice CRI.}
\label{nkabs_par}      
\centering
\setlength{\tabcolsep}{3pt} 
\renewcommand{\arraystretch}{1.5} 
\begin{tabular}{l c c}        
\hline\hline                 
Parameters & Values & Reference\\
\hline
Thickness$^a$ (cm) & 1.0 $\times$ 10$^{-4}$ & \citet{Oberg2007}\\
$n_0^{30K}$ & 1.16 & \citet{He2022}\\
$n_0^{75K}$ & 1.21 & \citet{He2022}\\
$n_0^{105K}$ & 1.23 & \citet{He2022}\\
$n_0^{135K}$ & 1.25 & \citet{He2022}\\
$n_{\rm{substrate}}$ & 1.73 & \citet{Querry1987}\\
Initial MAPE & 0.1\% & ...\\
\hline
\multicolumn{3}{c}{After calculation}\\
\hline
Final MAPE & $\leq$ 0.03\% & ...\\
Iterations & 5 & ...\\
\hline
\end{tabular}
\tablefoot{$^a$Assuming 1~ML $\sim$ 3.4$\AA$ \citep{Gonzalez2019}.}
\end{table}

\section{Future upgrades}
\label{future}
LIDA has already relevant data for all molecules securely or tentatively detected toward protostars (e.g., Tables~\ref{icedb_list} and \ref{analogue_list}), and covers the most abundant species for JWST, but more data are necessary to further boost the interpretation of upcoming observations. Table~\ref{missing_data} lists the molecules that are missing in the current version of LIDA, but are being measured or will be targeted in future experiments. Similarly, temperature and wavelength-dependent refractive index values of CO, CO$_2$, NH$_3$ and CH$_3$OH were recently measured on the OASIS setup, and will be added to the database after data reduction (Rachid et al. {\it in prep.}). We also mention that LIDA is available to host data from other astrochemistry groups. Here we hope that one central point to search for ice properties will be considered helpful for the full ice community.
\begin{table}
\caption{List of missing molecules in LIDA, which will be included via new measurements or data sharing from other laboratories.}
\label{missing_data}
\centering
 \begin{tabular}{cc}
\hline\hline
\multicolumn{2}{c}{\bf{< 6 atoms}}\\
Molecule & Name\\
\hline
C$_2$H$_2$ & Acetylene\\
H$_2$S & Hydrogen sulfide\\
HNCO & Isocyanic acid\\
HCN & Hydrogen cyanide\\
\hline
\multicolumn{2}{c}{\bf{> 6 atoms}}\\
\hline
NH$_2$OH & Hydroxylamine\\
NH$_2$CHO & Formamide\\
H$_2$CO$_3$ & Carbonic acid\\
HOCH$_2$CHO & Glycolaldehyde\\
\hline
\end{tabular}
\end{table}

The online tools will also be further developed to support astronomical data interpretation. With this goal, the effect of grain shape will be available when simulating synthetic spectra of protostars. Additionally, the UV/vis $n$ and $k$ values of difference ices ans ice mixtures will be included in LIDA. Another forthcoming upgrade on LIDA is the inclusion of diagnostic plots relating the peak position and full width at half maximum (FWHM) of ice features that can be compared with the similar information from different astronomical observations. 

\section{Summary and outlook}
\label{summary}
The Leiden Ice Database has served the astronomical community for more than 20 years by providing IR spectra of ice samples. In 2015, all ice IR spectra were assembled in one server and visualization tools were developed. In this paper, we present the most recent version of LIDA that includes over 1100 IR spectra of ice samples in astrophysically relevant conditions, as well as the UV/vis and mid-IR refractive index of H$_2$O at different temperatures. In addition to the large ensemble of experimental data, the current upgrade includes astronomy oriented online tools to help the interpretation of observations provided by JWST, in general, as well as the past ice observations. Both data and tools are offered in a user-friendly format to boost the usability of the database. It is worth mentioning that LIDA is a specific deliverable within ICE AGE, an ERS JWST program.  

The database is under expansion, and spectra of several COMs and refractive index values of other ices will become publicly available in the next months and years. It is also hoped that other laboratory groups will make their ice spectra available through LIDA. Also, the online tools in the database will be further developed to attend to the needs of interpretation of ice observations in the upcoming years with the JWST, the METIS (Mid-Infrared Extremely Large Telescope Imager and Spectrograph) on the Extremely Large Telescope (ELT), and the SPHEREx (Spectro-Photometer for the History of the Universe, Epoch of Reionization and Ices Explorer). More information about LIDA can be found in the public-access online documentation: \url{https://leiden-ice-database.readthedocs.io}.

\begin{acknowledgements}
We thank the thoughtful comments of an anonymous referee on both manuscript and the LIDA website. WRMR thanks Leiden Observatory for financial support. We thank the many (under)graduates, postdocs and staff who have been contributing over many years to the data available in LIDA. We furthermore acknowledge the ICE AGE team whose JWST observing plans have been the trigger for updating the ``old'' Leiden Ice Database. We specifically mention Dr. Adwin Boogert for many useful discussions. LIDA is currently also at the base of interpreting data from MIRI GTO protostar program. We are grateful for continuing support through NOVA, the Netherlands Research School for Astronomy, the NWO through its Dutch Astrochemistry Program (DANII), and the NWO VICI grant ``Unlocking the chemistry of the heavens''. The present work is closely connected to ongoing research within INTERCAT, the Center for Interstellar Catalysis located in Aarhus, Denmark. This project has received funding from the European Research Council (ERC) under the European Union’s Horizon 2020 research and innovation programme (grant agreement No. 291141 MOLDISK). We also acknowledge the technical support of the Computer group at Leiden Observatory.
\end{acknowledgements}

\bibliographystyle{aa}
\bibliography{References}

\begin{thebibliography}{184}
\expandafter\ifx\csname natexlab\endcsname\relax\def\natexlab#1{#1}\fi

\bibitem[{Balkanski(1989)}]{BALKANSKI1989729}
Balkanski, M. 1989, Infrared Physics, 29, 729

\bibitem[{{Baratta} \& {Palumbo}(1998)}]{Baratta1998}
{Baratta}, G.~A. \& {Palumbo}, M.~E. 1998, Journal of the Optical Society of
  America A, 15, 3076

\bibitem[{{Bauschlicher} {et~al.}(2010){Bauschlicher}, {Boersma}, {Ricca},
  {Mattioda}, {Cami}, {Peeters}, {S{\'a}nchez de Armas}, {Puerta Saborido},
  {Hudgins}, \& {Allamandola}}]{Bauschlicher2010}
{Bauschlicher}, C.~W., J., {Boersma}, C., {Ricca}, A., {et~al.} 2010, \apjs,
  189, 341

\bibitem[{{Becke}(1993)}]{Becke1993}
{Becke}, A.~D. 1993, \jcp, 98, 5648

\bibitem[{{Belloche} {et~al.}(2020){Belloche}, {Maury}, {Maret}, {Anderl},
  {Bacmann}, {Andr{\'e}}, {Bontemps}, {Cabrit}, {Codella}, {Gaudel}, {Gueth},
  {Lef{\`e}vre}, {Lefloch}, {Podio}, \& {Testi}}]{Belloche2020}
{Belloche}, A., {Maury}, A.~J., {Maret}, S., {et~al.} 2020, \aap, 635, A198

\bibitem[{Beltr{\'a}n {et~al.}(2015)Beltr{\'a}n, Molina, Aznar, Molt{\'o}, \&
  Verd{\'u}}]{beltran2015double}
Beltr{\'a}n, M.~D., Molina, R.~L., Aznar, M. {\'A}.~S., Molt{\'o}, C.~S., \&
  Verd{\'u}, C.~M. 2015, Sensors, 15, 25123

\bibitem[{{Bernstein} {et~al.}(1995){Bernstein}, {Sandford}, {Allamandola},
  {Chang}, \& {Scharberg}}]{Bernstein1995}
{Bernstein}, M.~P., {Sandford}, S.~A., {Allamandola}, L.~J., {Chang}, S., \&
  {Scharberg}, M.~A. 1995, \apj, 454, 327

\bibitem[{{Bisschop} {et~al.}(2007){Bisschop}, {Fuchs}, {Boogert}, {van
  Dishoeck}, \& {Linnartz}}]{Bisschop2007}
{Bisschop}, S.~E., {Fuchs}, G.~W., {Boogert}, A.~C.~A., {van Dishoeck}, E.~F.,
  \& {Linnartz}, H. 2007, \aap, 470, 749

\bibitem[{{Blake} {et~al.}(1987){Blake}, {Sutton}, {Masson}, \&
  {Phillips}}]{Blake1987}
{Blake}, G.~A., {Sutton}, E.~C., {Masson}, C.~R., \& {Phillips}, T.~G. 1987,
  \apj, 315, 621

\bibitem[{{Boersma} {et~al.}(2014){Boersma}, {Bauschlicher}, {Ricca},
  {Mattioda}, {Cami}, {Peeters}, {S{\'a}nchez de Armas}, {Puerta Saborido},
  {Hudgins}, \& {Allamandola}}]{Boersma2014}
{Boersma}, C., {Bauschlicher}, C.~W., J., {Ricca}, A., {et~al.} 2014, \apjs,
  211, 8

\bibitem[{{Bokeh Development Team}(2018)}]{bokeh2018}
{Bokeh Development Team}. 2018, Bokeh: Python library for interactive
  visualization

\bibitem[{Bolton {et~al.}(2011)Bolton, Chen, Kim, Han, He, Shi, Simonyan, Sun,
  Thiessen, Wang, Yu, Zhang, \& Bryant}]{Pubchem2011}
Bolton, E.~E., Chen, J., Kim, S., {et~al.} 2011, Journal of cheminformatics, 3,
  1

\bibitem[{{Boogert} {et~al.}(2002{\natexlab{a}}){Boogert}, {Blake}, \&
  {Tielens}}]{Boogert2002_isotoplogue}
{Boogert}, A.~C.~A., {Blake}, G.~A., \& {Tielens}, A.~G.~G.~M.
  2002{\natexlab{a}}, \apj, 577, 271

\bibitem[{{Boogert} {et~al.}(2013){Boogert}, {Chiar}, {Knez}, {{\"O}berg},
  {Mundy}, {Pendleton}, {Tielens}, \& {van Dishoeck}}]{Boogert2013}
{Boogert}, A.~C.~A., {Chiar}, J.~E., {Knez}, C., {et~al.} 2013, \apj, 777, 73

\bibitem[{{Boogert} {et~al.}(2015){Boogert}, {Gerakines}, \&
  {Whittet}}]{Boogert2015}
{Boogert}, A.~C.~A., {Gerakines}, P.~A., \& {Whittet}, D. C.~B. 2015, \araa,
  53, 541

\bibitem[{{Boogert} {et~al.}(2002{\natexlab{b}}){Boogert}, {Hogerheijde},
  {Ceccarelli}, {Tielens}, {van Dishoeck}, {Blake}, {Latter}, \&
  {Motte}}]{Boogert2002}
{Boogert}, A.~C.~A., {Hogerheijde}, M.~R., {Ceccarelli}, C., {et~al.}
  2002{\natexlab{b}}, \apj, 570, 708

\bibitem[{{Boogert} {et~al.}(2008){Boogert}, {Pontoppidan}, {Knez}, {Lahuis},
  {Kessler-Silacci}, {van Dishoeck}, {Blake}, {Augereau}, {Bisschop},
  {Bottinelli}, {Brooke}, {Brown}, {Crapsi}, {Evans}, {Fraser}, {Geers},
  {Huard}, {J{\o}rgensen}, {{\"O}berg}, {Allen}, {Harvey}, {Koerner}, {Mundy},
  {Padgett}, {Sargent}, \& {Stapelfeldt}}]{Boogert2008}
{Boogert}, A.~C.~A., {Pontoppidan}, K.~M., {Knez}, C., {et~al.} 2008, \apj,
  678, 985

\bibitem[{{Boogert} {et~al.}(1997){Boogert}, {Schutte}, {Helmich}, {Tielens},
  \& {Wooden}}]{Boogert1997}
{Boogert}, A.~C.~A., {Schutte}, W.~A., {Helmich}, F.~P., {Tielens},
  A.~G.~G.~M., \& {Wooden}, D.~H. 1997, \aap, 317, 929

\bibitem[{{Boogert} {et~al.}(2000){Boogert}, {Tielens}, {Ceccarelli},
  {Boonman}, {van Dishoeck}, {Keane}, {Whittet}, \& {de Graauw}}]{Boogert2000}
{Boogert}, A.~C.~A., {Tielens}, A.~G.~G.~M., {Ceccarelli}, C., {et~al.} 2000,
  \aap, 360, 683

\bibitem[{{Bossa} {et~al.}(2014){Bossa}, {Isokoski}, {Paardekooper}, {Bonnin},
  {van der Linden}, {Triemstra}, {Cazaux}, {Tielens}, \&
  {Linnartz}}]{Bossa2014}
{Bossa}, J.~B., {Isokoski}, K., {Paardekooper}, D.~M., {et~al.} 2014, \aap,
  561, A136

\bibitem[{{Bossa} {et~al.}(2015){Bossa}, {Mat{\'e}}, {Fransen}, {Cazaux},
  {Pilling}, {Robson Monteiro Rocha}, {Ortigoso}, \& {Linnartz}}]{Bossa2015}
{Bossa}, J.-B., {Mat{\'e}}, B., {Fransen}, C., {et~al.} 2015, \apj, 814, 47

\bibitem[{{Bottinelli} {et~al.}(2010){Bottinelli}, {Boogert}, {Bouwman},
  {Beckwith}, {van Dishoeck}, {{\"O}berg}, {Pontoppidan}, {Linnartz}, {Blake},
  {Evans}, \& {Lahuis}}]{Bottinelli2010}
{Bottinelli}, S., {Boogert}, A.~C.~A., {Bouwman}, J., {et~al.} 2010, \apj, 718,
  1100

\bibitem[{{Boudin} {et~al.}(1998){Boudin}, {Schutte}, \&
  {Greenberg}}]{Boudin1998}
{Boudin}, N., {Schutte}, W.~A., \& {Greenberg}, J.~M. 1998, \aap, 331, 749

\bibitem[{{Bouilloud} {et~al.}(2015){Bouilloud}, {Fray}, {B{\'e}nilan},
  {Cottin}, {Gazeau}, \& {Jolly}}]{Bouilloud2015}
{Bouilloud}, M., {Fray}, N., {B{\'e}nilan}, Y., {et~al.} 2015, \mnras, 451,
  2145

\bibitem[{{Bouwman} {et~al.}(2007){Bouwman}, {Ludwig}, {Awad}, {{\"O}berg},
  {Fuchs}, {van Dishoeck}, \& {Linnartz}}]{Bouwman2007}
{Bouwman}, J., {Ludwig}, W., {Awad}, Z., {et~al.} 2007, \aap, 476, 995

\bibitem[{{Brown} {et~al.}(1982){Brown}, {Lanzerotti}, \&
  {Johnson}}]{Brown1982}
{Brown}, W.~L., {Lanzerotti}, L.~J., \& {Johnson}, R.~E. 1982, Science, 218,
  525

\bibitem[{{Bruggeman}(1935)}]{Bruggeman1935}
{Bruggeman}, D.~A.~G. 1935, Annalen der Physik, 416, 665

\bibitem[{{Bruggeman}(1936)}]{Bruggeman1936}
{Bruggeman}, D.~A.~G. 1936, Annalen der Physik, 417, 645

\bibitem[{{Brunken} {et~al.}(2022){Brunken}, {Booth}, {Leemker}, {Nazari}, {van
  der Marel}, \& {van Dishoeck}}]{Brunken2022}
{Brunken}, N. G.~C., {Booth}, A.~S., {Leemker}, M., {et~al.} 2022, \aap, 659,
  A29

\bibitem[{{Bulak} {et~al.}(2021){Bulak}, {Paardekooper}, {Fedoseev}, \&
  {Linnartz}}]{Bulak2021}
{Bulak}, M., {Paardekooper}, D.~M., {Fedoseev}, G., \& {Linnartz}, H. 2021,
  \aap, 647, A82

\bibitem[{{Ciaravella} {et~al.}(2019){Ciaravella}, {Jim{\'e}nez-Escobar},
  {Cecchi-Pestellini}, {Huang}, {Sie}, {Mu{\~n}oz Caro}, \&
  {Chen}}]{Ciaravella2019}
{Ciaravella}, A., {Jim{\'e}nez-Escobar}, A., {Cecchi-Pestellini}, C., {et~al.}
  2019, \apj, 879, 21

\bibitem[{{Clark} {et~al.}(2012){Clark}, {Cruikshank}, {Jaumann}, {Brown},
  {Stephan}, {Dalle Ore}, {Eric Livo}, {Pearson}, {Curchin}, {Hoefen},
  {Buratti}, {Filacchione}, {Baines}, \& {Nicholson}}]{Clark2012}
{Clark}, R.~N., {Cruikshank}, D.~P., {Jaumann}, R., {et~al.} 2012, \icarus,
  218, 831

\bibitem[{Coblentz(1905)}]{Coblentz1905}
Coblentz, W.~W. 1905, Phys. Rev. (Series I), 20, 337

\bibitem[{{Cuppen} {et~al.}(2011){Cuppen}, {Penteado}, {Isokoski}, {van der
  Marel}, \& {Linnartz}}]{Cuppen2011}
{Cuppen}, H.~M., {Penteado}, E.~M., {Isokoski}, K., {van der Marel}, N., \&
  {Linnartz}, H. 2011, \mnras, 417, 2809

\bibitem[{{D'Alessio} {et~al.}(2006){D'Alessio}, {Calvet}, {Hartmann},
  {Franco-Hern{\'a}ndez}, \& {Serv{\'\i}n}}]{Dalessio2006}
{D'Alessio}, P., {Calvet}, N., {Hartmann}, L., {Franco-Hern{\'a}ndez}, R., \&
  {Serv{\'\i}n}, H. 2006, \apj, 638, 314

\bibitem[{{Dalle Ore} {et~al.}(2015){Dalle Ore}, {Cruikshank}, {Mastrapa},
  {Lewis}, \& {White}}]{dalle2015}
{Dalle Ore}, C.~M., {Cruikshank}, D.~P., {Mastrapa}, R. M.~E., {Lewis}, E., \&
  {White}, O.~L. 2015, \icarus, 261, 80

\bibitem[{{Danger} {et~al.}(2011){Danger}, {Borget}, {Chomat}, {Duvernay},
  {Theul{\'e}}, {Guillemin}, {Le Sergeant D'Hendecourt}, \&
  {Chiavassa}}]{Danger2011}
{Danger}, G., {Borget}, F., {Chomat}, M., {et~al.} 2011, \aap, 535, A47

\bibitem[{{Dartois} \& {d'Hendecourt}(2001)}]{Dartois2001}
{Dartois}, E. \& {d'Hendecourt}, L. 2001, \aap, 365, 144

\bibitem[{{Dartois} {et~al.}(2002){Dartois}, {d'Hendecourt}, {Thi},
  {Pontoppidan}, \& {van Dishoeck}}]{Dartois2002}
{Dartois}, E., {d'Hendecourt}, L., {Thi}, W., {Pontoppidan}, K.~M., \& {van
  Dishoeck}, E.~F. 2002, \aap, 394, 1057

\bibitem[{{Dartois} {et~al.}(2022){Dartois}, {Noble}, {Ysard}, {Demyk}, \&
  {Chabot}}]{Dartois2022}
{Dartois}, E., {Noble}, J.~A., {Ysard}, N., {Demyk}, K., \& {Chabot}, M. 2022,
  arXiv e-prints, arXiv:2207.09411

\bibitem[{{de Graauw} {et~al.}(1996){de Graauw}, {Whittet}, {Gerakines},
  {Bauer}, {Beintema}, {Boogert}, {Boxhoorn}, {Chiar}, {Ehrenfreund},
  {Feuchtgruber}, {Helmich}, {Heras}, {Huygen}, {Kester}, {Kunze}, {Lahuis},
  {Leech}, {Lutz}, {Morris}, {Prusti}, {Roelfsema}, {Salama}, {Schaeidt},
  {Schutte}, {Spoon}, {Tielens}, {Valentijn}, {Vandenbusshe}, {van Dishoeck},
  {Wesselius}, {Wieprecht}, \& {Wright}}]{deGraauw1996CO2}
{de Graauw}, T., {Whittet}, D.~C.~B., {Gerakines}, P.~A., {et~al.} 1996, \aap,
  315, L345

\bibitem[{{D'Hendecourt} \& {Allamandola}(1986)}]{dHendecourt1986}
{D'Hendecourt}, L.~B. \& {Allamandola}, L.~J. 1986, \aaps, 64, 453

\bibitem[{{Domaracka} {et~al.}(2010){Domaracka}, {Seperuelo Duarte}, {Boduch},
  {Rothard}, {Ramillon}, {Dartois}, {Pilling}, {Farenzena}, \& {da
  Silveira}}]{Domaracka2010}
{Domaracka}, A., {Seperuelo Duarte}, E., {Boduch}, P., {et~al.} 2010, Nuclear
  Instruments and Methods in Physics Research B, 268, 2960

\bibitem[{{Dominik} {et~al.}(2021){Dominik}, {Min}, \& {Tazaki}}]{Dominik2021}
{Dominik}, C., {Min}, M., \& {Tazaki}, R. 2021, {OpTool: Command-line driven
  tool for creating complex dust opacities}

\bibitem[{{Dubernet} {et~al.}(2013){Dubernet}, {Alexander}, {Ba},
  {Balakrishnan}, {Balan{\c{c}}a}, {Ceccarelli}, {Cernicharo}, {Daniel},
  {Dayou}, {Doronin}, {Dumouchel}, {Faure}, {Feautrier}, {Flower}, {Grosjean},
  {Halvick}, {K{\l}os}, {Lique}, {McBane}, {Marinakis}, {Moreau}, {Moszynski},
  {Neufeld}, {Roueff}, {Schilke}, {Spielfiedel}, {Stancil}, {Stoecklin},
  {Tennyson}, {Yang}, {Vasserot}, \& {Wiesenfeld}}]{Dubernet2013}
{Dubernet}, M.~L., {Alexander}, M.~H., {Ba}, Y.~A., {et~al.} 2013, \aap, 553,
  A50

\bibitem[{{Dubernet} {et~al.}(2006){Dubernet}, {Grosjean}, {Flower}, {Roueff},
  {Daniel}, {Moreau}, \& {Debray}}]{Dubernet2006}
{Dubernet}, M.-L., {Grosjean}, A., {Flower}, D., {et~al.} 2006, Journal of
  Plasma Research SERIES, 7, 356

\bibitem[{{Ehrenfreund} {et~al.}(1996){Ehrenfreund}, {Boogert}, {Gerakines},
  {Jansen}, {Schutte}, {Tielens}, \& {van Dishoeck}}]{Ehrenfreund1996}
{Ehrenfreund}, P., {Boogert}, A.~C.~A., {Gerakines}, P.~A., {et~al.} 1996,
  \aap, 315, L341

\bibitem[{{Ehrenfreund} {et~al.}(1997){Ehrenfreund}, {Boogert}, {Gerakines},
  {Tielens}, \& {van Dishoeck}}]{Ehrenfreund1997}
{Ehrenfreund}, P., {Boogert}, A.~C.~A., {Gerakines}, P.~A., {Tielens},
  A.~G.~G.~M., \& {van Dishoeck}, E.~F. 1997, \aap, 328, 649

\bibitem[{{Ehrenfreund} {et~al.}(1999){Ehrenfreund}, {Kerkhof}, {Schutte},
  {Boogert}, {Gerakines}, {Dartois}, {D'Hendecourt}, {Tielens}, {van Dishoeck},
  \& {Whittet}}]{Ehrenfreund1999}
{Ehrenfreund}, P., {Kerkhof}, O., {Schutte}, W.~A., {et~al.} 1999, \aap, 350,
  240

\bibitem[{{Endres} {et~al.}(2016){Endres}, {Schlemmer}, {Schilke}, {Stutzki},
  \& {M{\"u}ller}}]{Endres2016}
{Endres}, C.~P., {Schlemmer}, S., {Schilke}, P., {Stutzki}, J., \&
  {M{\"u}ller}, H. S.~P. 2016, Journal of Molecular Spectroscopy, 327, 95

\bibitem[{{Fayolle} {et~al.}(2011){Fayolle}, {{\"O}berg}, {Cuppen}, {Visser},
  \& {Linnartz}}]{Fayolle2011}
{Fayolle}, E.~C., {{\"O}berg}, K.~I., {Cuppen}, H.~M., {Visser}, R., \&
  {Linnartz}, H. 2011, \aap, 529, A74

\bibitem[{{Fedoseev} {et~al.}(2017){Fedoseev}, {Chuang}, {Ioppolo}, {Qasim},
  {van Dishoeck}, \& {Linnartz}}]{Fedoseev2017}
{Fedoseev}, G., {Chuang}, K.~J., {Ioppolo}, S., {et~al.} 2017, \apj, 842, 52

\bibitem[{{Fraser} \& {van Dishoeck}(2004)}]{Fraser2004}
{Fraser}, H.~J. \& {van Dishoeck}, E.~F. 2004, Advances in Space Research, 33,
  14

\bibitem[{Fuchs {et~al.}(2006)Fuchs, Acharyya, Bisschop, Öberg, van
  Broekhuizen, Fraser, Schlemmer, van Dishoeck, \& Linnartz}]{Fuchs2006}
Fuchs, G.~W., Acharyya, K., Bisschop, S.~E., {et~al.} 2006, Faraday Discuss.,
  133, 331

\bibitem[{{Fuchs} {et~al.}(2009){Fuchs}, {Cuppen}, {Ioppolo}, {Romanzin},
  {Bisschop}, {Andersson}, {van Dishoeck}, \& {Linnartz}}]{Fuchs2009}
{Fuchs}, G.~W., {Cuppen}, H.~M., {Ioppolo}, S., {et~al.} 2009, \aap, 505, 629

\bibitem[{{Garnett}(1904)}]{Garnett1904}
{Garnett}, J.~C.~M. 1904, Philosophical Transactions of the Royal Society of
  London Series A, 203, 385

\bibitem[{{Garnett}(1906)}]{Garnett1906}
{Garnett}, J.~C.~M. 1906, Philosophical Transactions of the Royal Society of
  London Series A, 205, 237

\bibitem[{{Gerakines} \& {Hudson}(2020)}]{Gerakines2020}
{Gerakines}, P.~A. \& {Hudson}, R.~L. 2020, \apj, 901, 52

\bibitem[{{Gerakines} {et~al.}(1996){Gerakines}, {Schutte}, \&
  {Ehrenfreund}}]{Gerakines1996}
{Gerakines}, P.~A., {Schutte}, W.~A., \& {Ehrenfreund}, P. 1996, \aap, 312, 289

\bibitem[{{Gerakines} {et~al.}(1995){Gerakines}, {Schutte}, {Greenberg}, \&
  {van Dishoeck}}]{Gerakines1995}
{Gerakines}, P.~A., {Schutte}, W.~A., {Greenberg}, J.~M., \& {van Dishoeck},
  E.~F. 1995, \aap, 296, 810

\bibitem[{{Gibb} \& {Whittet}(2002)}]{Gibb2002}
{Gibb}, E.~L. \& {Whittet}, D.~C.~B. 2002, \apjl, 566, L113

\bibitem[{{Gibb} {et~al.}(2004){Gibb}, {Whittet}, {Boogert}, \&
  {Tielens}}]{Gibb2004}
{Gibb}, E.~L., {Whittet}, D.~C.~B., {Boogert}, A.~C.~A., \& {Tielens},
  A.~G.~G.~M. 2004, \apjs, 151, 35

\bibitem[{{Gillett} \& {Forrest}(1973)}]{Gillett1973}
{Gillett}, F.~C. \& {Forrest}, W.~J. 1973, \apj, 179, 483

\bibitem[{{Glasse} {et~al.}(2015){Glasse}, {Rieke}, {Bauwens},
  {Garc{\'\i}a-Mar{\'\i}n}, {Ressler}, {Rost}, {Tikkanen}, {Vandenbussche}, \&
  {Wright}}]{Glasse2015}
{Glasse}, A., {Rieke}, G.~H., {Bauwens}, E., {et~al.} 2015, \pasp, 127, 686

\bibitem[{{Gonz{\'a}lez D{\'\i}az} {et~al.}(2019){Gonz{\'a}lez D{\'\i}az},
  {Carrascosa de Lucas}, {Aparicio}, {Mu{\~n}oz Caro}, {Sie}, {Hsiao},
  {Cazaux}, \& {Chen}}]{Gonzalez2019}
{Gonz{\'a}lez D{\'\i}az}, C., {Carrascosa de Lucas}, H., {Aparicio}, S.,
  {et~al.} 2019, \mnras, 486, 5519

\bibitem[{{Grim} {et~al.}(1989){Grim}, {Greenberg}, {de Groot}, {Baas},
  {Schutte}, \& {Schmitt}}]{Grim1989}
{Grim}, R.~J.~A., {Greenberg}, J.~M., {de Groot}, M.~S., {et~al.} 1989, \aaps,
  78, 161

\bibitem[{Grinberg(2018)}]{flask2018}
Grinberg, M. 2018, Flask web development: developing web applications with
  python (" O'Reilly Media, Inc.")

\bibitem[{{Hagen} {et~al.}(1979){Hagen}, {Allamandola}, \&
  {Greenberg}}]{Hagen1979}
{Hagen}, W., {Allamandola}, L.~J., \& {Greenberg}, J.~M. 1979, \apss, 65, 215

\bibitem[{He {et~al.}(2022)He, Diamant, Wang, Yu, Rocha, Rachid, \&
  Linnartz}]{He2022}
He, J., Diamant, S.~J., Wang, S., {et~al.} 2022, The Astrophysical Journal,
  925, 179

\bibitem[{{Heays} {et~al.}(2017){Heays}, {Bosman}, \& {van
  Dishoeck}}]{Heays2017}
{Heays}, A.~N., {Bosman}, A.~D., \& {van Dishoeck}, E.~F. 2017, \aap, 602, A105

\bibitem[{{Henning} {et~al.}(1999){Henning}, {Il'In}, {Krivova}, {Michel}, \&
  {Voshchinnikov}}]{Henning1999}
{Henning}, T., {Il'In}, V.~B., {Krivova}, N.~A., {Michel}, B., \&
  {Voshchinnikov}, N.~V. 1999, \aaps, 136, 405

\bibitem[{{Herbst} \& {van Dishoeck}(2009)}]{Herbst2009}
{Herbst}, E. \& {van Dishoeck}, E.~F. 2009, \araa, 47, 427

\bibitem[{{Hudgins} {et~al.}(1993){Hudgins}, {Sandford}, {Allamandola}, \&
  {Tielens}}]{Hudgins1993}
{Hudgins}, D.~M., {Sandford}, S.~A., {Allamandola}, L.~J., \& {Tielens},
  A.~G.~G.~M. 1993, \apjs, 86, 713

\bibitem[{{Hudson}(2017)}]{Hudson2017}
{Hudson}, R.~L. 2017, Spectrochimica Acta Part A: Molecular Spectroscopy, 187,
  82

\bibitem[{{Hudson} {et~al.}(2018){Hudson}, {Gerakines}, \&
  {Ferrante}}]{Hudson2018}
{Hudson}, R.~L., {Gerakines}, P.~A., \& {Ferrante}, R.~F. 2018, Spectrochimica
  Acta Part A: Molecular Spectroscopy, 193, 33

\bibitem[{{Hudson} {et~al.}(2017){Hudson}, {Loeffler}, \&
  {Gerakines}}]{Hudson2017bs}
{Hudson}, R.~L., {Loeffler}, M.~J., \& {Gerakines}, P.~A. 2017, \jcp, 146,
  024304

\bibitem[{{Hudson} {et~al.}(2001){Hudson}, {Moore}, \&
  {Gerakines}}]{Hudson2001}
{Hudson}, R.~L., {Moore}, M.~H., \& {Gerakines}, P.~A. 2001, \apj, 550, 1140

\bibitem[{{Ioppolo} {et~al.}(2021){Ioppolo}, {Fedoseev}, {Chuang}, {Cuppen},
  {Clements}, {Jin}, {Garrod}, {Qasim}, {Kofman}, {van Dishoeck}, \&
  {Linnartz}}]{Ioppolo2021}
{Ioppolo}, S., {Fedoseev}, G., {Chuang}, K.~J., {et~al.} 2021, Nature
  Astronomy, 5, 197

\bibitem[{{Ishibashi} {et~al.}(2021){Ishibashi}, {Hidaka}, {Oba}, {Kouchi}, \&
  {Watanabe}}]{Ishibashi2021}
{Ishibashi}, A., {Hidaka}, H., {Oba}, Y., {Kouchi}, A., \& {Watanabe}, N. 2021,
  \apjl, 921, L13

\bibitem[{{Isokoski} {et~al.}(2013){Isokoski}, {Poteet}, \&
  {Linnartz}}]{Isokoski2013}
{Isokoski}, K., {Poteet}, C.~A., \& {Linnartz}, H. 2013, \aap, 555, A85

\bibitem[{{J{\"a}ger} {et~al.}(2003){J{\"a}ger}, {Il'in}, {Henning},
  {Mutschke}, {Fabian}, {Semenov}, \& {Voshchinnikov}}]{Jager2003}
{J{\"a}ger}, C., {Il'in}, V.~B., {Henning}, T., {et~al.} 2003, \jqsrt, 79-80,
  765

\bibitem[{{Jmol development team:}(Accessed in June 2021)}]{jmol}
{Jmol development team:}. Accessed in June 2021, Jmol: an open-source Java
  viewer for chemical structures in 3D.

\bibitem[{{J{\o}rgensen} {et~al.}(2020){J{\o}rgensen}, {Belloche}, \&
  {Garrod}}]{Jorgensen2020}
{J{\o}rgensen}, J.~K., {Belloche}, A., \& {Garrod}, R.~T. 2020, \araa, 58, 727

\bibitem[{{J{\o}rgensen} {et~al.}(2012){J{\o}rgensen}, {Favre}, {Bisschop},
  {Bourke}, {van Dishoeck}, \& {Schmalzl}}]{Jorgensen2012}
{J{\o}rgensen}, J.~K., {Favre}, C., {Bisschop}, S.~E., {et~al.} 2012, \apjl,
  757, L4

\bibitem[{{Keane} {et~al.}(2001){Keane}, {Tielens}, {Boogert}, {Schutte}, \&
  {Whittet}}]{Keane2001}
{Keane}, J.~V., {Tielens}, A.~G.~G.~M., {Boogert}, A.~C.~A., {Schutte}, W.~A.,
  \& {Whittet}, D.~C.~B. 2001, \aap, 376, 254

\bibitem[{{Kemper} {et~al.}(2004){Kemper}, {Vriend}, \& {Tielens}}]{Kemper2004}
{Kemper}, F., {Vriend}, W.~J., \& {Tielens}, A.~G.~G.~M. 2004, \apj, 609, 826

\bibitem[{{Kerkhof} {et~al.}(1999){Kerkhof}, {Schutte}, \&
  {Ehrenfreund}}]{Kerkhof1999}
{Kerkhof}, O., {Schutte}, W.~A., \& {Ehrenfreund}, P. 1999, \aap, 346, 990

\bibitem[{Kim {et~al.}(2020)Kim, Chen, Cheng, Gindulyte, He, He, Li, Shoemaker,
  Thiessen, Yu, Zaslavsky, Zhang, \& Bolton}]{Pubchem2021}
Kim, S., Chen, J., Cheng, T., {et~al.} 2020, Nucleic Acids Research, 49, D1388

\bibitem[{{Knez} {et~al.}(2012){Knez}, {Moore}, {Ferrante}, \&
  {Hudson}}]{Knez2012}
{Knez}, C., {Moore}, M.~H., {Ferrante}, R.~F., \& {Hudson}, R.~L. 2012, \apj,
  748, 95

\bibitem[{Kofman {et~al.}(2019)Kofman, He, ten Kate, \& Linnartz}]{kofman2019}
Kofman, V., He, J., ten Kate, I.~L., \& Linnartz, H. 2019, The Astrophysical
  Journal, 875, 131

\bibitem[{Kramers(1927)}]{Kramers1927}
Kramers, H.~A. 1927, Atti Cong. Intern. Fisica (Transactions of Volta Centenary
  Congress) Como, 2, 545

\bibitem[{{Kronig}(1926)}]{Kronig1926}
{Kronig}, R. D.~L. 1926, Journal of the Optical Society of America (1917-1983),
  12, 547

\bibitem[{{Lacy} {et~al.}(1984){Lacy}, {Baas}, {Allamandola}, {Persson},
  {McGregor}, {Lonsdale}, {Geballe}, \& {van de Bult}}]{Lacy1984}
{Lacy}, J.~H., {Baas}, F., {Allamandola}, L.~J., {et~al.} 1984, \apj, 276, 533

\bibitem[{{Lacy} {et~al.}(1991){Lacy}, {Carr}, {Evans}, {Baas}, {Achtermann},
  \& {Arens}}]{Lacy1991}
{Lacy}, J.~H., {Carr}, J.~S., {Evans}, Neal~J., I., {et~al.} 1991, \apj, 376,
  556

\bibitem[{{Lacy} {et~al.}(1998){Lacy}, {Faraji}, {Sandford}, \&
  {Allamandola}}]{Lacy1998}
{Lacy}, J.~H., {Faraji}, H., {Sandford}, S.~A., \& {Allamandola}, L.~J. 1998,
  \apjl, 501, L105

\bibitem[{{Langmuir}(1938)}]{Langmuir1938}
{Langmuir}, I. 1938, Science, 87, 493

\bibitem[{{Ligterink} {et~al.}(2018){Ligterink}, {Walsh}, {Bhuin},
  {Vissapragada}, {Terwisscha van Scheltinga}, \& {Linnartz}}]{Ligterink2018}
{Ligterink}, N.~F.~W., {Walsh}, C., {Bhuin}, R.~G., {et~al.} 2018, \aap, 612,
  A88

\bibitem[{Linnartz {et~al.}(2015)Linnartz, Ioppolo, \& Fedoseev}]{Linnartz2015}
Linnartz, H., Ioppolo, S., \& Fedoseev, G. 2015, International Reviews in
  Physical Chemistry, 34, 205

\bibitem[{{Mastrapa} {et~al.}(2008){Mastrapa}, {Bernstein}, {Sandford},
  {Roush}, {Cruikshank}, \& {Dalle Ore}}]{Mastrapa2008}
{Mastrapa}, R.~M., {Bernstein}, M.~P., {Sandford}, S.~A., {et~al.} 2008,
  \icarus, 197, 307

\bibitem[{{Mastrapa} {et~al.}(2009){Mastrapa}, {Sandford}, {Roush},
  {Cruikshank}, \& {Dalle Ore}}]{Mastrapa2009}
{Mastrapa}, R.~M., {Sandford}, S.~A., {Roush}, T.~L., {Cruikshank}, D.~P., \&
  {Dalle Ore}, C.~M. 2009, \apj, 701, 1347

\bibitem[{{Mat{\'e}} {et~al.}(2009){Mat{\'e}}, {G{\'a}lvez}, {Herrero},
  {Fern{\'a}ndez-Torre}, {Moreno}, \& {Escribano}}]{Mate2009}
{Mat{\'e}}, B., {G{\'a}lvez}, O., {Herrero}, V.~J., {et~al.} 2009, \apjl, 703,
  L178

\bibitem[{{Mat{\'e}} {et~al.}(2012){Mat{\'e}}, {Herrero},
  {Rodr{\'\i}guez-Lazcano}, {Fern{\'a}ndez-Torre}, {Moreno}, {G{\'o}mez}, \&
  {Escribano}}]{Mate2012}
{Mat{\'e}}, B., {Herrero}, V.~J., {Rodr{\'\i}guez-Lazcano}, Y., {et~al.} 2012,
  \apj, 759, 90

\bibitem[{{Materese} {et~al.}(2015){Materese}, {Cruikshank}, {Sandford},
  {Imanaka}, \& {Nuevo}}]{Materese2015}
{Materese}, C.~K., {Cruikshank}, D.~P., {Sandford}, S.~A., {Imanaka}, H., \&
  {Nuevo}, M. 2015, \apj, 812, 150

\bibitem[{{Mattioda} {et~al.}(2020){Mattioda}, {Hudgins}, {Boersma},
  {Bauschlicher}, {Ricca}, {Cami}, {Peeters}, {S{\'a}nchez de Armas}, {Puerta
  Saborido}, \& {Allamandola}}]{Mattioda2020}
{Mattioda}, A.~L., {Hudgins}, D.~M., {Boersma}, C., {et~al.} 2020, \apjs, 251,
  22

\bibitem[{{McElroy} {et~al.}(2013){McElroy}, {Walsh}, {Markwick}, {Cordiner},
  {Smith}, \& {Millar}}]{McElroy2013}
{McElroy}, D., {Walsh}, C., {Markwick}, A.~J., {et~al.} 2013, \aap, 550, A36

\bibitem[{{McGuire}(2021)}]{McGuire2021}
{McGuire}, B.~A. 2021, arXiv e-prints, arXiv:2109.13848

\bibitem[{{McGuire} {et~al.}(2016){McGuire}, {Carroll}, {Loomis}, {Finneran},
  {Jewell}, {Remijan}, \& {Blake}}]{McGuire2016}
{McGuire}, B.~A., {Carroll}, P.~B., {Loomis}, R.~A., {et~al.} 2016, Science,
  352, 1449

\bibitem[{{McLean} {et~al.}(1998){McLean}, {Becklin}, {Bendiksen}, {Brims},
  {Canfield}, {Figer}, {Graham}, {Hare}, {Lacayanga}, {Larkin}, {Larson},
  {Levenson}, {Magnone}, {Teplitz}, \& {Wong}}]{McLean1998}
{McLean}, I.~S., {Becklin}, E.~E., {Bendiksen}, O., {et~al.} 1998, in Society
  of Photo-Optical Instrumentation Engineers (SPIE) Conference Series, Vol.
  3354, Infrared Astronomical Instrumentation, ed. A.~M. {Fowler}, 566--578

\bibitem[{{Meinert} {et~al.}(2016){Meinert}, {Myrgorodska}, {de Marcellus},
  {Buhse}, {Nahon}, {Hoffmann}, {d'Hendecourt}, \&
  {Meierhenrich}}]{Meinert2016}
{Meinert}, C., {Myrgorodska}, I., {de Marcellus}, P., {et~al.} 2016, Science,
  352, 208

\bibitem[{{Merrill} {et~al.}(1976){Merrill}, {Russell}, \&
  {Soifer}}]{Merrill1976}
{Merrill}, K.~M., {Russell}, R.~W., \& {Soifer}, B.~T. 1976, \apj, 207, 763

\bibitem[{{Mifsud} {et~al.}(2021){Mifsud}, {Juh{\'a}sz}, {Herczku},
  {Kov{\'a}cs}, {Ioppolo}, {Ka{\r{A}}uchov{\'a}}, {Czentye}, {Hailey},
  {Mui{\~n}a}, {Mason}, {McCullough}, {Parip{\'a}s}, \& {Sulik}}]{Mifsud2021}
{Mifsud}, D.~V., {Juh{\'a}sz}, Z., {Herczku}, P., {et~al.} 2021, European
  Physical Journal D, 75, 182

\bibitem[{{Modica} \& {Palumbo}(2010)}]{Modica2010}
{Modica}, P. \& {Palumbo}, M.~E. 2010, \aap, 519, A22

\bibitem[{{Moore} {et~al.}(2010){Moore}, {Ferrante}, {Moore}, \&
  {Hudson}}]{Moore2010}
{Moore}, M.~H., {Ferrante}, R.~F., {Moore}, W.~J., \& {Hudson}, R. 2010, \apjs,
  191, 96

\bibitem[{{Mu{\~n}oz Caro} {et~al.}(2002){Mu{\~n}oz Caro}, {Meierhenrich},
  {Schutte}, {Barbier}, {Arcones Segovia}, {Rosenbauer}, {Thiemann}, {Brack},
  \& {Greenberg}}]{Caro2002}
{Mu{\~n}oz Caro}, G.~M., {Meierhenrich}, U.~J., {Schutte}, W.~A., {et~al.}
  2002, \nat, 416, 403

\bibitem[{{Mu{\~n}oz Caro} \& {Schutte}(2003)}]{MunozCaro2003}
{Mu{\~n}oz Caro}, G.~M. \& {Schutte}, W.~A. 2003, \aap, 412, 121

\bibitem[{{M{\"u}ller} {et~al.}(2005){M{\"u}ller}, {Schl{\"o}der}, {Stutzki},
  \& {Winnewisser}}]{Muller2005}
{M{\"u}ller}, H. S.~P., {Schl{\"o}der}, F., {Stutzki}, J., \& {Winnewisser}, G.
  2005, Journal of Molecular Structure, 742, 215

\bibitem[{{M{\"u}ller} {et~al.}(2001){M{\"u}ller}, {Thorwirth}, {Roth}, \&
  {Winnewisser}}]{Muller2001}
{M{\"u}ller}, H.~S.~P., {Thorwirth}, S., {Roth}, D.~A., \& {Winnewisser}, G.
  2001, \aap, 370, L49

\bibitem[{{Nazari} {et~al.}(2021){Nazari}, {van Gelder}, {van Dishoeck},
  {Tabone}, {van't Hoff}, {Ligterink}, {Beuther}, {Boogert}, {Caratti o
  Garatti}, {Klaassen}, {Linnartz}, {Taquet}, \& {Tychoniec}}]{Nazari2021}
{Nazari}, P., {van Gelder}, M.~L., {van Dishoeck}, E.~F., {et~al.} 2021, \aap,
  650, A150

\bibitem[{Neese(2012)}]{Orca2012}
Neese, F. 2012, WIREs Computational Molecular Science, 2, 73

\bibitem[{Neese(2018)}]{Orca2018}
Neese, F. 2018, WIREs Computational Molecular Science, 8, e1327

\bibitem[{Neese {et~al.}(2020)Neese, Wennmohs, Becker, \& Riplinger}]{Orca2020}
Neese, F., Wennmohs, F., Becker, U., \& Riplinger, C. 2020, The Journal of
  Chemical Physics, 152, 224108

\bibitem[{{Novozamsky} {et~al.}(2001){Novozamsky}, {Schutte}, \&
  {Keane}}]{Novozamsky2001}
{Novozamsky}, J.~H., {Schutte}, W.~A., \& {Keane}, J.~V. 2001, \aap, 379, 588

\bibitem[{{Nuevo} {et~al.}(2018){Nuevo}, {Cooper}, \& {Sandford}}]{Nuevo2018}
{Nuevo}, M., {Cooper}, G., \& {Sandford}, S.~A. 2018, Nature Communications, 9,
  5276

\bibitem[{{{\"O}berg} {et~al.}(2011){{\"O}berg}, {Boogert}, {Pontoppidan}, {van
  den Broek}, {van Dishoeck}, {Bottinelli}, {Blake}, \& {Evans}}]{Oberg2011}
{{\"O}berg}, K.~I., {Boogert}, A.~C.~A., {Pontoppidan}, K.~M., {et~al.} 2011,
  \apj, 740, 109

\bibitem[{{{\"O}berg} {et~al.}(2007){{\"O}berg}, {Fraser}, {Boogert},
  {Bisschop}, {Fuchs}, {van Dishoeck}, \& {Linnartz}}]{Oberg2007}
{{\"O}berg}, K.~I., {Fraser}, H.~J., {Boogert}, A.~C.~A., {et~al.} 2007, \aap,
  462, 1187

\bibitem[{{{\"O}berg} {et~al.}(2009){{\"O}berg}, {Garrod}, {van Dishoeck}, \&
  {Linnartz}}]{Oberg2009}
{{\"O}berg}, K.~I., {Garrod}, R.~T., {van Dishoeck}, E.~F., \& {Linnartz}, H.
  2009, \aap, 504, 891

\bibitem[{{Ohta} \& {Ishida}(1988)}]{Ohta1988}
{Ohta}, K. \& {Ishida}, H. 1988, Applied Spectroscopy, 42, 952

\bibitem[{{Onaka} {et~al.}(2021){Onaka}, {Kimura}, {Sakon}, \&
  {Shimonishi}}]{Onaka2021}
{Onaka}, T., {Kimura}, T., {Sakon}, I., \& {Shimonishi}, T. 2021, \apj, 916, 75

\bibitem[{{Palumbo}(2006)}]{Palumbo2006_oh}
{Palumbo}, M.~E. 2006, \aap, 453, 903

\bibitem[{{Palumbo} {et~al.}(1998){Palumbo}, {Baratta}, {Brucato}, {Castorina},
  {Satorre}, \& {Strazzulla}}]{Palumbo1998}
{Palumbo}, M.~E., {Baratta}, G.~A., {Brucato}, J.~R., {et~al.} 1998, \aap, 334,
  247

\bibitem[{{Palumbo} {et~al.}(2006){Palumbo}, {Baratta}, {Collings}, \&
  {McCoustra}}]{palumbo2006}
{Palumbo}, M.~E., {Baratta}, G.~A., {Collings}, M.~P., \& {McCoustra}, M.~R.~S.
  2006, Physical Chemistry Chemical Physics (Incorporating Faraday
  Transactions), 8, 279

\bibitem[{{Palumbo} {et~al.}(1995){Palumbo}, {Tielens}, \&
  {Tokunaga}}]{Palumbo1995}
{Palumbo}, M.~E., {Tielens}, A.~G.~G.~M., \& {Tokunaga}, A.~T. 1995, \apj, 449,
  674

\bibitem[{{Pearson} {et~al.}(2010){Pearson}, {M{\"u}ller}, {Pickett}, {Cohen},
  \& {Drouin}}]{Pearson2010}
{Pearson}, J.~C., {M{\"u}ller}, H.~S.~P., {Pickett}, H.~M., {Cohen}, E.~A., \&
  {Drouin}, B.~J. 2010, \jqsrt, 111, 1614

\bibitem[{{Penteado} {et~al.}(2015){Penteado}, {Boogert}, {Pontoppidan},
  {Ioppolo}, {Blake}, \& {Cuppen}}]{Penteado2015}
{Penteado}, E.~M., {Boogert}, A.~C.~A., {Pontoppidan}, K.~M., {et~al.} 2015,
  \mnras, 454, 531

\bibitem[{{Perotti} {et~al.}(2020){Perotti}, {Rocha}, {J{\o}rgensen},
  {Kristensen}, {Fraser}, \& {Pontoppidan}}]{Perotti2020}
{Perotti}, G., {Rocha}, W.~R.~M., {J{\o}rgensen}, J.~K., {et~al.} 2020, \aap,
  643, A48

\bibitem[{{Pickett} {et~al.}(1998){Pickett}, {Poynter}, {Cohen}, {Delitsky},
  {Pearson}, \& {M{\"u}ller}}]{Pickett1998}
{Pickett}, H.~M., {Poynter}, R.~L., {Cohen}, E.~A., {et~al.} 1998, \jqsrt, 60,
  883

\bibitem[{{Pilling} \& {Bergantini}(2015)}]{Pilling2015}
{Pilling}, S. \& {Bergantini}, A. 2015, \apj, 811, 151

\bibitem[{{Pilling} {et~al.}(2010){Pilling}, {Seperuelo Duarte}, {Domaracka},
  {Rothard}, {Boduch}, \& {da Silveira}}]{Pilling2010}
{Pilling}, S., {Seperuelo Duarte}, E., {Domaracka}, A., {et~al.} 2010, \aap,
  523, A77

\bibitem[{{Pontoppidan} {et~al.}(2008){Pontoppidan}, {Boogert}, {Fraser}, {van
  Dishoeck}, {Blake}, {Lahuis}, {{\"O}berg}, {Evans}, \&
  {Salyk}}]{Pontoppidan2008}
{Pontoppidan}, K.~M., {Boogert}, A.~C.~A., {Fraser}, H.~J., {et~al.} 2008,
  \apj, 678, 1005

\bibitem[{{Pontoppidan} {et~al.}(2005){Pontoppidan}, {Dullemond}, {van
  Dishoeck}, {Blake}, {Boogert}, {Evans}, {Kessler-Silacci}, \&
  {Lahuis}}]{Pontoppidan2005}
{Pontoppidan}, K.~M., {Dullemond}, C.~P., {van Dishoeck}, E.~F., {et~al.} 2005,
  \apj, 622, 463

\bibitem[{{Pontoppidan} {et~al.}(2003){Pontoppidan}, {Fraser}, {Dartois},
  {Thi}, {van Dishoeck}, {Boogert}, {d'Hendecourt}, {Tielens}, \&
  {Bisschop}}]{Pontoppidan2003}
{Pontoppidan}, K.~M., {Fraser}, H.~J., {Dartois}, E., {et~al.} 2003, \aap, 408,
  981

\bibitem[{{Pontoppidan} {et~al.}(2016){Pontoppidan}, {Pickering}, {Laidler},
  {Gilbert}, {Sontag}, {Slocum}, {Sienkiewicz}, {Hanley}, {Earl}, {Pueyo},
  {Ravindranath}, {Karakla}, {Robberto}, {Noriega-Crespo}, \&
  {Barker}}]{Pontoppidan2016}
{Pontoppidan}, K.~M., {Pickering}, T.~E., {Laidler}, V.~G., {et~al.} 2016, in
  Society of Photo-Optical Instrumentation Engineers (SPIE) Conference Series,
  Vol. 9910, Observatory Operations: Strategies, Processes, and Systems VI, ed.
  A.~B. {Peck}, R.~L. {Seaman}, \& C.~R. {Benn}, 991016

\bibitem[{{Potapov} {et~al.}(2021){Potapov}, {Bouwman}, {J{\"a}ger}, \&
  {Henning}}]{Potapov2021}
{Potapov}, A., {Bouwman}, J., {J{\"a}ger}, C., \& {Henning}, T. 2021, Nature
  Astronomy, 5, 78

\bibitem[{{Poteet} {et~al.}(2013){Poteet}, {Pontoppidan}, {Megeath}, {Watson},
  {Isokoski}, {Bjorkman}, {Sheehan}, \& {Linnartz}}]{Poteet2013}
{Poteet}, C.~A., {Pontoppidan}, K.~M., {Megeath}, S.~T., {et~al.} 2013, \apj,
  766, 117

\bibitem[{Querry(1987)}]{Querry1987}
Querry, M.~R. 1987, in Optical constants of minerals and other materials from
  the millimeter to the ultraviolet

\bibitem[{{Rachid} {et~al.}(2021){Rachid}, {Brunken}, {de Boe}, {Fedoseev},
  {Boogert}, \& {Linnartz}}]{Rachid2021}
{Rachid}, M.~G., {Brunken}, N., {de Boe}, D., {et~al.} 2021, \aap, 653, A116

\bibitem[{{Rachid} {et~al.}(2022){Rachid}, {Rocha}, \& {Linnartz}}]{Rachid2022}
{Rachid}, M.~G., {Rocha}, W., \& {Linnartz}, H. 2022, arXiv e-prints,
  arXiv:2207.12502

\bibitem[{{Rachid} {et~al.}(2020){Rachid}, {Terwisscha van Scheltinga},
  {Koletzki}, \& {Linnartz}}]{Rachid2020}
{Rachid}, M.~G., {Terwisscha van Scheltinga}, J., {Koletzki}, D., \&
  {Linnartz}, H. 2020, \aap, 639, A4

\bibitem[{{Rivilla} {et~al.}(2021){Rivilla}, {Jim{\'e}nez-Serra},
  {Mart{\'\i}n-Pintado}, {Briones}, {Rodr{\'\i}guez-Almeida}, {Rico-Villas},
  {Tercero}, {Zeng}, {Colzi}, {de Vicente}, {Mart{\'\i}n}, \&
  {Requena-Torres}}]{Rivilla2021}
{Rivilla}, V.~M., {Jim{\'e}nez-Serra}, I., {Mart{\'\i}n-Pintado}, J., {et~al.}
  2021, Proceedings of the National Academy of Science, 118, 2101314118

\bibitem[{Rocha \& Pilling(2014)}]{Rocha2014}
Rocha, W. \& Pilling, S. 2014, Spectrochimica Acta Part A: Molecular and
  Biomolecular Spectroscopy, 123, 436

\bibitem[{{Rocha} {et~al.}(2021){Rocha}, {Perotti}, {Kristensen}, \&
  {J{\o}rgensen}}]{Rocha2021}
{Rocha}, W.~R.~M., {Perotti}, G., {Kristensen}, L.~E., \& {J{\o}rgensen}, J.~K.
  2021, \aap, 654, A158

\bibitem[{{Rocha} \& {Pilling}(2015)}]{Rocha2015}
{Rocha}, W.~R.~M. \& {Pilling}, S. 2015, \apj, 803, 18

\bibitem[{{Rocha} \& {Pilling}(2018)}]{Rocha2018}
{Rocha}, W.~R.~M. \& {Pilling}, S. 2018, \mnras, 478, 5190

\bibitem[{{Rocha} {et~al.}(2020){Rocha}, {Pilling}, {Domaracka}, {Rothard}, \&
  {Boduch}}]{Rocha2020}
{Rocha}, W.~R.~M., {Pilling}, S., {Domaracka}, A., {Rothard}, H., \& {Boduch},
  P. 2020, Spectrochimica Acta Part A: Molecular Spectroscopy, 228, 117826

\bibitem[{{Schmitt} {et~al.}(2018){Schmitt}, {Bollard}, {Garenne}, {Albert},
  {Bonal}, \& {Poch}}]{Schmitt2018}
{Schmitt}, B., {Bollard}, P., {Garenne}, A., {et~al.} 2018, in European
  Planetary Science Congress, EPSC2018--529

\bibitem[{{Schmitt} {et~al.}(1989){Schmitt}, {Grim}, \&
  {Greenberg}}]{Schmitt1989}
{Schmitt}, B., {Grim}, R., \& {Greenberg}, M. 1989, in Infrared Spectroscopy in
  Astronomy, ed. E.~{B{\"o}hm-Vitense}, 213

\bibitem[{{Sch{\"o}ier} {et~al.}(2005){Sch{\"o}ier}, {van der Tak}, {van
  Dishoeck}, \& {Black}}]{Schoier2005}
{Sch{\"o}ier}, F.~L., {van der Tak}, F.~F.~S., {van Dishoeck}, E.~F., \&
  {Black}, J.~H. 2005, \aap, 432, 369

\bibitem[{{Schutte}(1999)}]{Schutte1999}
{Schutte}, W.~A. 1999, in NATO Advanced Study Institute (ASI) Series C, Vol.
  523, Formation and Evolution of Solids in Space, ed. J.~M. {Greenberg} \&
  A.~{Li}, 177

\bibitem[{{Schutte} {et~al.}(1999){Schutte}, {Boogert}, {Tielens}, {Whittet},
  {Gerakines}, {Chiar}, {Ehrenfreund}, {Greenberg}, {van Dishoeck}, \& {de
  Graauw}}]{Schutte1999_weak}
{Schutte}, W.~A., {Boogert}, A.~C.~A., {Tielens}, A.~G.~G.~M., {et~al.} 1999,
  \aap, 343, 966

\bibitem[{{Schutte} \& {Greenberg}(1997)}]{Schutte1997}
{Schutte}, W.~A. \& {Greenberg}, J.~M. 1997, \aap, 317, L43

\bibitem[{{Schutte} \& {Khanna}(2003)}]{Schutte2003}
{Schutte}, W.~A. \& {Khanna}, R.~K. 2003, \aap, 398, 1049

\bibitem[{{Schutte} {et~al.}(1996){Schutte}, {Tielens}, {Whittet}, {Boogert},
  {Ehrenfreund}, {de Graauw}, {Prusti}, {van Dishoeck}, \&
  {Wesselius}}]{Schutte1996}
{Schutte}, W.~A., {Tielens}, A.~G.~G.~M., {Whittet}, D.~C.~B., {et~al.} 1996,
  \aap, 315, L333

\bibitem[{{Smith} {et~al.}(1989){Smith}, {Sellgren}, \& {Tokunaga}}]{Smith1989}
{Smith}, R.~G., {Sellgren}, K., \& {Tokunaga}, A.~T. 1989, \apj, 344, 413

\bibitem[{Stephens {et~al.}(1994)Stephens, Devlin, Chabalowski, \&
  Frisch}]{Stephens1994}
Stephens, P.~J., Devlin, F.~J., Chabalowski, C.~F., \& Frisch, M.~J. 1994, The
  Journal of Physical Chemistry, 98, 11623

\bibitem[{{Strazzulla} {et~al.}(1984){Strazzulla}, {Cataliotti}, {Calcagno}, \&
  {Foti}}]{Strazzulla1984}
{Strazzulla}, G., {Cataliotti}, R.~S., {Calcagno}, L., \& {Foti}, G. 1984,
  \aap, 133, 77

\bibitem[{Tempelmeyer \& Mills~Jr(1968)}]{tempelmeyer1968refractive}
Tempelmeyer, K. \& Mills~Jr, D. 1968, Journal of Applied Physics, 39, 2968

\bibitem[{{Terwisscha van Scheltinga} {et~al.}(2018){Terwisscha van
  Scheltinga}, {Ligterink}, {Boogert}, {van Dishoeck}, \&
  {Linnartz}}]{Scheltinga2018}
{Terwisscha van Scheltinga}, J., {Ligterink}, N.~F.~W., {Boogert}, A.~C.~A.,
  {van Dishoeck}, E.~F., \& {Linnartz}, H. 2018, \aap, 611, A35

\bibitem[{{Terwisscha van Scheltinga} {et~al.}(2021){Terwisscha van
  Scheltinga}, {Marcandalli}, {McClure}, {Hogerheijde}, \&
  {Linnartz}}]{Scheltinga2021}
{Terwisscha van Scheltinga}, J., {Marcandalli}, G., {McClure}, M.~K.,
  {Hogerheijde}, M.~R., \& {Linnartz}, H. 2021, arXiv e-prints,
  arXiv:2105.02226

\bibitem[{Theul{\'e} {et~al.}(2013)Theul{\'e}, Duvernay, Danger, Borget, Bossa,
  Vinogradoff, Mispelaer, \& Chiavassa}]{theule2013thermal}
Theul{\'e}, P., Duvernay, F., Danger, G., {et~al.} 2013, Advances in Space
  Research, 52, 1567

\bibitem[{{Urso} {et~al.}(2020){Urso}, {Vuitton}, {Danger}, {Le Sergeant
  d'Hendecourt}, {Flandinet}, {Djouadi}, {Mivumbi}, {Orthous-Daunay}, {Ruf},
  {Vinogradoff}, {Wolters}, \& {Brunetto}}]{Urso2020}
{Urso}, R.~G., {Vuitton}, V., {Danger}, G., {et~al.} 2020, \aap, 644, A115

\bibitem[{{van Broekhuizen}(2005)}]{vanBroekhuizen2005phd}
{van Broekhuizen}, F.~A. 2005, PhD thesis, Leiden Observatory, Leiden
  University, P.O. Box 9513, 2300 RA Leiden, The Netherlands

\bibitem[{{van Broekhuizen} {et~al.}(2006){van Broekhuizen}, {Groot}, {Fraser},
  {van Dishoeck}, \& {Schlemmer}}]{vanBroekhuizen2006}
{van Broekhuizen}, F.~A., {Groot}, I.~M.~N., {Fraser}, H.~J., {van Dishoeck},
  E.~F., \& {Schlemmer}, S. 2006, \aap, 451, 723

\bibitem[{{van Broekhuizen} {et~al.}(2005){van Broekhuizen}, {Pontoppidan},
  {Fraser}, \& {van Dishoeck}}]{vanBroekhuizen2005}
{van Broekhuizen}, F.~A., {Pontoppidan}, K.~M., {Fraser}, H.~J., \& {van
  Dishoeck}, E.~F. 2005, \aap, 441, 249

\bibitem[{{van der Tak} {et~al.}(2020){van der Tak}, {Lique}, {Faure}, {Black},
  \& {van Dishoeck}}]{Tak2020}
{van der Tak}, F. F.~S., {Lique}, F., {Faure}, A., {Black}, J.~H., \& {van
  Dishoeck}, E.~F. 2020, Atoms, 8, 15

\bibitem[{{van Dishoeck}(1988)}]{vanDishoeck1988}
{van Dishoeck}, E.~F. 1988, in Astrophysics and Space Science Library, Vol.
  146, Rate Coefficients in Astrochemistry, ed. T.~J. {Millar} \& D.~A.
  {Williams}, 49

\bibitem[{{van Dishoeck} {et~al.}(2006){van Dishoeck}, {Jonkheid}, \& {van
  Hemert}}]{vanDishoeck2006}
{van Dishoeck}, E.~F., {Jonkheid}, B., \& {van Hemert}, M.~C. 2006, Faraday
  Discussions, 133, 231

\bibitem[{{van Gelder} {et~al.}(2020){van Gelder}, {Tabone}, {Tychoniec}, {van
  Dishoeck}, {Beuther}, {Boogert}, {Caratti o Garatti}, {Klaassen}, {Linnartz},
  {M{\"u}ller}, \& {Taquet}}]{vanGelder2020}
{van Gelder}, M.~L., {Tabone}, B., {Tychoniec}, {\L}., {et~al.} 2020, \aap,
  639, A87

\bibitem[{{Vinogradoff} {et~al.}(2015){Vinogradoff}, {Duvernay}, {Fray},
  {Bouilloud}, {Chiavassa}, \& {Cottin}}]{Vinogradoff2015}
{Vinogradoff}, V., {Duvernay}, F., {Fray}, N., {et~al.} 2015, \apjl, 809, L18

\bibitem[{{Wakelam} {et~al.}(2012){Wakelam}, {Herbst}, {Loison}, {Smith},
  {Chandrasekaran}, {Pavone}, {Adams}, {Bacchus-Montabonel}, {Bergeat},
  {B{\'e}roff}, {Bierbaum}, {Chabot}, {Dalgarno}, {van Dishoeck}, {Faure},
  {Geppert}, {Gerlich}, {Galli}, {H{\'e}brard}, {Hersant}, {Hickson},
  {Honvault}, {Klippenstein}, {Le Picard}, {Nyman}, {Pernot}, {Schlemmer},
  {Selsis}, {Sims}, {Talbi}, {Tennyson}, {Troe}, {Wester}, \&
  {Wiesenfeld}}]{Wakelam2012}
{Wakelam}, V., {Herbst}, E., {Loison}, J.~C., {et~al.} 2012, \apjs, 199, 21

\bibitem[{{Warren}(1984)}]{Warren1984}
{Warren}, S.~G. 1984, \ao, 23, 1206

\bibitem[{{Watanabe} \& {Kouchi}(2002)}]{Watanabe2002}
{Watanabe}, N. \& {Kouchi}, A. 2002, \apjl, 571, L173

\bibitem[{{Weingartner} \& {Draine}(2001)}]{Weingartner2001}
{Weingartner}, J.~C. \& {Draine}, B.~T. 2001, \apj, 548, 296

\bibitem[{{Zasowski} {et~al.}(2009){Zasowski}, {Kemper}, {Watson}, {Furlan},
  {Bohac}, {Hull}, \& {Green}}]{Zasowski2009}
{Zasowski}, G., {Kemper}, F., {Watson}, D.~M., {et~al.} 2009, \apj, 694, 459

\bibitem[{Öberg(2016)}]{Oberg2016}
Öberg, K.~I. 2016, Chemical Reviews, 116, 9631, pMID: 27099922

\end{thebibliography}

\appendix
\section{List of ice samples in LIDA.}
\label{Laboratory_data_list}
The current version of LIDA contains the IR spectrum of over 1100 ice samples, which are listed in Table~\ref{analogue_list}. They are categorized by ices in pure samples, and mixtures with, two, three, four and five components. This also includes warmed-up samples or processed by UV radiation.

\longtab[1]{
\begin{landscape}
\begin{longtable}{lccccrrrr}
\caption{\label{analogue_list} Ice analogues hosted in LIDA. Irradiated samples are indicated by the symbol $\leadsto$. ``s'' and ``h'' indicate seconds and hour, respectively.}\\
\hline
\hline
Sample & Thickness & $N_{\rm{ice}}$ $^a$ & Resolution & Ratios & Temperature (K)/ & Substrate/ & Reference\\
 & (ML)  & (cm$^{-2}$) & (cm$^{-1}$) &  & UV radiation (time) & $n_{\rm{substrate}}$ & \\
\hline
\endfirsthead
\caption{Continued.}\\
\hline
Sample & Thickness & $N_{\rm{ice}}$ $^a$ & Resolution & Ratios & Temperature (K)/ & Substrate/ & Reference\\
 & (ML)  & (cm$^{-2}$) & (cm$^{-1}$) &  & UV radiation (time) & $n_{\rm{substrate}}$ & \\
\hline
\endhead
\hline
\endfoot
\hline
\endlastfoot
\multicolumn{8}{c}{\bf{Pure ices}}\\
\hline
H$_2$O & 500 & 5e17 & 1.0 & ... & 10$-$160 & CsI/1.73 & \citet{Gerakines1996}\\
H$_2$O & 3000 & 1e17 & 2.0 & ... & 15$-$135 & CsI/1.73 & \citet{Oberg2007}\\
H$_2$O & 10000 & 3.5e17 & 2.0 & ... & 15$-$135 & CsI/1.73 & \citet{Oberg2007}\\
H$_2$O ($\leadsto$) & 500 & 4.4e17 & 1.0 & ... & 10/5s$-$1h & CsI/1.73 & \citet{Gerakines1996} \\
H$_2$O ($\leadsto$) & 500 & 4.4e17& 1.0 & ... & 25$-$160/1h & CsI/1.73 & \citet{Gerakines1996} \\
CO & 600 & 6e17 & 0.5 & ... & 15$-$45 & CsI/1.73 & \citet{vanBroekhuizen2006} \\
CO ($\leadsto$) & 500 & 5e17& 1.0 & ... & 10/5s$-$1h & CsI/1.73 & \citet{Gerakines1996} \\
CO ($\leadsto$) & 500 & 5e17& 1.0 & ... & 26$-$105/1h & CsI/1.73 & \citet{Gerakines1996} \\
CO$_2$ & 600 & 6e17 & 0.5 & ... & 15$-$130 & CsI/1.73 & \citet{vanBroekhuizen2006} \\
CO$_2$ ($\leadsto$) & 500 & 5e17 & 1.0 & ... & 10/5s$-$1h & CsI/1.73  & \citet{Gerakines1996} \\
CO$_2$ ($\leadsto$) & 500 & 5e17 & 1.0 & ... & 30,70/1h & CsI/1.73 & \citet{Gerakines1996} \\
CH$_4$ ($\leadsto$) & 500 & 5e17 & 1.0 & ... & 10/5s$-$1h & CsI/1.73 & \citet{Gerakines1996} \\
CH$_4$ ($\leadsto$) & 500 & 5e17 & 1.0 & ... & 25$-$280/1h & CsI/1.73 & \citet{Gerakines1996} \\
NH$_3$ & 500 & 5e17 & 1.0 & ... & 10 & CsI/1.73 & \citet{Gerakines1996} \\
NH$_3$ ($\leadsto$) & 500 & 5e17 & 1.0 & ... & 25$-$280/1h & CsI/1.73 & \citet{Gerakines1996} \\
NH$_3$ ($\leadsto$) & 500 & 5e17 & 1.0 & ... & 27,60/5s$-$1h & CsI/1.73 & \citet{Gerakines1996} \\
SO$_2$ & 4000 & 4.5e17 & 1.0 & ... & 10 & CsI/1.73 & \citet{Boogert1997} \\
H$_2$CO & 500 & 3.7e18 & 1.0 & ... & 10 & CsI/1.73 & \citet{Gerakines1996} \\
H$_2$CO ($\leadsto$) & 500 & 5.1e17 & 1.0 & ... & 10/5s$-$1h & CsI/1.73 & \citet{Gerakines1996} \\
H$_2$CO ($\leadsto$) & 500 & 1.1e17 & 1.0 & ... & 30$-$275/1h & CsI/1.73 & \citet{Gerakines1996} \\
NH$_4^+$ & <500 & 3e18 & 1.0 & ... & 80 & CsI/1.73 & \citet{Novozamsky2001}\\
OCN$^-$ & <500 & <5e17& 1.0 & ... & 80 & CsI/1.73 & \citet{Novozamsky2001}\\
OCS & 620 & 6.2e17 & 0.9 & ... & 10$-$170 & KBr/1.54 & Rachid et al. (in prep.)\\
CH$_3$OH  & 3400 & 3.4e18 & 0.5 & ... & 15 & ZnSe/2.54 & \citet{Scheltinga2018}\\
CH$_3$OH  & 4000 & 6.1e16 & 0.5 & ... & 15$-$160 & ... & \citet{Fraser2004}\\
CH$_3$OH ($\leadsto$) & 500 & 5e17& 1.0 & ... & 10/5s$-$1h & CsI/1.73 & \citet{Gerakines1996} \\
CH$_3$OH ($\leadsto$) & 500 & 5e17& 1.0 & ... & 25$-$230/1h & CsI/1.73 & \citet{Gerakines1996} \\
HCOOH  & 900 & 1.4e17 & 1.0 & ... & 30$-$105 & CsI/1.73 & \citet{Bisschop2007}\\
HCOOH  & 900 & 1.4e17& 1.0 & ... & 145 (deposition) & CsI/1.73 & \citet{Bisschop2007}\\
CH$_3$CHO  & 4500 & 9.7e18 & 1.0 & ... & 30$-$120 & ZnSe/2.54 & \citet{Scheltinga2018} \\
CH$_3$OCH$_3$  & 4500 & 2.9e18 & 1.0 & ... & 30$-$100 & ZnSe/2.54 & \citet{Scheltinga2018}\\
CH$_3$CH$_2$OH  & 4500 & 3e18 & 1.0 & ... &  30$-$150 & ZnSe/2.54 & \citet{Scheltinga2018}\\
CH$_3$COCH$_3$ & 2800 & 2.1e18 & 0.5 & ... & 15$-$140 & ZnSe/2.54 & \citet{Rachid2020}\\
CH$_3$OCHO & 2000 & 2.2e18 & 0.5 & ... & 15$-$120 & ZnSe/2.54 & \citet{Scheltinga2021}\\
CH$_3$NH$_2$ & 850 & 1.6e18 & 0.5 & ... & 15$-$140 & ZnSe/2.54 & \citet{Rachid2021}\\
CH$_3$CN & 5000 &3.0e18 & 1.0 & ... & 15$-$150 & KBr/1.54 & \citet{Rachid2022}\\
\hline
\multicolumn{8}{c}{\bf{Binary mixtures}}\\
\hline
H$_2$O:CO & 2500 & 5.1e17 & 1.0 & 1:100 & 15,30 & CsI/1.73 & \citet{Ehrenfreund1997}\\
H$_2$O:CO & 2500 & 4.2e17 & 1.0 & 100:14 & 10 & CsI/1.73 & \citet{Ehrenfreund1997}\\
CO:OCS & ... & ... & 0.9 & 20:1 & 11,20 & KBr/1.54 & Rachid et al. (in prep.)\\
CO$_2$:OCS & ... & ... & 0.9 & 24:1 & 11$-$70 & KBr/1.54 & Rachid et al. (in prep.)\\
H$_2$O:OCS & ... & ... & 0.9 & 20:1 & 11$-$120 & KBr/1.54 & Rachid et al. (in prep.)\\
CO:O$_2$ & 2500 & 1.2e17 & 1.0 & 100:50 & 10,35 & CsI/1.73 & \citet{Ehrenfreund1997}\\
CO:O$_2$ & 2500 & 1e17 & 1.0 & 100:70 & 10 & CsI/1.73 & \citet{Ehrenfreund1997}\\
H$_2$O:CO$_2$ & 2500 &3.9e17 & 1.0 & 1:100 & 10,30 & CsI/1.73 & \citet{Ehrenfreund1997}\\
H$_2$O:CO$_2$ & 2500 &1.7e17 & 1.0 & 1:10 & 10,80 & CsI/1.73 & \citet{Ehrenfreund1997}\\
H$_2$O:CO$_2$ & 2500 &1.2e17 & 1.0 & 1:6 & 10$-$75 &CsI/1.73 & \citet{Ehrenfreund1997}\\
H$_2$O:CO$_2$ & 2500 &4.6e17 & 1.0 & 100:14 & 10 & CsI/1.73& \citet{Ehrenfreund1997}\\
CO:CO$_2$ & 2500 &2e17 & 1.0 & 100:4 & 10,30 & CsI/1.73& \citet{Ehrenfreund1997}\\
CO:CO$_2$ & 2500 & 1.5e17& 1.0 & 100:8 & 10,30 & CsI/1.73& \citet{Ehrenfreund1997}\\
CO:CO$_2$ & 2500 & 1.4e17 & 1.0 & 100:16 & 10,30 &CsI/1.73 & \citet{Ehrenfreund1997}\\
CO:CO$_2$ & 2500 & 1.4e17 & 1.0 & 100:21 & 10,30 &CsI/1.73 & \citet{Ehrenfreund1997}\\
CO:CO$_2$ & 2500 & 1e17 & 1.0 & 100:23 & 10,30 &CsI/1.73 & \citet{Ehrenfreund1997}\\
CO:CO$_2$ & 2500 & 1.3e17 & 1.0 & 100:26 & 10,30 & CsI/1.73& \citet{Ehrenfreund1997}\\
CO$_2$:O$_2$ & 2500 &1e17 & 1.0 & 1:1 & 10 & CsI/1.73& \citet{Ehrenfreund1997}\\
HCOOH:CH$_3$OH & 1800 & 8.3e17 & 1.0 & 1:9 & 15$-$75 & CsI/1.73 & \citet{Bisschop2007}\\
HCOOH:CO & 1800 & 1.25e18 & 1.0 & 1:9 & 15$-$165 & CsI/1.73 & \citet{Bisschop2007}\\
HCOOH:H$_2$O & 900 & 5.5e17 & 1.0 & 0.25:1 & 15$-$165 & CsI/1.73 & \citet{Bisschop2007}\\
HCOOH:H$_2$O & 900 & 3.7e17 & 1.0 & 0.5:1 & 15$-$165 & CsI/1.73 & \citet{Bisschop2007}\\
HCOOH:H$_2$O & 900 & 5.5e17 & 1.0 & 1:1 & 15$-$165 & CsI/1.73 & \citet{Bisschop2007}\\
HCOOH:H$_2$O & 900 & 6e17 & 1.0 & 0.1:1 & 15$-$165 & CsI/1.73 & \citet{Bisschop2007}\\
HCOOH:CO$_2$ & 900 & 6e17 & 1.0 & 0.1:1 & 15$-$165 & CsI/1.73 & \citet{Bisschop2007}\\
H$_2$O:C$^{18}$O$_2$ & 2000 & 1.1e17 & 2.0 & 1:1 & 15$-$135 & CsI/1.73 & \citet{Oberg2007}\\
H$_2$O:C$^{18}$O$_2$ & 6000 & 1.95e17 & 2.0 & 1:1 & 15$-$135 & CsI/1.73 & \citet{Oberg2007}\\
H$_2$O:C$^{18}$O$_2$ & 20000 & 6.2e17 & 2.0 & 1:1 & 15$-$135 & CsI/1.73 & \citet{Oberg2007}\\
H$_2$O:C$^{18}$O$_2$ & 9000 & 1.3e17 & 2.0 & 1:2 & 15$-$135 & CsI/1.73 & \citet{Oberg2007}\\
H$_2$O:C$^{18}$O$_2$ & 30000 & 6.4e17 & 2.0 & 1:2 & 15$-$135 & CsI/1.73 & \citet{Oberg2007}\\
H$_2$O:C$^{18}$O$_2$ & 15000 & 1e18 & 2.0 & 2:1 & 15$-$135 & CsI/1.73 & \citet{Oberg2007}\\
H$_2$O:C$^{18}$O$_2$ & 4500 & 2.8e17 & 2.0 & 2:1 & 15$-$135 & CsI/1.73 & \citet{Oberg2007}\\
H$_2$O:C$^{18}$O$_2$ & 3750 & 1.3e17 & 2.0 & 4:1 & 15$-$135 & CsI/1.73 & \citet{Oberg2007}\\
H$_2$O:C$^{18}$O$_2$ & 15000 & 2.76e18 & 2.0 & 1:4 & 15$-$135 & CsI/1.73 & \citet{Oberg2007}\\
H$_2$O:CO$_2$ & 500 &1.65e18 & 1.0 & 10:1 & 10$-$185 & CsI/1.73 & \citet{Ehrenfreund1999} \\
H$_2$O:CO$_2$ & 500 & 5.75e17 & 1.0 & 1:1 & 10$-$187 & CsI/1.73 & \citet{Ehrenfreund1999} \\
CO$_2$:CH$_3$OH & 500 & 6e17 & 1.0 & 10:1 & 10$-$75 & CsI/1.73 & \citet{Ehrenfreund1999} \\
CO$_2$:CH$_3$OH & 500 & 4.6e17 & 1.0 & 3:1 & 10$-$130 & CsI/1.73 & \citet{Ehrenfreund1999} \\
CO$_2$:CH$_3$OH & 500 & 3.7e17 & 1.0 & 2:1 & 10$-$145 & CsI/1.73 & \citet{Ehrenfreund1999} \\
CO$_2$:CH$_3$OH & 500 & 2.5e17 & 1.0 & 1:1 & 10$-$145 & CsI/1.73 & \citet{Ehrenfreund1999} \\
CO$_2$:CH$_3$OH & 500 & 4.9e17 & 1.0 & 1:2 & 10$-$155 & CsI/1.73 & \citet{Ehrenfreund1999} \\
CO$_2$:CH$_3$OH & 500 & 5.75e17 & 1.0 & 1:3 & 10$-$160 & CsI/1.73 & \citet{Ehrenfreund1999} \\
CO$_2$:CH$_3$OH & 500 &6.4e17 & 1.0 & 1:10 & 10$-$180 & CsI/1.73 & \citet{Ehrenfreund1999} \\
CH$_3$OH:SO$_2$ & ... & 2.7e17 & 1.0 & 1:1 & 10 & CsI/1.73& \citet{Boogert1997} \\
CH$_3$OH:SO$_2$ & ... & 5e17 & 1.0 & 11:1 & 10 & CsI/1.73& \citet{Boogert1997} \\
CO:CO$_2$ & 8000 & 1e17 & 0.5 & 1:1 & 15$-$100 & ... & \citet{Fraser2004} \\
CO over HCOOH & 16000 & 9.7e16 & 0.5 & ... & 15$-$160 & ... & \citet{Fraser2004} \\
CO under HCOOH & 16000 & 9.7e16 & 0.5 & ... & 15$-$160 & ... & \citet{Fraser2004} \\
CO over CO$_2$ & 16000 & 3.1e17 & 0.5 & ... & 15$-$160 & ... & \citet{Fraser2004} \\
CO under CO$_2$ & 16000 & 4.3e17 & 0.5 & ... & 15$-$100 & ... & \citet{Fraser2004} \\
CO:HCOOH & 8000 & 1.2e17 &  0.5 & 1:1 & 15$-$160 & ... & \citet{Fraser2004} \\
CO under CH$_3$OH & 16000 & 7.4e17 & 0.5 & ... & 15$-$160 & ... & \citet{Fraser2004} \\
CO over CH$_3$OH & 16000 & 7.4e17 & 0.5 & ... & 15$-$160 & ... & \citet{Fraser2004} \\
CO over CH$_4$ & 8000 & 2.4e17 & 0.5 & ... & 15$-$40 & ... & \citet{Fraser2004} \\
CO under CH$_4$ & 8000 & 2.4e17 & 0.5 & ... & 15$-$40 & ... & \citet{Fraser2004} \\
CO over CO$_2$ & 1200 & 1e17 & 0.5 & 1:1 & 15$-$110 & CsI/1.73 & \citet{vanBroekhuizen2006} \\
CO over CO$_2$ & 1800 & 1.6e17 & 0.5 & 1:2 & 15$-$110 & CsI/1.73 & \citet{vanBroekhuizen2006} \\
CO$_2$ over CO & 1200 & 9.8e16 & 0.5 & 1:1 & 15$-$110 & CsI/1.73 & \citet{vanBroekhuizen2006} \\
CO$_2$ over CO & 1800 & 3e17 & 0.5 & 2:1 & 15$-$110 & CsI/1.73 & \citet{vanBroekhuizen2006} \\
CO$_2$ over CO & 2400 & 1.9e17 & 0.5 & 3:1 & 15$-$110 & CsI/1.73 & \citet{vanBroekhuizen2006} \\
CO$_2$ over CO & 6600 & 6.7e17 & 0.5 & 10:1 & 15$-$110 & CsI/1.73 & \citet{vanBroekhuizen2006} \\
CO:CO$_2$ & 1200 & 4.4e17 & 0.5 & 1:1 & 15$-$130 & CsI/1.73 & \citet{vanBroekhuizen2006} \\
CO:CO$_2$ & 1800 & 1.9e17 & 0.5 & 2:1 & 15$-$130 & CsI/1.73 & \citet{vanBroekhuizen2006} \\
CO:CO$_2$ & 6600 & 5.5e17 & 0.5 & 1:10 & 15$-$130 & CsI/1.73 & \citet{vanBroekhuizen2006} \\
H$_2$O:CH$_3$CHO & 2000 & 3.4e18 & 1.0 & 20:1 & 15$-$160 & ZnSe/2.54 & \citet{Scheltinga2018} \\
CO:CH$_3$CHO & 2000 & 3.1e18 & 1.0 & 20:1 & 15$-$160 & ZnSe/2.54 & \citet{Scheltinga2018} \\
CH$_3$OH:CH$_3$CHO & 2000 & 1.4e18 & 1.0 & 20:1 & 15$-$140 & ZnSe/2.54 & \citet{Scheltinga2018} \\
H$_2$O:CH$_3$CH$_2$OH & 2000 & 2.5e18 & 1.0 & 20:1 & 15$-$160 & ZnSe/2.54 & \citet{Scheltinga2018} \\
CO:CH$_3$CH$_2$OH & 2000 & 2.6e18 & 1.0 & 20:1 & 15, 30 & ZnSe/2.54 & \citet{Scheltinga2018} \\
CH$_3$OH:CH$_3$CH$_2$OH & 2000 & 4.6e18 & 1.0 & 20:1 & 15$-$150 & ZnSe/2.54 & \citet{Scheltinga2018} \\
H$_2$O:CH$_3$OCH$_3$ & 2000 & 2.3e18 & 1.0 & 20:1 & 15$-$160 & ZnSe/2.54 & \citet{Scheltinga2018} \\
CO:CH$_3$OCH$_3$ & 2000 & 2.7e18 & 1.0 & 20:1 & 15$-$120 & ZnSe/2.54 & \citet{Scheltinga2018} \\
CH$_3$OH:CH$_3$OCH$_3$ & 2000 & 3.8e18 & 1.0 & 20:1 & 15$-$120 & ZnSe/2.54 & \citet{Scheltinga2018} \\
H$_2$O:CH$_3$COCH$_3$ & 3500 &2e18 & 0.5 & 5:1 & 15$-$160 & ZnSe/2.54 & \citet{Rachid2020} \\
H$_2$O:CH$_3$COCH$_3$ & 3500 &2.6e18 & 0.5 & 20:1 & 15$-$160 & ZnSe/2.54 & \citet{Rachid2020} \\
CO:CH$_3$COCH$_3$ & 3500 & 1.9e18 & 0.5 & 5:1 & 15, 30 & ZnSe/2.54 & \citet{Rachid2020} \\
CO:CH$_3$COCH$_3$ & 3500 & 2e18 & 0.5 & 20:1 & 15, 30 & ZnSe/2.54 & \citet{Rachid2020} \\
CO$_2$:CH$_3$COCH$_3$ & 3500 & 1.3e18 & 0.5 & 5:1 & 15$-$100 & ZnSe/2.54 & \citet{Rachid2020} \\
CO$_2$:CH$_3$COCH$_3$ & 3500 & 1.4e18 & 0.5 & 20:1 & 15$-$100 & ZnSe/2.54 & \citet{Rachid2020} \\
CH$_3$OH:CH$_3$COCH$_3$ & 3500 & 2.6e18 & 0.5 & 5:1 & 15$-$140 & ZnSe/2.54 & \citet{Rachid2020} \\
CH$_3$OH:CH$_3$COCH$_3$ & 3500 & 5.5e18 & 0.5 & 20:1 & 15$-$140 & ZnSe/2.54 & \citet{Rachid2020} \\
CH$_3$NH$_2$:H$_2$O & 3500 & 2.2e18 & 0.5 & 1:5 & 15$-$150 & ZnSe/2.54 & \citet{Rachid2021} \\
CH$_3$NH$_2$:H$_2$O & 3500 & 2.5e18 & 0.5 & 1:10 & 15$-$150 & ZnSe/2.54 & \citet{Rachid2021} \\
CH$_3$NH$_2$:H$_2$O & 3500 &2.6e18 & 0.5 & 1:20 & 15$-$150 & ZnSe/2.54 & \citet{Rachid2021} \\
CH$_3$NH$_2$:CH$_4$ & 3500 &1.9e18 & 0.5 & 1:5 & 15$-$45 & ZnSe/2.54 & \citet{Rachid2021} \\
CH$_3$NH$_2$:CH$_4$ & 3500 &2e18 & 0.5 & 1:10 & 15$-$45 & ZnSe/2.54 & \citet{Rachid2021} \\
CH$_3$NH$_2$:CH$_4$ & 3500 &2e18 & 0.5 & 1:20 & 15$-$45 & ZnSe/2.54 & \citet{Rachid2021} \\
CH$_3$NH$_2$:NH$_3$ & 3500 &2.2e18 & 0.5 & 1:5 & 15$-$115 & ZnSe/2.54 & \citet{Rachid2021} \\
CH$_3$NH$_2$:NH$_3$ & 3500 &2.4e18 & 0.5 & 1:10 & 15$-$115 & ZnSe/2.54 & \citet{Rachid2021} \\
CH$_3$NH$_2$:NH$_3$ & 3500 &2.6e18 & 0.5 & 1:20 & 15$-$115 & ZnSe/2.54 & \citet{Rachid2021} \\
CH$_3$OCHO:CO & 2000 & 1.8e18 & 0.5 & 1:20 & 15$-$120 & ZnSe/2.54 & \citet{Scheltinga2021}\\
CH$_3$OCHO:H$_2$CO & 2000 & 2.2e18 & 0.5 & 1:20 & 15$-$120 & ZnSe/2.54 & \citet{Scheltinga2021}\\
CH$_3$OCHO:CH$_3$OH & 2000 & 1.6e18 & 0.5 & 1:20 & 15$-$120 & ZnSe/2.54 & \citet{Scheltinga2021}\\
CH$_3$OCHO:H$_2$O & 2000 & 1.5e18 & 0.5 & 1:20 & 15$-$120 & ZnSe/2.54 & \citet{Scheltinga2021}\\
CH$_3$CN:H$_2$O & 5000 & 1.7e18 & 1.0 & 1:5 & 15$-$150 & Ge/4.0& \citet{Rachid2022}\\
CH$_3$CN:H$_2$O & 5000 & 2.3e18 & 1.0 & 1:10 & 15$-$150 & Ge/4.0& \citet{Rachid2022}\\
CH$_3$CN:H$_2$O & 5000 & 1.6e18 & 1.0 & 1:20 & 15$-$150 & Ge/4.0& \citet{Rachid2022}\\
CH$_3$CN:CO & 5000 & 1.3e18 & 1.0 & 1:5 & 15,30 & Ge/4.0& \citet{Rachid2022}\\
CH$_3$CN:CO & 5000 & 1.6e18 & 1.0 & 1:10 & 15,30 & Ge/4.0& \citet{Rachid2022}\\
CH$_3$CN:CO$_2$ & 5000 & 1.2e18 & 1.0 & 1:5 & 15$-$150 & Ge/4.0& \citet{Rachid2022}\\
CH$_3$CN:CO$_2$ & 5000 & 1.8e18 & 1.0 & 1:10 & 15$-$150 &Ge/4.0 & \citet{Rachid2022}\\
CH$_3$CN:NH$_3$ & 5000 & 2e18 & 1.0 & 1:5 & 15$-$150 & Ge/4.0& \citet{Rachid2022}\\
CH$_3$CN:NH$_3$ & 5000 & 2.1e18 & 1.0 & 1:10 & 15$-$150 & Ge/4.0& \citet{Rachid2022}\\
CH$_3$CN:NH$_3$ & 5000 & 2.1e18 & 1.0 & 1:20 & 15$-$150 & Ge/4.0& \citet{Rachid2022}\\
\hline
\multicolumn{8}{c}{\bf{Tertiary mixtures}}\\
\hline
HCOOH:H$_2$O:CO$_2$ & 1800 & 1.5e18 &1.0  & 0.1:1:0.4 & 15 & CsI/1.73 & \citet{Bisschop2007}\\
HCOOH:H$_2$O:CH$_3$OH & 1800 & 1.6e18 & 1.0 & 0.1:1:0.4 & 15 & CsI/1.73 & \citet{Bisschop2007}\\
HCOOH:H$_2$O:CO & 1800 & 1.74e18 & 1.0 & 0.1:1:0.4 & 15 & CsI/1.73 & \citet{Bisschop2007}\\
H$_2$O:CO:O$_2$ & 2500 & 3.8e17 & 1.0 & 1:80:20 & 10,30 & CsI/1.73& \citet{Ehrenfreund1997}\\
H$_2$O:CO:CO$_2$ & 2500 & 3.3e17 & 1.0 & 1:50:50 & 10,30 & CsI/1.73& \citet{Ehrenfreund1997}\\
H$_2$O:CO:CO$_2$ & 2500 & 3.7e18 & 1.0 & 1:50:56 & 10,45 & CsI/1.73& \citet{Ehrenfreund1997}\\
CO:O$_2$:CO$_2$ & 2500 & 1.2e17 & 1.0 & 100:50:4 & 10,30 & CsI/1.73& \citet{Ehrenfreund1997}\\
CO:O$_2$:CO$_2$ & 2500 & 1.2e17 & 1.0 & 100:50:8 & 10 & CsI/1.73& \citet{Ehrenfreund1997}\\
CO:O$_2$:CO$_2$ & 2500 & 1.2e17 & 1.0 & 100:50:16 & 10,30 &CsI/1.73 & \citet{Ehrenfreund1997}\\
CO:O$_2$:CO$_2$ & 2500 & 1.2e17 & 1.0 & 100:50:21 & 10,30 &CsI/1.73 & \citet{Ehrenfreund1997}\\
CO:O$_2$:CO$_2$ & 2500 & 1.2e17 & 1.0 & 100:50:32 & 10 & CsI/1.73& \citet{Ehrenfreund1997}\\
CO:O$_2$:CO$_2$ & 2500 & 1.2e17 & 1.0 & 100:54:10 & 10,30 &CsI/1.73 & \citet{Ehrenfreund1997}\\
CO:O$_2$:CO$_2$ & 2500 &1.2e17 & 1.0 & 100:20:11 & 10,30 & CsI/1.73& \citet{Ehrenfreund1997}\\
CO:O$_2$:CO$_2$ & 2500 &1.2e17 & 1.0 & 100:11:20 & 10,30 & CsI/1.73& \citet{Ehrenfreund1997}\\
CO:O$_2$:CO$_2$ & 2500 & 1.2e17 & 1.0 & 100:10:23 & 10,30 & CsI/1.73& \citet{Ehrenfreund1997}\\
H$_2$O:CO:N$_2$ & 2500 & 2e17 & 1.0 & 1:40:50 & 10,30 & CsI/1.73& \citet{Ehrenfreund1997}\\
CO:N$_2$:CO$_2$ & 2500 &1.2e17 & 1.0 & 100:50:20 & 10,30 & CsI/1.73& \citet{Ehrenfreund1997}\\
H$_2$O:CO$_2$:CO & 2500 &2.7e17 & 1.0 & 100:20:3 & 20 & CsI/1.73& \citet{Ehrenfreund1997}\\
H$_2$O:CH$_3$OH:CO$_2$ & 500 &1.2e18 & 1.0 & 9:1:2 & 10$-$185 & CsI/1.73 & \citet{Ehrenfreund1999}\\
H$_2$O:CH$_3$OH:CO$_2$ & 500 &2.8e17 & 1.0 & 0.2:0.6:1 & 10$-$140 & CsI/1.73 & \citet{Ehrenfreund1999}\\
H$_2$O:CH$_3$OH:CO$_2$ & 500 &2.8e17 & 1.0 & 0.4:0.6:1 & 10$-$140 & CsI/1.73 & \citet{Ehrenfreund1999}\\
H$_2$O:CH$_3$OH:CO$_2$ & 500 & 6e17& 1.0 & 1:0.6:1 & 10$-$180 & CsI/1.73 & \citet{Ehrenfreund1999}\\
H$_2$O:CH$_3$OH:CO$_2$ & 500 &9.8e17 & 1.0 & 0.7:0.7:1 & 10$-$146 & CsI/1.73 & \citet{Ehrenfreund1999}\\
H$_2$O:CH$_3$OH:CO$_2$ & 500 &1.5e18 & 1.0 & 0.8:0.9:1 & 10$-$135 & CsI/1.73 & \citet{Ehrenfreund1999}\\
H$_2$O:CH$_3$OH:CO$_2$ & 500 &2.7e18 & 1.0 & 1:1:1 & 10$-$145 & CsI/1.73 & \citet{Ehrenfreund1999}\\
H$_2$O:CH$_3$OH:CO$_2$ & 500 &4.9e17 & 1.0 & 0.7:1:1 & 10$-$120 & CsI/1.73 & \citet{Ehrenfreund1999}\\
H$_2$O:CH$_3$OH:CO$_2$ & 500 &1.8e16 & 1.0 & 0.6:1:0.8 & 10$-$121 & CsI/1.73 & \citet{Ehrenfreund1999}\\
H$_2$O:CH$_3$OH:CO$_2$ & 500 &7.8e17 & 1.0 & 1.2:0.7:1.0 & 10$-$119 & CsI/1.73 & \citet{Ehrenfreund1999}\\
H$_2$O:CH$_3$OH:CO$_2$ & 500 &3e17 & 1.0 & 0.7:0.9:1.0 & 10$-$134 & CsI/1.73 & \citet{Ehrenfreund1999}\\
H$_2$O:CH$_3$OH:CO$_2$ & 500 &5e16 & 1.0 & 0.5:1:1 & 10 & CsI/1.73 & \citet{Ehrenfreund1999}\\
H$_2$O:CH$_3$OH:CO$_2$ & 500 &9.7e16 & 1.0 & 0.9:1.4:1 & 10$-$125 &CsI/1.73 & \citet{Ehrenfreund1999}\\
H$_2$O:CH$_3$OH:CO$_2$ & 500 &7.9e16 & 1.0 & 0.2:0.5:1 & 10, 98 & CsI/1.73 & \citet{Ehrenfreund1999}\\
H$_2$O:CH$_3$OH:CO$_2$ & 500 &7.7e16 & 1.0 & 0.3:0.5:1 & 10$-$95 & CsI/1.73 & \citet{Ehrenfreund1999}\\
H$_2$O:CH$_3$OH:CO$_2$ & 500 &7.1e16 & 1.0 & 0.3:0.7:1 & 10$-$82 & CsI/1.73 & \citet{Ehrenfreund1999}\\
H$_2$O:CH$_3$OH:CO$_2$ & 500 &1.2e18 & 1.0 & 1.1:1.2:1 & 10$-$131 & CsI/1.73 & \citet{Ehrenfreund1999}\\
H$_2$O:CH$_3$OH:CO$_2$ & 500 & 3e17 & 1.0 & 0.7:0.9:1 & 10$-$134 & CsI/1.73 & \citet{Ehrenfreund1999}\\
H$_2$O:CH$_3$OH:CO$_2$ & 500 &7.2e17 & 1.0 & 0.9:0.3:1 & 10$-$115 & CsI/1.73 & \citet{Ehrenfreund1999}\\
CO:CH$_3$OH:CH$_3$CHO & 2000 &1.8e18 & 1.0 & 20:20:1 & 15$-$120 & ZnSe/2.54 & \citet{Scheltinga2018}\\
CO:CH$_3$OH:CH$_3$CH$_2$OH & 2000 &1.8e18 & 1.0 & 20:20:1 & 15$-$150 & ZnSe/2.54 & \citet{Scheltinga2018}\\
CO:CH$_3$OH:CH$_3$OCH$_3$ & 2000 &1.7e18 & 1.0 & 20:20:1 & 15$-$100 & ZnSe/2.54 & \citet{Scheltinga2018}\\
CH$_3$COCH$_3$:H$_2$O:CO$_2$ & 3500 & 7.2e17 & 0.5 & 1:2.5:2.5 & 15$-$160 & ZnSe/2.54 & \citet{Rachid2020}\\
CH$_3$COCH$_3$:H$_2$O:CO$_2$ & 3500 & 6.8e17 & 0.5 & 1:10:10 & 15$-$160 & ZnSe/2.54 & \citet{Rachid2020}\\
CH$_3$COCH$_3$:CO:CH$_3$OH & 3500 & 9.7e17  & 0.5 & 1:2.5:2.5 & 15$-$140 & ZnSe/2.54 & \citet{Rachid2020}\\
CH$_3$COCH$_3$:CO:CH$_3$OH & 3500 & 1e18 & 0.5 & 1:10:10 & 15$-$140 & ZnSe/2.54 & \citet{Rachid2020}\\
CH$_3$NH$_2$:H$_2$O:CH$_4$ & 2500 &1.1e18 & 0.5 & 1:5:5 & 15$-$150 & ZnSe/2.54 & \citet{Rachid2021}\\
CH$_3$NH$_2$:H$_2$O:CH$_4$ & 2500 &1.2e18 & 0.5 & 1:10:10 & 15$-$150 & ZnSe/2.54 & \citet{Rachid2021}\\
CH$_3$NH$_2$:H$_2$O:NH$_3$ & 2500 &1.2e18 & 0.5 & 1:5:5 & 15$-$150 & ZnSe/2.54 & \citet{Rachid2021}\\
CH$_3$NH$_2$:H$_2$O:NH$_3$ & 2500 &1.3e18 & 0.5 & 1:10:10 & 15$-$150 & ZnSe/2.54 & \citet{Rachid2021}\\
CH$_3$NH$_2$:CH$_4$:NH$_3$ & 2500 &1.1e18 & 0.5 & 1:5:5 & 15$-$115 & ZnSe/2.54 & \citet{Rachid2021}\\
CH$_3$NH$_2$:CH$_4$:NH$_3$ & 2500 &1.2e18 & 0.5 & 1:10:10 & 15$-$115 & ZnSe/2.54 & \citet{Rachid2021}\\
CH$_3$CN:H$_2$O:CO$_2$ & 5000 & 1.3e18 & 1.0 & 1:5:2 & 15$-$150 & Ge/4.0 & \citet{Rachid2022}\\
\hline
\multicolumn{8}{c}{\bf{Quaternary mixtures}}\\
\hline
H$_2$O:CO:O$_2$:N$_2$ & 2500 &2.6e17 & 1.0 & 1:40:40:15 & 10,30 &CsI/1.73 & \citet{Ehrenfreund1997}\\
H$_2$O:CH$_3$OH:CO$_2$:NH$_3$ & 500 &3.1e17 & 1.0 & 0.7:0.7:1:0.7 & 10$-$104 & CsI/1.73 & \citet{Ehrenfreund1999}\\
H$_2$O:CH$_3$OH:CO$_2$:CH$_4$ & 500 &3.1e17 & 1.0 & 0.6:0.7:1:0.1 & 10$-$119 & CsI/1.73 & \citet{Ehrenfreund1999}\\
H$_2$O:CH$_3$OH:CO$_2$:CH$_4$ & 500 & 1e17 & 1.0 & 0.4:0.6:1:0.23 & 10 & CsI/1.73 & \citet{Ehrenfreund1999}\\
CO:O$_2$:N$_2$:CO$_2$ & 2500 &4.4e18 & 1.0 & 1:50:25:32 & 10,30 & CsI/1.73 & \citet{Ehrenfreund1997}\\
CH$_3$OCHO:CO:H$_2$CO:CH$_3$OH & 2000 & 8.9e17 & 0.5 & 1:20:20:20 & 15$-$120 & ZnSe/2.54 & \citet{Scheltinga2021}\\
CH$_3$NH$_2$:H$_2$O:CH$_4$:NH$_3$ & 3400 & 9.3e17 & 0.5 & 3:10:10:10 & 15$-$120 & ZnSe/2.54& \citet{Rachid2021}\\
CH$_3$CN:H$_2$O:CH$_4$:NH$_3$ & 5000 & 2.3e18 & 1.0 & 1:20:2:2 & 15$-$150 & Ge/4.0& \citet{Rachid2022}\\
\hline
\multicolumn{6}{c}{\bf{Five components mixture}}\\
\hline
H$_2$O:CO:O$_2$:N$_2$:CO$_2$ & 2500 & 3.8e17 & 1.0 & 1:50:35:15:3 & 10 & CsI/1.73& \citet{Ehrenfreund1997}\\
H$_2$O:CO:O$_2$:N$_2$:CO$_2$ & 500 & 8e16 & 1.0 & 1:50:35:15:3 & 10 & CsI/1.73& \citet{Ehrenfreund1997}\\
\hline
\end{longtable}
\tablefoot{$^a$ Except in the case of pure ices, $N_{\rm{ice}}$ values correspond to the major ice component.
}
\end{landscape}
}

\twocolumn

\section{Database design and back-end information}
\label{DB_design}
The structure of LIDA is built with \texttt{Flask}\footnote{\url{https://flask.palletsprojects.com/en/2.0.x/}} \citep{flask2018}, an open-source web framework written in Python. \texttt{Flask} is widely extensible in the sense that external software can be embedded in the web application. LIDA has two major interfaces that provide access to administrators and users, respectively. The user interface is described in Section~\ref{us_interface}. Here, we provide details about the administrator interface, which obviously only accessible via login and is restricted to collaborators and developers.

The administrator interface that provides access to all information hosted in the database, as well as the capability to add and modify data. In this module, the database is structured in a relation design between {\bf Analogues} and {\bf Spectrum}. {\it Analogues} are the name of the ice sample (e.g., Pure H$_2$O), whereas {\it Spectrum} is the IR spectrum of the analogue at a specific temperature (e.g., Pure H$_2$O at 15~K). Table~\ref{database_design} shows a scheme of the information contained in the database. All this information is also visible in the user interface, which is introduced in Section~\ref{us_interface}.


\begin{table*}
\caption{\label{database_design} Example of relational database for pure H$_2$O ice. All information is visible in the user interface.}
\renewcommand{\arraystretch}{1.1}
\scalebox{0.9}{
\centering 
\begin{tabular}{lccccccc}
\hline\hline
\multicolumn{7}{c}{\bf{Analogue}}\\
ID$^a$ & Name$^b$ & Deposition & Author & DOI & Upload date & Annotation$^b$\\
& & temperature (K)& & & &\\
\hline
14 & Pure H$_2$O~3000~ML & 15 & {\"O}berg et al. & 10.1051/0004-6361:20065881 & 2021-10-27 & Pure\_H2O\_3000\_ML.csv \\

\hline
\multicolumn{7}{c}{\bf{Spectrum}}\\
ID$^a$ & Temperature & Column density & Ice thickness & Resolution & Wavenumber range & File name$^c$\\
& (K) & (molec./cm$^{2}$) & ($\mu$m) & (cm$^{-1}$) & (cm$^{-1}$) & \\
\hline
14 & 15 & 1e17 & 3000 & 2.0 & 500$-$4000 & 96\_15.0K\\
\hline
\end{tabular}}
\tablefoot{$^a$ identifier number that relates the analogue to the spectrum.\\
$^b$ Annotation file containing the respective position and assignments of the vibration modes.\\
$^c$ Name of the file containing the wavenumber and absorbance of the ice sample when uploaded to the database. This file is stored in HDF5 format in LIDA, but available for download as \texttt{ascii} file to the user.\\
}
\end{table*}

In addition to the IR spectra of ice samples, the database hosts data of experimentally derived UV-vis refractive index values (optical constants), and calculated values for the mid-IR. The continuum spectral energy distribution (SED) of protostars is also hosted, but only accessible via the online tool \texttt{SPECFY}. The database files containing this information, are also structured in a relational design as used for {\it Analogues} and {\it Spectrum}. In common, the files containing the spectral data, optical constants, and continuum SED, are stored on the server using HDF5 (Hierarchical Data Format) format, that has been designed to store large amount of data. The web interface allows the administrator to upload the data as a simple two-column file. The first column (X-axis) is the wavenumber (cm$^{-1}$) for the absorbance and refractive index data, whereas wavelength ($\mu$m) is used for continuum SED data. Likewise, the second column (Y-axis) gives the physical quantities, such as absorbance, refractive index and flux in Jy, respectively. Python has a package supporting HDF5 called H5PY\footnote{\url{https://docs.h5py.org/en/stable/}}, which is used to generate compressed files in LIDA to improve the efficiency of the database. Despite the files being stored in HDF5 format, they are available for download by the user in ASCII files (\texttt{.txt} extension). For security reasons, LIDA performs a check when uploading data which consists of validating the file extension, structure and size. Absorbance data can be uploaded under the category of warm-up or irradiation time (exposition). Similarly, the refractive index is uploaded under the category of real or imaginary values. The continuum SED data is uploaded in the categories of polynomial or blackbody. 

The administrator module also contains the {\it access information tracker} that allows to track the number of accesses and downloads over the months and years. The goal of this feature is to check the impact of LIDA in providing the astronomical community with essential and accurate data to interpret telescope observations.

\section{Web interfaces of the online tools}
\label{app_specfy}
\begin{figure*}
   \centering
   \includegraphics[width=\hsize]{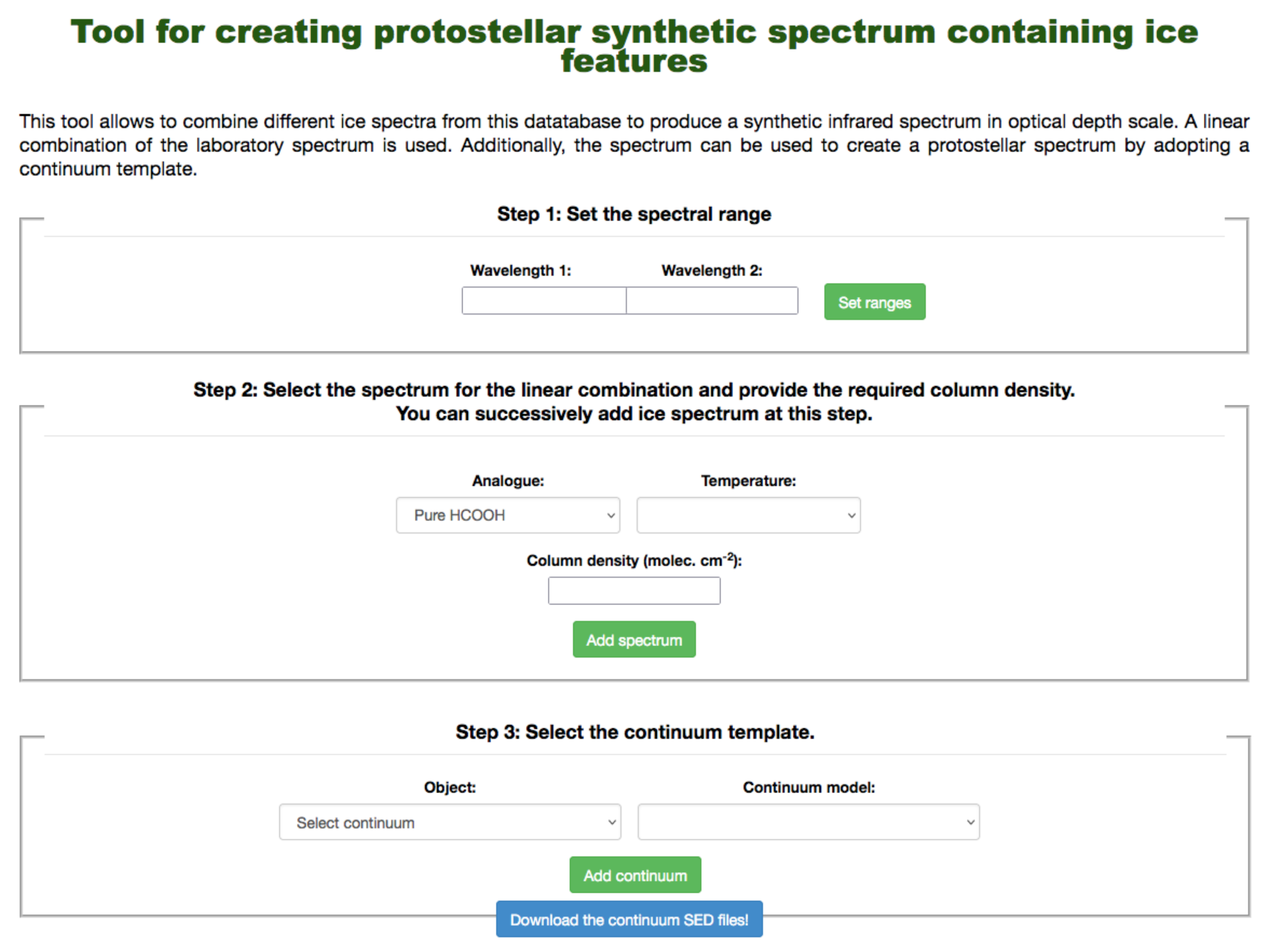}
      \caption{Screenshot of the web interface of SPECFY showing the three steps to create a synthetic protostar spectrum. The green buttons submit the information added to the white rectangles. In steps 2 and 3, the user can scroll and search for ice analogues and temperatures, and for continuum models, respectively. The blue button allows the user to download the continuum SED files.}
         \label{Specfy}
   \end{figure*}

\begin{figure*}
   \centering
   \includegraphics[width=\hsize]{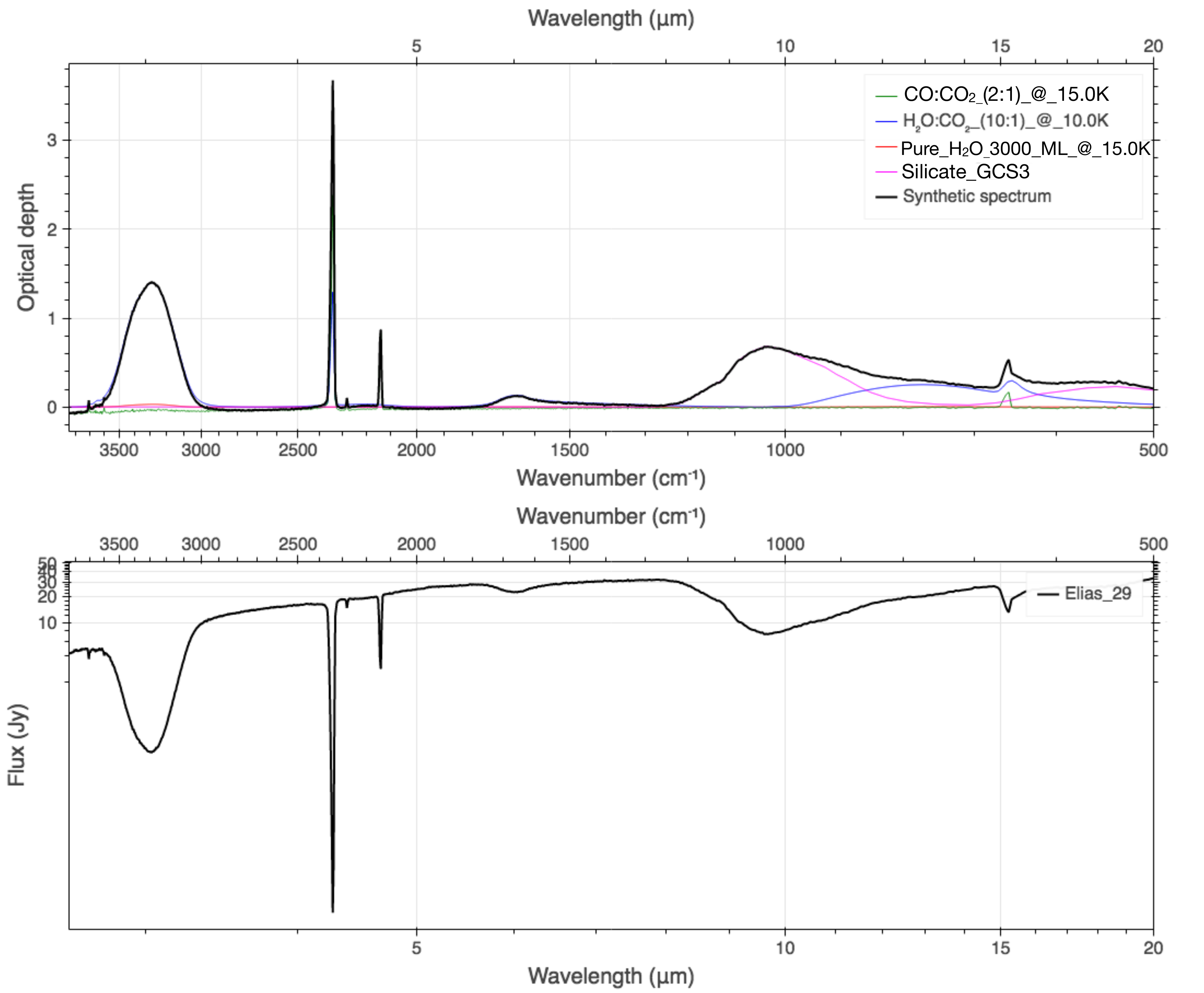}
      \caption{Screenshot of the outputs of the \texttt{SPECFY} online tool. {\it Top:} Synthetic spectrum in optical depth scale composed by the linear combination of three ice spectra (Pure H$_2$O, H$_2$O:CO$_2$, CO:CO$_2$) and silicate template from GCS~3 source. {\it Bottom:} Synthetic spectrum in flux scale adopting the Elias~29 protostar continuum SED, taken from \citet{Boogert2008}.}
         \label{synhtetic}
   \end{figure*}

The web interface of \texttt{SPECFY} is show in Figure~\ref{Specfy}. In ``Step 1'', \texttt{SPECFY} sets the wavelength range to create the synthetic spectrum. This step is crucial because some absorbance spectra in the database have different ranges. By setting the range, all the absorbance spectra selected in step 2 are evenly interpolated to ensure that all spectral components have the same range. Next, in ``Step 2''the laboratory ice spectrum from LIDA can be selected to be converted to an optical depth scale and combined to other ice spectrum. This step can be repeated multiple times. Finally, in ``Step 3'', the optical depth scale spectrum is converted to a spectrum in flux units based on the object and the continuum SED model adopted by the user. An example of the output files are shown in Figure~\ref{synhtetic}. The {\it top} panel shows the combined spectrum used to match the AFGL~989 {\it ISO} spectrum. The {\it bottom} panel displays the synthetic spectrum in flux scale that adopts the continuum SED of Elias~29 protostar.

\begin{figure*}
   \centering
   \includegraphics[width=\hsize]{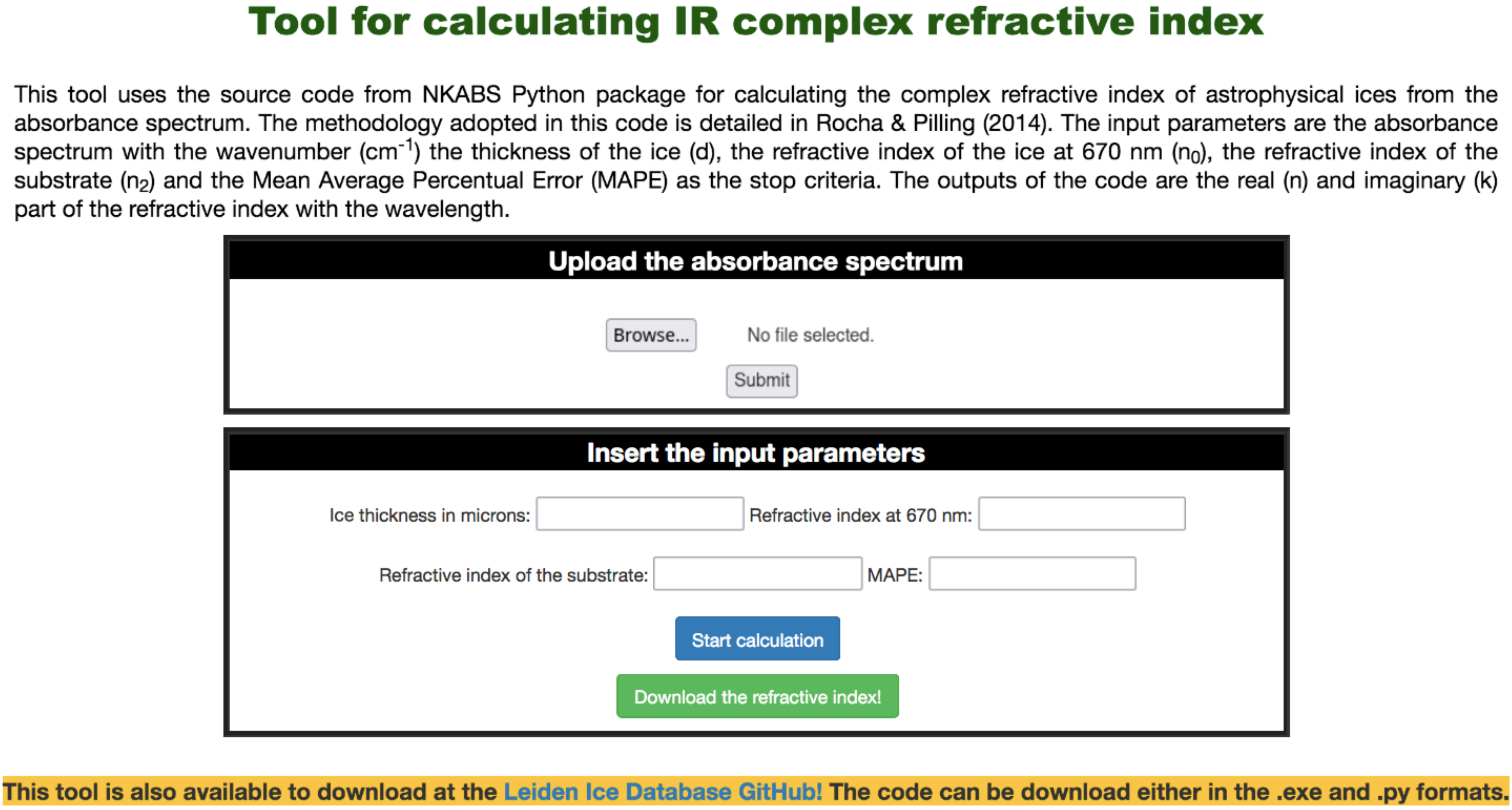}
      \caption{Screenshot of the web interface of the online tool for calculating the refractive index of ices. The user can upload the absorbance spectrum as input data, and provide the ice parameters. The calculations start by clicking on the blue button ``Start calculation'', and the files with the results are downloaded by clicking on the green button ``'Download the refractive index.' The bottom yellow box informs that other formats of this tool are available for download as well.}
         \label{icenk_page}
   \end{figure*}

Figure~\ref{icenk_page} shows the web interface of the refractive index calculator. This tool requires the upload of an external file containing the absorbance spectrum, which is done via the button ``Submit''. Next, the user is asked to parse the values of three physical parameters (ice thickness, $n_{670~\rm{nm}}$, $n_{\rm{subs}}$) and the stop criteria (MAPE). The calculations can be started by clicking on the blue button ``Start calculation'', and they roughly last 1$-$4 seconds for a spectrum with 18000 rows. The output data can be download by clicking on the green button ``Download the refractive index''. One of the outputs is called ``\texttt{lnk\_optool}''. This file contains the real and imaginary parts of the refractive index and are formatted to be used as input in the computational code \texttt{optool}\footnote{\url{https://github.com/cdominik/optool}} \citep{Dominik2021}, a command-line tool written in \texttt{Fortran}, which is dedicated to derive opacity of ice and bare grains.

\end{document}